\documentclass{iopart}
\usepackage{graphicx}
\usepackage{amssymb}
\usepackage{amsbsy}

\newcommand{\bds}{\boldsymbol}
\newcommand{\acrit}{a_{\rm crit}}
\newcommand{\chr}{$^{52}${\rm Cr}}
\newcommand{\kb}[0]{k_{\mathrm{B}}}
\newcommand{\mub}[0]{\mu_{\mathrm{B}}}
\newcommand{\cdd}[0]{C_{\rm dd}}
\newcommand{\add}[0]{a_{\rm dd}}
\newcommand{\edd}[0]{\varepsilon_{\rm dd}}
\newcommand{\udd}[0]{U_{\rm dd}}
\newcommand{\fdd}[0]{\Phi_{\rm dd}}
\newcommand{\ket}[1]{|#1\rangle}
\newcommand{\ddi}[0]{dipole-dipole interaction}

\eqnobysec

\begin{document}

\title[The physics of dipolar bosonic quantum gases]{The physics of dipolar bosonic quantum gases}

\author{T Lahaye$^{1,2,3}$, C Menotti$^4$, L Santos$^5$, M Lewenstein$^{6,7}$ and T Pfau$^1$}

\address{$^1$ 5. Physikalisches Institut, Universit\"at Stuttgart, Pfaffenwaldring 57, D-70550 Stuttgart, Germany}
\address{$^2$ Universit\'e de Toulouse, UPS, Laboratoire Collisions Agr\'egats R\'eactivit\'e, IRSAMC, F-31062 Toulouse, France}
\address{$^3$ CNRS, UMR 5589, F-31062 Toulouse, France}
\address{$^4$ CNR-INFM BEC and Dipartimento di Fisica, Universit\`a di Trento, I-38050 Povo, Italy}
\address{$^5$ Institut f\"ur Theoretische Physik, Leibniz Universit\"at Hannover, Appelstr. 2, D-30167, Hannover, Germany}
\address{$^6$ ICFO--Institut de Ci\`{e}ncies Fot\`{o}niques, 08860 Castelldefels (Barcelona), Spain}
\address{$^7$ ICREA-Instituci\'o Catalana de Recerca i Estudis Avan\c cats, 08010 Barcelona, Spain}

\ead{t.pfau@physik.uni-stuttgart.de}

\begin{abstract}
This article reviews the recent theoretical and experimental advances in the study of ultracold gases made of bosonic particles interacting via the long-range, anisotropic dipole-dipole interaction, in addition to the short-range and isotropic contact interaction usually at work in ultracold gases. The specific properties emerging from the dipolar interaction are emphasized, from the mean-field regime valid for dilute Bose-Einstein condensates, to the strongly correlated regimes reached for dipolar bosons in optical lattices.
\end{abstract}

\submitto{\RPP}
\maketitle
\tableofcontents

\section{Introduction}

\subsection{Bose-Einstein condensation and a new quantum era}

The achievement of Bose-Einstein condensation (BEC) of dilute gases in 1995~\cite{anderson1995,bradley1995,davis1995}
has marked the beginning of a new era in atomic, molecular and optical physics  and quantum optics. For the AMO community it was immediately clear that the specific experimental techniques of these fields could be used to study problems usually encountered in condensed matter physics: degenerate quantum many body systems. The condensed matter community remained at this stage much more skeptical and argued that at the very end what was achieved experimentally was the regime of weakly interacting Bose gases, that had been thoroughly investigated by condensed matter theorists in the 50's and 60's~\cite{mahan1993,fetter71}. For solid state/condensed matter experts the very fact that the AMO experiments dealt with confined systems of finite size and typically inhomogeneous density was of technical, rather than fundamental importance. Nevertheless, the Nobel foundation decided to give its yearly prize in 2001 to E. A. Cornell, C. E. Wieman and W. Ketterle ``for the achievement of Bose-Einstein condensation in dilute gases of alkali atoms, and for early fundamental studies of the properties of the condensate''~\cite{cornell02,ketterle02}. Today, from the perspective of some years, we see that due to the efforts of the whole community these fundamental studies have enriched amazingly the standard ``condensed matter'' understanding  of static and dynamical properties of weakly interacting Bose gases~\cite{pitaevskii2003}.

At the same time, the AMO community continued the efforts to extend the BEC physics toward new regimes and new challenges. The progress in this directions was indeed spectacular and in the beginning of the third millennium is clear both for AMO and condensed matter communities that we are entering a truly new quantum era with unprecedented possibilities of control on many body systems. In particular it became clear that the regime of strongly correlated systems may be reached with ultracold atoms and/or molecules. Few years after the first observation of BEC, atomic degenerate Fermi gases~\cite{demarco1999,truscott2001,schreck2001,hadzibabic02} have been achieved.
This has  paved the way toward the observations of Fermi superfluidity (described in the weak interaction limit by Bardeen-Cooper-Schrieffer (BCS) theory~\cite{fetter71}), and the so called BEC-BCS crossover in the limit of strong correlations (for recent reviews of an enormous activity in this field see~\cite{giorgini08,bloch2008}). Even earlier,  following the seminal proposal by Jaksch {\it et al.}~\cite{jaksch1998}, Greiner {\it et al.}~\cite{greiner2002} observed the signatures of the quantum phase transition from the superfluid to the so-called Mott insulator state for bosons confined in an optical lattice.

Nowadays, ultracold atomic and molecular systems are at the frontier of modern quantum physics, and are seriously considered as ones that offer more control than solid state systems. It is generally believed that these systems will find highly nontrivial applications in quantum information (either as quantum simulators, \emph{i.e.} quantum computers for a special purpose, or as universal ones) or quantum metrology. At the level of theory a fascinating ``grand unification'' takes place: AMO, condensed matter, nuclear physics, and even high energy physics theorists join their efforts to work on ultracold gases (for recent reviews see~\cite{giorgini08,bloch2008,lewenstein2007}).

\subsection{Interactions}

Although quantum gases are very dilute systems (with densities typically ranging from $10^{14}$ to $10^{15}$~cm$^{-3}$), most of their properties are governed by the interaction between particles. Usually, in the ultracold regime characteristic of quantum gases (temperatures in the nanoKelvin range), only $s$-wave scattering between particles can take place. This allows one to replace the real interatomic potential (which at long distances is the usual van der Waals interaction) by a pseudo-potential, which is short range, isotropic, and characterized by a single parameter, the $s$-wave scattering length $a$. This \emph{contact interaction} potential reads
\begin{equation}
U_{\rm contact}({\bds r})=\frac{4\pi\hbar^2 a}{m}\delta({\bds r})\equiv g\delta({\bds r}),
\label{eq:contact}
\end{equation}
where $m$ is the atomic mass.
For a large number of atomic species, the magnitude and even the sign of the scattering length $a$ can be tuned by means of an external magnetic field. This phenomenon called Feshbach resonance~\cite{kohler2006,chin2009} has found multiple applications since its first experimental observation in cold gases~\cite{inouye1998,courteille1998}.

Despite its simplicity, the interaction potential (\ref{eq:contact}) is responsible for an extremely rich variety of physical properties of quantum gases. As we mentioned above, already for weakly interacting Bose gases, well described by mean-field theory, the interactions play a crucial role in the static and dynamic properties of Bose-Einstein condensates~\cite{pitaevskii2003,pethick2002}; one of the most fascinating properties they are responsible for is the superfluid character of those gases. The interactions between particles play obviously an even more crucial role in the very active field concerning the study of strongly correlated systems realized with ultracold atoms~\cite{bloch2008,lewenstein2007}.

For all those reasons, in the last few years, there has been a quest for realizing quantum gases with different, richer interactions, in order to obtain even more interesting properties. Several researchers, among which\footnote{We apologize in advance if we have forgotten someone.} K. Rz\c a{\.z}ewski, G.~V. Shlyapnikov, P. Zoller, G. Kurizki, L. You, D. DeMille, M. A. Baranov, P. Meystre, H. Pu, and some of us, have pointed out  that the dipole-dipole interaction, acting between particles having a permanent electric or magnetic dipole moment, should lead to a novel kind of degenerate quantum gases already in the weakly interacting limit. Its effects should be even more pronounced in the strongly correlated regime.

\begin{figure}
\begin{center}
\includegraphics[width=5cm]{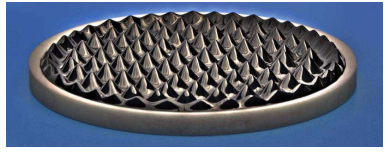}
\end{center}
\caption{The Rosensweig instability~\cite{rosensweig} of a ferrofluid (a colloidal dispersion in a carrier liquid of subdomain ferromagnetic particles, with typical dimensions of 10~nm) in a magnetic field perpendicular to its surface is a fascinating example of the novel physical phenomena appearing in classical physics due to long range, anisotropic interactions. Figure reprinted with permission from~\cite{knieling2007}. Copyright 2007 by the American Physical Society.}
\label{fig:ferro}
\end{figure}

The dipole--dipole interaction has attracted a huge interest for two reasons:
\begin{itemize}
\item Significant experimental progress was made in recent years in the cooling and trapping of polar molecules~\cite{doyle2004}, and of atomic species having a large magnetic moment. For the case of polar molecules, a very promising technique is to associate ultracold atoms by means of Feshbach resonances, and then to use photoassociation to bring the weakly bound Feshbach molecules to their ground state~\cite{ospelkaus2008}. In practice this technique requires  a very good phase stability of the involved lasers. A few months ago D. Jin and J. Ye groups at JILA have been able to create a gas of motionally ultracold Rubidium-Potassium molecules in their ground rotovibrational state~\cite{ni2008}; similar work was done in the group of M. Weidem\"uller with LiCs molecules~\cite{deiglmayr2008}. These amazing achievements open the way toward degenerate gases with dominant dipole-dipole interactions. For the case of magnetic dipoles, Bose-Einstein condensation of $^{52}$Cr, a species with a large magnetic moment of $6\,\mub$, was achieved in 2004~\cite{griesmaier2005}, and has since then allowed for the first experimental investigations of the unique properties of dipolar quantum gases~\cite{lahaye2009}. Although the relative effect of the dipole forces in Chromium can be tuned using the Feshbach resonance technique, they are typically smaller, or at most on the same order as the van der Waals forces.  Nevertheless their influence on the physics of the Chromium BEC is stunning, as we shall see in the following.

\item The properties of the {\ddi} are radically different from the ones of the contact interaction~\cite{menotti2008}. Indeed, one directly sees from expression (\ref{eq:udd:polarized}) below, giving the interaction energy between two dipoles polarized along the same direction, that the {\ddi} is \emph{long-range} (it decays like $1/r^3$, where $r$ is the distance between the particles), and \emph{anisotropic} (the strength and sign of the interaction depend on the angle $\theta$ between the polarization direction and the relative position of the particles). Note that if one limits oneself to neutral particles, the {\ddi} is the only interaction between electric or magnetic multipole moments which is long-range (interactions between higher order multipoles decay fast enough at large distances so that they can be replaced by a short range contact pseudo-potential at low temperatures). Long range and/or anisotropic interactions are known, already in classical fluids, to lead to completely new physical phenomena (see for example the case of ferrofluids~\cite{rosensweig} in figure~\ref{fig:ferro}). Similarly, anisotropy of interactions lies behind the fascinating physics of liquid crystals~\cite{liquidcrystals}. As we will argue in this review, dipole interactions in quantum gases lead also to a variety of novel, fascinating, and sometimes completely unexpected  effects.
\end{itemize}

This review is organized as follows. After a brief description of the main properties of the {\ddi} (section \ref{sec:ddi:properties}), and of the systems in which it can be studied (section \ref{sec:systems}), the mean field theory for a weakly interacting, polarized dipolar condensate is presented (section \ref{sec:4}). We derive here the non-local Gross-Pitaevski equation and discuss its applicability.  In subsequent sections we describe a number of properties of dipolar BECs, in particular their static and dynamic properties (sections \ref{sec:groundstate} and \ref{sec:expansion}). In section~\ref{sec:nlin:ao} we enter the very rich field of non-linear, non-local atom optics with dipolar gases.   Section \ref{sec:spinor} is devoted to the physics of dipolar spinor condensates. Finally, strongly correlated systems obtained by loading a dipolar BEC into an optical lattice are described in section \ref{sec:lattices}. Because of the lack of space, some very interesting topics are not addressed here; in particular, for a review of the properties of dipolar Fermi gases, and of (strongly correlated) rapidly rotating dipolar condensates, the reader is referred to the recent review article~\cite{baranov2008}.

\section{Dipole-dipole interaction}
\label{sec:ddi:properties}
\subsection{Properties of the dipole-dipole interaction}

For two particles 1 and 2 with dipole moments along the unit vectors ${\bds e}_1$ and ${\bds e}_2$, and whose relative position is ${\bds r}$ (see figure 1a), the energy due to the dipole-dipole interaction reads
\begin{equation}
U_{\rm dd}({\bds r})=\frac{\cdd}{4\pi} \frac{\left({\bds e}_1\cdot{\bds e}_2\right)r^2-3\left({\bds e}_1\cdot{\bds r}\right)\left({\bds e}_2\cdot{\bds r}\right)}{r^5}.
\label{eq:udd:general}
\end{equation}
The coupling constant $\cdd$ is $\mu_0\mu^2$ for particles having a permanent magnetic dipole moment $\mu$ ($\mu_0$ is the permeability of vacuum) and $d^2/\varepsilon_0$ for particles having a permanent electric dipole moment $d$ ($\varepsilon_0$ is the permittivity of vacuum).
For a polarized sample where all dipoles point in the same direction $z$ (figure 1b), this expression simplifies to
\begin{equation}
U_{\rm dd}({\bds r})=\frac{\cdd}{4\pi}  \frac{1-3\cos^2\theta}{r^3},
\label{eq:udd:polarized}
\end{equation}
where $\theta$ is the angle between the direction of polarization and the relative position of the particles. Two main properties of the {\ddi}, namely its long-range ($\sim 1/r^3$) and anisotropic character, are obvious from (\ref{eq:udd:general}) and (\ref{eq:udd:polarized}), and contrast strongly with the short-range, isotropic contact interaction \eref{eq:contact} usually at work between particles in ultracold atom clouds.

\begin{figure}
\begin{center}
\includegraphics[width=6.5cm]{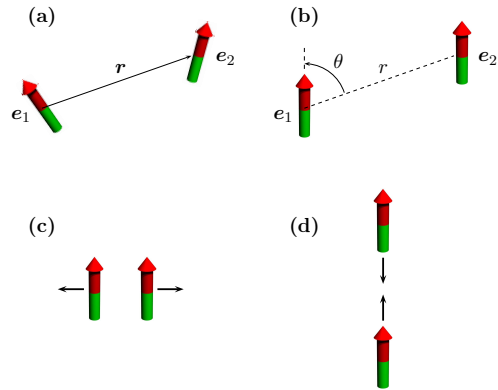}
\end{center}
\caption{Two particles interacting via the {\ddi}. (a) Non-polarized case; (b) Polarized case; (c) Two polarized dipoles side by side repel each other (black arrows); (d) Two polarized dipoles in a `head to tail' configuration attract each other (black arrows).}
\label{fig:ddi}
\end{figure}

\paragraph{Long-range character ---}

In a system of particles interacting via short-range interactions, the energy is extensive in the thermodynamic limit. On the contrary, in systems with long-range interactions, the energy per particle does not depend only on the density, but also on the total number of particles. It is easy to see that a necessary condition for obtaining an extensive energy is that the integral of the interaction potential $U({\bds r})$
\begin{equation}
\int_{r_0}^\infty U({\bds r})\,{\rm d}^D r,
\end{equation}
where $D$ is the dimensionality of the system and $r_0$ some short-distance cutoff, converges at large distances. For interactions decaying at large distances as $1/r^n$, this implies that one needs to have $D<n$ in order to consider the interaction to be short-range.  Therefore, the {\ddi} ($n=3$) is long-range in three dimensions, and short range in one and two dimensions. For a more detailed discussion, including alternative definitions of the long-range character of a potential, the reader is referred to~\cite{astrakharchik2008}.

\paragraph{Anisotropy ---}

The {\ddi} has the angular symmetry of the Legendre polynomial of second order $P_2(\cos\theta)$, \emph{i.e.} $d$-wave. As $\theta$ varies between $0$ and $\pi/2$, the factor $1-3\cos^2\theta$ varies between $-2$ and $1$, and thus the {\ddi} is repulsive for particles sitting side by side, while it is attractive (with twice the strength of the previous case) for dipoles in a `head-to-tail' configuration (see figure \ref{fig:ddi}(c) and (d)). For the special value $\theta_{\rm m}=\arccos\left(1/\sqrt{3}\right)\simeq54.7^\circ$ --- the so-called `magic-angle' used in high resolution solid-state nuclear magnetic resonance~\cite{nmr,nmr2}~---, the {\ddi} vanishes.

\paragraph{Scattering properties ---}

Usually, the interaction potential between two atoms separated by a distance $r$ behaves like $-C_6/r^6$ at large distances. For such a van der Waals potential, one can show that in the limit of a vanishing collision energy, only the $s$-wave scattering plays a role. This comes from the general result stating that for a central potential falling off at large distances like $1/r^n$, the scattering phase shifts $\delta_{\ell}(k)$ scale, for $k\to 0$, like $k^{2\ell+1}$ if $\ell<(n-3)/2$, and like $k^{n-2}$ otherwise~\cite{landau1977}.  In the ultracold regime, the scattering is thus fully characterized by the scattering length $a$. In the study of quantum gases, the true interaction potential between the atoms can then be replaced by a pseudo-potential having the same scattering length, the so-called contact interaction given by~(\ref{eq:contact}).

In the case of the {\ddi}, the slow decay as $1/r^3$ at large distances implies that for all $\ell$, $\delta_\ell\sim k$ at low momentum, and all partial waves contribute to the scattering amplitude. Moreover, due to the anisotropy of the {\ddi}, partial waves with different angular momenta couple with each other. Therefore, one cannot replace the true potential by a short-range, isotropic contact interaction. This specificity of the dipolar interaction has an interesting consequence in the case of a polarized Fermi gas: contrary to the case of a short-range interaction, which freezes out at low temperature, the collision cross section for identical fermions interacting via the {\ddi} does not vanish even at zero temperature. This could be used to perform evaporative cooling of polarized fermions, without the need for sympathetic cooling \emph{via} a bosonic species.

Dipolar interactions also play an important role in determining inelastic scattering properties. In particular, because of its anisotropy, the {\ddi} can induce spin-flips, leading to dipolar relaxation. The cross-section for dipolar relaxation scales with the cube of the dipole moment~\cite{hensler2003}, and therefore plays a crucial role in strongly dipolar systems (see section \ref{sec:crbec}). Dipolar relaxation is usually a nuisance, but can in fact be used to implement novel cooling schemes inspired by adiabatic demagnetization as described in section \ref{sec:demag}.

\paragraph{Fourier transform ---}

In view of studying the elementary excitations in a dipolar condensate, as well as for numerical calculations, it is convenient to use the Fourier transform of the {\ddi}. The Fourier transform
\begin{equation}
\widetilde{\udd}({\bds k})=\int U_{\rm dd}({\bds r}) {\rm e}^{-i{\bds k}\cdot{\bds r}}\,{\rm d}^3r
\end{equation}
of (\ref{eq:udd:polarized}) reads
\begin{equation}
\widetilde{\udd}({\bds k})=\cdd\left(\cos^2\alpha-1/3\right),
\label{eq:tf:dd}
\end{equation}
where $\alpha$ is the angle between ${\bds k}$ and the polarization direction (see~\ref{sec:append:fourier}). Remarkably, in three dimensions, the Fourier transform of the {\ddi} does not depend on the modulus of the wavevector ${\bds k}$, a feature which is shared by the contact interaction (\ref{eq:contact}), whose Fourier transform is simply $g$.

\subsection{Tuning of the dipole-dipole interaction}\label{sec:tuning}

\begin{figure}
\begin{center}
\includegraphics[width=6cm]{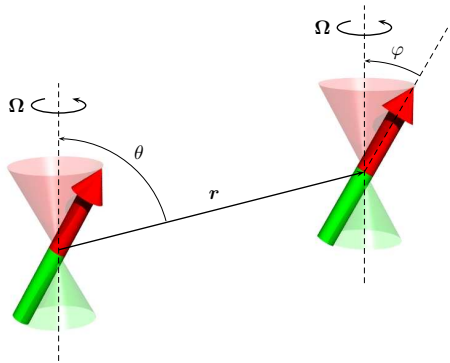}
\end{center}
\caption{Tuning of the {\ddi} can be obtained by making the dipoles precess around $z$ using a rotating field.}
\label{fig:tune:ddi}
\end{figure}

By using a rotating polarizing field, it is possible, by time-averaging, to tune the {\ddi}, namely to reduce its effective strength and even change its sign~\cite{giovanazzi2002a}. For definiteness we consider here the case of magnetic dipoles ${\bds \mu}$ in a magnetic field ${\bds B}(t)=B{\bds e}(t)$ (see figure~\ref{fig:tune:ddi}). The unit vector ${\bds e}(t)=\cos\varphi {\bds e}_z+\sin\varphi\left[\cos(\Omega t) {\bds e}_x+\sin(\Omega t) {\bds e}_y\right]$ is rotated about the $z$-axis on a cone of aperture $2\varphi$ at an angular frequency $\Omega$ which is small compared to the Larmor frequency $\mu B/\hbar$, but much larger than the trapping frequencies. Then, only the time-average over a period $2\pi/\Omega$ of the {\ddi} (\ref{eq:udd:general}) with ${\bds e}_1={\bds e}_2={\bds e}(t)$ plays a role to determine the properties of the gas. This time-averaged potential reads
\begin{equation}
\left\langle\udd (t)\right\rangle=\frac{\cdd}{4\pi}  \frac{1-3\cos^2\theta}{r^3}\left[\frac{3\cos^2\varphi-1}{2}\right].
\end{equation}
The last factor between brackets decreases from $1$ to $-1/2$ when the tilt angle $\varphi$ varies from 0 to $\pi/2$, and vanishes when $\varphi$ is equal to the magic angle $\theta_{\rm m}$. The `inverted' configuration ($\varphi>\theta_{\rm m}$) in which the averaged {\ddi} is attractive for particles sitting side by side, allows to explore otherwise inaccessible physics (see section~\ref{sec:nlin:ao} for some examples of applications).

\section{Creation of a dipolar gas}
\label{sec:systems}
In order to realize a quantum gas with significant {\ddi}s, one can use particles having either an electric dipole moment $d$, or a magnetic dipole moment $\mu$. Usually, the dipolar coupling is much higher in the electric case. Indeed, the typical order of magnitude of $d$ for an atomic or molecular system is $d\sim q_{\rm e} a_0$, where $q_{\rm e}$ is the electron charge and $a_0$ the Bohr radius, while the magnetic moments are on the order of the Bohr magneton $\mu_{\rm B}$. Using the definitions of $a_0$ and $\mu_{\rm B}$ in terms of fundamental constants, one sees that the ratio of magnetic to electric dipolar coupling constants is
\begin{equation}
\frac{\mu_0\mu^2}{d^2/\varepsilon_0}\sim\alpha^2\sim10^{-4},
\end{equation}
where $\alpha\simeq1/137$ is the fine structure constant.

For a given species, it is convenient to define various quantities to quantify the strength of the dipolar interaction. From the dipole moment (\emph{i.e.} the dipolar coupling constant $\cdd$) and the mass $m$ of the particle, one can define the following length:
\begin{equation}
\add\equiv\frac{\cdd m}{12\pi\hbar^2}.
\label{eq:def:add}
\end{equation}
This 'dipolar length' is a measure of the absolute strength of the {\ddi}. However, in some circumstances, it is the ratio
\begin{equation}
\edd\equiv\frac{\add}{a}=\frac{\cdd}{3g}
\label{eq:def:edd}
\end{equation}
of the dipolar length to the $s$-wave scattering length, comparing the relative strength of the dipolar and contact interactions, which determines the physical properties of the system. This dipolar parameter needs to be non negligible if one wants to observe dipolar effects. The numerical factors in (\ref{eq:def:add}) are chosen in such a way that for $\edd\geqslant 1$ a \emph{homogeneous} condensate is unstable against 3D collapse (see section \ref{sec:phonon:instability}). Table~\ref{tab:species:ddi} summarizes some typical numerical values of the dipolar constants for various atomic and molecular species.

In this section, we review the different systems that can be used in principle to study experimentally the effect of the {\ddi} in degenerate quantum gases. We first address the various candidates having an \emph{electric} dipole moment, either static or induced by a laser. The case of \emph{magnetic} dipoles (the only system to date in which strong dipolar effects in a quantum gas have been observed) is then described, with an emphasis on the experimental techniques used to achieve Bose-Einstein condensation of Chromium.

\begin{table}[t]
\caption{\label{tab:species:ddi}Dipolar constants for various atomic and molecular species. For the molecular species, the (yet unknown) scattering length is assumed to be $100\,a_0$ (as the $C_6$ coefficient of the dimer is comparable to the one of a single atom, the order of magnitude of the scattering length is similar, but obviously the actual value highly depends on the details of the potential).}
\begin{indented}
\item[] \begin{tabular}{@{}llll}
\br
\;Species\; & \;Dipole moment\; & $a_{\rm dd}$ & $\edd$ \\
\mr
$^{87}$Rb & $1.0\,\mub$ & $0.7\, a_0$ & \lineup\0\00.007 \\
\chr & $6.0\,\mub$ & $16\,a_0$ &  \lineup\0\00.16 \\
KRb & $0.6$ D & $2.0 \times 10^3 a_0$ & \lineup\020 \\
ND$_3$ & $1.5$ D & $3.6 \times10^3\,a_0$ & \lineup\036 \\
HCN & $3.0$ D & $2.4 \times10^4\, a_0$ & 240 \\
\br
\end{tabular}
\end{indented}
\end{table}

\subsection{Polar molecules}\label{sec:systems:mol}

Due to their strong electric dipole moment, polar molecules are ideal candidates to show dipolar effects. Three requirements need to be fulfilled in order for a molecule to have a significant dipole moment:
\begin{enumerate}
\item[(i)] a \emph{heteronuclear} molecule, having a permanent dipole moment, is needed\footnote{Note however that exotic \emph{homonuclear} molecules, such as the ultra-long-range Rydberg molecules predicted in~\cite{greene2000} and recently observed in~\cite{bendkowski2009} can have a permanent dipole moment.};
\item[(ii)] the molecule must be in a low rovibrational state in order to have a dipole moment whose magnitude is not vanishingly small (as would be the case for a highly excited vibrational state, especially for Feshbach molecules; indeed, the dipole moment scales asymptotically as $R^{-7}$ with the internuclear separation $R$~\cite{kotochigova2002}.) and to be stable against collisional relaxation;
\item[(iii)] an external electric field (with a typical value on the order of $10^4$~V$/$cm) must be applied to \emph{orient} the molecule in the laboratory frame and approach the asymptotic value of the permanent electric dipole moment along the internuclear axis (indeed, the ground state $J=0$ is rotationally symmetric and therefore the dipole moment averages to zero; only \emph{via} a mixing with higher rotational levels, induced by the electric field, does the average dipole become non-zero, see \ref{sec:append:mol}). Note that this effect can be used to tune the strength of the {\ddi} (but not its sign, unlike the rotating field method described in section~\ref{sec:tuning}). Using additional microwave fields allows for advanced tailoring of the interactions between molecules~\cite{buechler2007}.
\end{enumerate}
If these requirements are met, the dipole moment is on the order of one Debye ($1\,{\rm D}\simeq 3.335 \times
10^{-30}\,{\rm C}\cdot{\rm m}$). Assuming that the order of magnitude for the scattering length is similar to that of atoms commonly used in BEC experiments (typically around $100\,a_0$), the corresponding value of $\edd$ is on the order of 100 (see Table~\ref{tab:species:ddi}), meaning that the properties of such a quantum gas would be dominated by the {\ddi}.

Quantum degenerate gases of polar molecules are a 'Holy Grail' of experimental molecular physics. Progress has been made recently in cooling of molecules, with techniques such as Stark deceleration (see e.g.~\cite{vandemeerakker2008} for a review) or buffer-gas cooling~\cite{doyle1995,weinstein1998,egorov2002}, but the densities and temperatures achieved so far are still orders of magnitude away from the quantum degenerate regime. A very promising approach to degeneracy, actively explored by several groups~\cite{ospelkaus2008,ni2008,deiglmayr2008} is to start from already ultracold atomic mixtures, and then use a Feshbach resonance to create heteronuclear molecules~\cite{kohler2006}. Created in a highly excited vibrational state, they must then be brought to the vibrational ground state, e.g. by photoassociation using STIRAP processes, as demonstrated recently~\cite{ni2008,deiglmayr2008}.

\subsection{Rydberg atoms}

Extraordinarily large electric dipole moments can be obtained for highly excited Rydberg atoms. As the Kepler radius ---~and thus the dipole moment~--- scales with $n^2$, where $n$ is the main quantum number, the dipolar interaction energy can in principle scale like $n^4$. Individual Rydberg atoms experience lifetimes which scale with $n^{-3}$. However due to the weak binding  of the valence electrons and the strong and partially attractive forces between Rydberg atoms, the lifetime of a dense gas is limited to time scales much shorter than the lifetime of a free Rydberg atom~\cite{li2004-TP}. Therefore Rydberg atoms in a BEC~\cite{heidemann2008} are currently investigated as a frozen gas. Collective behaviour in the excitation dynamics has been observed, as well as the excitation blockade due to dipolar interactions~\cite{vogt2006}. However hydrodynamic collective phenomena due to moving dipoles have not been observed to date. Besides the static dipolar interaction also van der Waals interactions ($\propto n^{11}$) and AC dipolar interactions can occur if neighboring energy levels allow for resonant energy transfer via a so-called F\"orster resonance.

\subsection{Light-induced dipoles}\label{sec:light:induced}

Atoms in their ground state, which is a parity eigenstate, do not possess an electric dipole moment. Their electric polarizability is usually very small, such that extreme electric field strengths would be necessary to induce a sizable dipolar interaction~\cite{marinescu1998,yi2000,yi2001}. Following G. Kurizki and coworkers, one might consider to use resonant excitation of a dipole optical allowed transition to induce an AC dipole moment on the order of one atomic unit $e a_0$. However as this dipole moment also couples to the vacuum modes of the radiation field, the spontaneous light forces scale just like the light induced dipolar interactions which makes their observation very difficult. Nevertheless, the anisotropic nature of the interaction might be used for a proof of principle experiment, which would allow to discriminate the spontaneous light forces from the dipolar forces~\cite{low2005}. Such interactions have the same form as retarded interactions between two dipoles~\cite{jackson1999}: they contain $1/r^3$, $1/r^2$ and radiative $1/r$ terms multiplied by the appropriate factors oscillating with the spatial period of the laser wavelength. Using an arrangement of several laser fields it has been proposed to cancel all anisotropic $1/r^3$ terms, leaving an effective isotropic, gravity-like $1/r$ potential~\cite{odell2000,odell2003b,giovanazzi2001,giovanazzi2002b}. In some situations this may lead to self-trapping of the BEC. Even before the discovery of the roton instability~\cite{santos2003} discussed in section \ref{sec:roton}, a similar effect was predicted in a gas with laser induced dipole-dipole interactions~\cite{odell2003b}.  Such interactions lead naturally to density modulations of the BEC as in supersolid~\cite{giovanazzi2002b}, and other effects, such as one dimensional compression of the condensate~\cite{giovanazzi2001}, or squeezing~\cite{odell2003a}.
Laser-induced interactions lead in particular to interesting density modulations in the condensate, somewhat analogous to self-assembled ``supersolid'' (see section \ref{sectextBH}).  Due to the above mentioned limitations caused by spontaneous emission, these proposals have not been realized yet. The situation in this respect might be more promising if one uses CO$_2$ lasers~\cite{dobrek2004}.

\subsection{Magnetic dipoles}

In alkali atoms, the maximum magnetic moment in the ground state is of one Bohr magneton ($\mub$), and thus the magnetic  dipolar effects are very weak. However, very recently, dipolar effects have been observed in spinor $^{87}$Rb condensates (see section \ref{sec:spinor} below) and in condensates with a very small scattering length (obtained using a Feshbach resonance), either by studying the size of the condensate (case of $^7$Li, see section \ref{sec:elong} below) or by using atom interferometry (case of $^{39}$K, see section \ref{sectblochdip} below).

Some other atoms, like Chromium, Erbium, Europium, Dysprosium, and others, have a large magnetic moment of several Bohr magnetons in their ground state, and thus experience significant magnetic {\ddi}. Among them, only {\chr} has been Bose-condensed to date~\cite{griesmaier2005,beaufils2008a}. Chromium has a magnetic dipole moment of $6\,\mub$, and a scattering length of about $100\,a_0$~\cite{schmidt2003a}. This gives $\edd\simeq0.16$~\cite{griesmaier2006a}, which allows to observe a perturbative effect of the dipolar interaction on the expansion dynamics of the cloud~\cite{stuhler2005}. Here we describe briefly the main steps leading to the creation of a {\chr} BEC, with a special emphasis on the specificities arising from the {\ddi}.

\subsubsection{Creation of a BEC of $^{52}${\rm Cr}}\label{sec:crbec}

\begin{figure}
\begin{center}
\includegraphics[width=6cm]{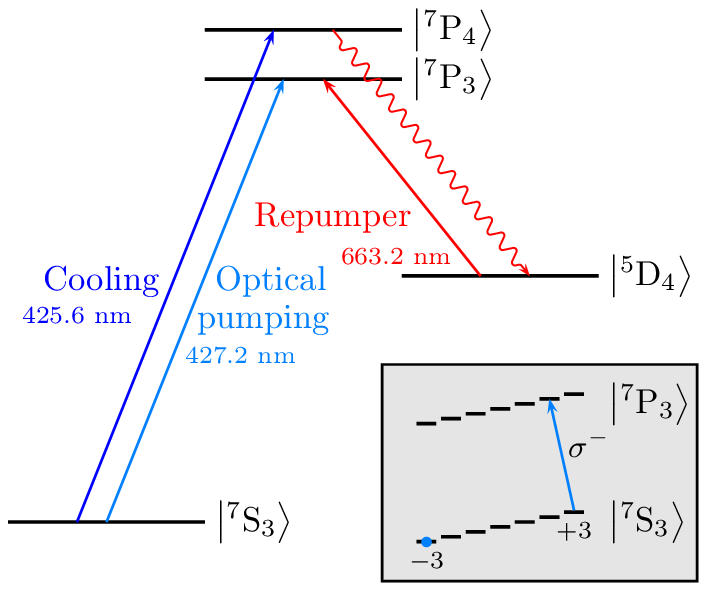}
\end{center}
\caption{Scheme of the energy levels of $^{52}$Cr relevant for the realization of a {\chr} BEC. The inset (in gray) gives details of the optical pumping.}
\label{fig:cr:levels}
\end{figure}

Chromium can be laser cooled using the $^7{\rm S_3}\leftrightarrow\,^7{\rm P_4}$ transition at 425.6 nm (see figure \ref{fig:cr:levels}). However, strong excited state collisions limit the density in a magneto-optical trap (MOT) to relatively low values~\cite{bell1999}. Therefore, in a typical MOT configuration, the steady state atom number in a Cr MOT is limited to a few $10^6$. In addition the cooling transition is not closed, as there is a decay channel from the $^7{\rm P}_4$ state to the metastable state $^5{\rm D}_4$ (via an intercombination transition) with a branching ratio of approximately $1:250,000$. These facts seem to rule out any hope to achieve Bose condensation of Chromium by standard methods. However, the atoms in the metastable state have a magnetic dipole moment of $6\,\mub$ which is strong enough so that they remain magnetically trapped in the inhomogeneous magnetic field configuration of the MOT. One can thus accumulate large atom numbers in the metastable state $^5{\rm D}_4$ (where they are decoupled from the MOT light and thus do not suffer from excited state collisions), and then, at the end of the MOT phase, repump them using light at 663.2~nm. In this way, one ends up with more than $10^8$ ground state atoms magnetically trapped. In~\cite{schmidt2003c}, the magnetic field configuration of the MOT was modified to the one of a Ioffe-Pritchard trap, allowing a continuous loading of a magnetic trap, in which rf evaporative cooling could be performed.

However, when the density in the magnetic trap becomes too high, one cannot gain anymore in phase-space density due to increasing \emph{dipolar relaxation}. This two-body loss mechanism, in which the spin of one of the colliding atoms is flipped (this is allowed as the {\ddi} does not conserve the spin, but only the \emph{total} angular momentum), is especially important for Chromium, compared to the case of alkalis, as its cross section scales as the cube of the magnetic dipole moment. Typical relaxation rates of $\beta\sim 10^{-12}$~cm$^{-3}$/s were measured in {\chr} at magnetic fields of about 1~G~\cite{hensler2003}, thus preventing the achievement of BEC in the (magnetically trapped) low-field seeking state $m_S=+3$.

The way to circumvent dipolar relaxation is to optically pump the atoms into the absolute ground state $m_S=-3$ (via the $^7{\rm S_3}\leftrightarrow\,^7{\rm P_3}$ transition at 427.6 nm, see figure~\ref{fig:cr:levels}) and hold them in an optical dipole trap. Then, in the presence of a magnetic field such that the Zeeman splitting between adjacent spin states is much higher than the thermal energy, dipolar relaxation is energetically forbidden. One can then perform evaporative cooling in the dipole trap and obtain a Chromium condensate~\cite{griesmaier2005}. Recently, an alternative method has been used to obtain {\chr} condensates, using direct loading of the optical dipole trap~\cite{beaufils2008a}.

\subsubsection{Feshbach resonances in {\chr}}

A very appealing feature of {\chr} is the existence of several Feshbach resonances. These allow to tune the scattering length $a$, which, close to resonance, varies with the applied external magnetic field $B$ as
\begin{equation}
a=a_{\rm bg}\left(1-\frac{\Delta}{B-B_0}\right),
\label{eq:fr:general}
\end{equation}
where $a_{\rm bg}$ is the background scattering length, $B_0$ is the resonance position (where $a$ diverges) and $\Delta$ the resonance width. In order to study the effect of the {\ddi} in a BEC, it is of course interesting to use the Feshbach resonance to \emph{reduce} the scattering length towards zero by approaching $B_0+\Delta$ from above, thus enhancing $\edd$.

In {\chr}, for magnetic fields $B$ below 600~G, a total of fourteen resonances were found by monitoring inelastic losses in a thermal cloud of atoms in the $\ket{^7{\rm S}_3,m_S=-3}$ state~\cite{werner2005}. An accurate assignment of the resonances was possible by considering the selections rules and the shifts of the resonances imposed by the dipole-dipole interaction only. In contrast to other atomic species, the dipolar contribution is therefore dominant as compared to other coupling mechanisms, like second order spin orbit coupling, which have the same symmetry. The inclusion of {\ddi} in multichannel calculations~\cite{werner2005} gave a theoretical understanding of the width $\Delta$ of the various resonances, which turn out to be relatively small (the broadest one, located at $B_0=589$~G, having a predicted width of $\Delta=1.7$~G only).

\begin{figure}
\begin{center}
\includegraphics[width=5.5cm]{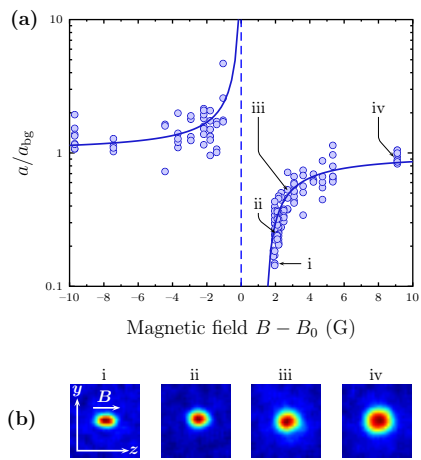}
\end{center}
\caption{(a) Measured variation of the scattering length across the 589~G Feshbach resonance in {\chr}~\cite{lahaye2007}. (b) Absorption images of the condensate after expansion for different values of the magnetic field. One clearly observes a decrease in size and an increase in ellipticity when $a$ decreases. }
\label{fig:feshbach:exp}
\end{figure}

In~\cite{lahaye2007}, this resonance was used to enhance dipolar effects in a BEC. An active control of the magnetic field at the level of $3\times10^{-5}$ in relative value was implemented, allowing for a control of $a$ at the level of $\sim a_0$.
Figure~\ref{fig:feshbach:exp} shows the measured variation of $a$, inferred from the released energy during expansion (see section \ref{sec:ferro}).

\subsubsection{Demagnetization cooling}\label{sec:demag}

The large magnetic dipole moment of {\chr} is responsible for strong spin-flip collisions, which, as we have seen above, prevent condensation of Cr in a magnetic trap. However, these inelastic collisions can be used to implement a novel cooling scheme, proposed in~\cite{hensler2005} and demonstrated experimentally in~\cite{fattori2006}. This technique is inspired from the well-known \emph{adiabatic demagnetization} used in solid state physics to cool paramagnetic salts~\cite{lounasmaa1974}. In the context of cold atoms this scheme has been proposed for the first time in~\cite{cirac1995}, and termed ``elevator cooling''. Particularly important was the analysis of the limitations of the scheme due to reabsorption effects in the Raman repumping process.

\begin{figure}
\begin{center}
\includegraphics[width=12cm]{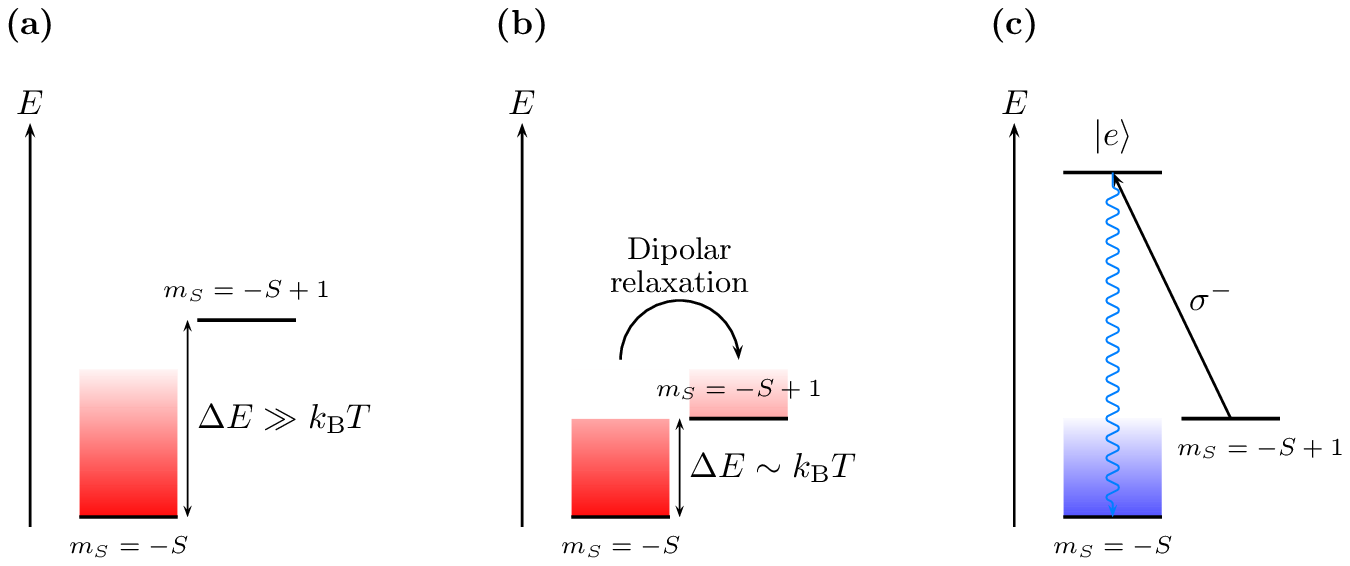}
\end{center}
\caption{Principle of demagnetization cooling. (a) If the magnetic field is large enough so that the Zeeman splitting $\Delta E$ between adjacent Zeeman sublevels is much higher than the thermal energy $k_{\rm B}T$, dipolar relaxation is energetically suppressed, and the system, fully polarized in $m_S=-S$, is stable. (b) If one reduces the magnetic field so that $\Delta E\sim k_{\rm B}T$, dipolar relaxation occurs, and some kinetic energy is converted into Zeeman energy. (c) By applying an optical pumping pulse of $\sigma^-$ polarized light, one can polarize the cloud again, but with a decrease of the temperature since the optical pumping process deposits an energy which is only on the order of the recoil energy. The excess Zeeman energy is taken away by the spontaneously emitted photons.}
\label{fig:demag}
\end{figure}

The principle of this novel cooling mechanism is represented schematically in figure~\ref{fig:demag}.
Dipolar relaxation introduces a coupling between spin and external degrees of freedom. It can thus be used to cool an atomic
cloud by letting a sample, initially polarized in the lowest energy state (in a field $B_0\gg\kb T_0/\mu$, where $T_0$ is the cloud temperature), relax towards full thermal equilibrium at a field $B_1\sim \kb T_0/\mu$: energy is then absorbed by the spin reservoir at the expense of the kinetic energy (see figure~\ref{fig:demag}). The temperature of the sample thus decreases, by an amount which can be up to a few tens of percents. By optical pumping, the sample can be polarized again, and a new cycle can begin. One can also use a continuous cooling scheme, with the optical pumping light always on, and a ramp in magnetic field. This scheme can be seen as an evaporation process (where one selects the most energetic particles in the cloud) in which the evaporated particles have their energy decreased and are then 'recycled' by injecting them back into the trap; it is therefore lossless.

Note that this scheme is applicable for all dipolar species with a large enough dipolar relaxation rate. This rate scales as the third power of the electronic spin. There could be even a variant of this cooling technique for electric dipole moments in heteronuclear molecules. As under optimized conditions any scattered photon in the cooling cycle takes more energy than the mean motional energy the number of required photons for a certain cooling rate is much lower than in regular laser cooling techniques. Therefore the requirements for the closedness of an optical transition used here are much less stringent as compared to regular laser cooling techniques.

This scheme has been successfully applied to Cr, allowing for a reduction of the cloud temperature by a factor of two (from 20 to 11 $\mu$K), with almost no atom loss~\cite{fattori2006}. This cooling technique is therefore much more efficient than evaporative cooling, where the decrease in atom number is large. An important figure of merit for cooling schemes in view of obtaining quantum degeneracy is the gain $\chi$ in phase-space density $\rho$ per atom loss:
\begin{equation}
\chi\equiv-\frac{{\rm d}\ln\rho}{{\rm d}\ln N}.
\end{equation}
For evaporative cooling, $\chi$ is limited in practice\footnote{Using a higher evaporation threshold increases $\chi$, but the evaporation time then increases prohibitively.} to values about~4. In~\cite{fattori2006}, the measured efficiency of
demagnetization cooling reached $\chi\simeq11$.

The practical limitations in view of achieving lower temperatures lie essentially in the control of the polarization of the optical pumping light, as any residual $\sigma^+$ component yields a heating of the cloud, and in the control of stray magnetic fields at the milligauss level. However, the recoil temperature should be attainable in principle with this technique~\cite{cirac1995}, which could be used, in the future, to realize dipolar condensates with large atom numbers. Note that the dipolar coupling mechanism between spin and motional degrees of freedom demonstrated in this cooling experiment is the same as the one employed in the proposals to observe the quantum version of the Einstein-de Haas effect, as explained in section~\ref{sec:spinor}.

\section{Non-local Gross-Pitaevskii equation}\label{sec:4}

Dipolar interactions are expected to change many of the properties of the gas, even in the non-degenerate case, where thermodynamical quantities can be affected. For example, dipolar interactions lead to a shift of the critical temperature for condensation~\cite{glaum2007b,glaum2007a}, which, although negligible for Cr condensates, could be significant for strongly dipolar systems made out of polar molecules. However, the most dramatic effects of the dipolar interactions arise for pure condensates. In the sections below, unless otherwise stated, we thus consider the case of a gas at zero temperature.

\subsection{Pseudo-potential and Gross-Pitaevskii equation}\label{sec:pseudo}

To describe dilute (and therefore weakly interacting) BECs at zero temperature, the mean field approach gives extremely good results~\cite{pitaevskii2003,pethick2002}. In the case of short range van der Waals interactions and low energy scattering further simplification can be made, namely the van der Waals interaction potential $V_{\rm vdW}({\bds r-\bds r'})$ may be replaced by the pseudo-potential
\begin{equation}
\frac{4\pi \hbar^2 a}{m}\delta({\bds r-\bds r'})\frac{\partial}{\partial|\bds r-\bds r'|}|\bds r- \bds r'|.
\label{eq:pseudo}
\end{equation}
This result has been obtained for the first time in the seminal paper by Huang and Yang~\cite{huang1957} for the case of a gas of hard spheres. It has however more general meaning and holds for arbitrary short range potentials. In the language of many-body theory it is the result of the $\bf T$-matrix, or ladder approximation applied to many-body systems~\cite{mahan1993}. It amounts to the resummation of diagrams corresponding to multiple two-body scattering. Note that when acting on a non-singular function, the pseudopotential is not different from the simple contact (Fermi) potential ${4\pi \hbar^2 a}\delta(\bds r- \bds r')/m$. This is however not true when we deal with singular, yet square integrable functions. In fact, strictly speaking, the contact potential  ${4\pi \hbar^2 a}\delta(\bds r-\bds r')/m$ is mathematically ill-defined~\cite{pascual1991}.

In the mean field theory, where the use of contact interactions is legitimate, the order parameter $\psi({\bds{r}},t)$ of the condensate is the solution of the Gross-Pitaevskii equation
(GPE)~\cite{pitaevskii2003}:
\begin{equation}
i\hbar\frac{\partial\psi}{\partial t}=-\frac{\hbar^2}{2m}\triangle\psi+\left(V_{\rm ext}+g|\psi|^2\right)\psi.
\label{eq:gpe:cont}
\end{equation}
The non-linear term proportional to $g$ accounts for the effect of interactions within the mean-field approximation, and $V_{\rm ext}$ denotes the external potential.  The normalization of $\psi$ chosen here is $\int|\psi|^2=N$, where $N$ is the total atom number.

\subsection{Validity of non-local Gross-Pitaevskii equation}\label{sec:validity}

In the simple man's approach, to include dipolar effects, one just needs to add an extra term to the mean-field potential $g|\psi|^2$ to account for the effect of the {\ddi}, and one gets
\begin{equation}
i\hbar\frac{\partial\psi}{\partial t}=-\frac{\hbar^2}{2m}\triangle\psi+\left(V_{\rm ext}+g|\psi|^2+\fdd\right)\psi.
\label{eq:gpe:dd}
\end{equation}
where $\fdd$ is the dipolar contribution to the mean field interaction
\begin{equation}
\fdd({\bds{r}},t)=\int|\psi({\bds{r}'},t)|^2\,U_{\rm dd}({\bds{r}}-{\bds{r}'})\,{\rm d}^3r'.
\label{eq:phidd}
\end{equation}
This term is \emph{non-local} (due to the long-range character of the dipolar interaction) and makes it much more complicated to solve the GPE, even numerically, as one now faces an integro-differential equation. In the time-independent case, the left-hand side of the above equation has to be replaced by $\mu\psi$, with $\mu$ the chemical potential, and the GPE becomes:
\begin{equation}
-\frac{\hbar^2}{2m}\triangle\psi+\left(V_{\rm ext}+g|\psi|^2+\fdd\right)\psi=\mu\psi.
\label{eq:gpe:dd:ti}
\end{equation}

It should be stressed that, due to the long range and anisotropic character of the dipole-dipole interactions, it is by no means obvious that one can put the pseudo-potential and the real potential into one single equation --- obviously this implies  that the long and short range physics can somehow be treated separately.

Questions concerning the validity of (\ref{eq:phidd}) were a subject of intensive studies in the recent years. In their pioneering papers, L. You and S. Yi~\cite{yi2000,yi2001} have constructed, in the spirit of the ladder approximation, a pseudo-potential for the general case of anisotropic potentials. Their results were rigorous, but perturbative, in the sense that the Born scattering amplitude from the pseudo-potential was reproducing the exact one. The conclusion was that, away from shape resonances, the generalized GPE (\ref{eq:phidd}) is valid, and the effective pseudo-potential has the form (assuming for instance that we deal with electric dipoles):
\begin{eqnarray}
V_{\rm eff}({\bds r- \bds r'})&=&\frac{4\pi \hbar^2 a(d)}{m}\delta({\bds r- \bds r'})\frac{\partial}{\partial|\bds r-\bds r'|}|\bds r-\bds r'|\nonumber\\
&&+ \frac{1}{4\pi\varepsilon_0}\frac{d^2-3({\bds n \cdot {\bds d}})^2}{|{\bds r-\bds r'}|^3},
\label{eq:pseudo1}
\end{eqnarray}
where ${\bds n}=({\bds r-\bds r'})/|{\bds r-\bds r'}|$, whereas $a(d)$ depends effectively on the strength of the dipole moment $d=|{\bds d}|$.

A. Derevianko succeeded to derive a more rigorous version of the pseudo-potential with a velocity dependence~\cite{derevianko2003}, which was then used in~\cite{yi2004a} to calibrate the dipole interactions. The author predicted that the effects of dipole interactions should be significantly enhanced due to the velocity dependent part of the pseudo-potential. Unfortunately, these  conclusions were too optimistic, due to some incorrect factors in the expressions of~\cite{derevianko2003, derevianko2005, derevianko2005a}.

Further important contributions to these questions came from J. L. Bohn and D. Blume~\cite{bortolotti2006,ronen2006}. These authors studied the instability and collapse of the trapped dipolar gas and compared the mean field (MF) results with diffusive Monte Carlo (DMC) calculations. The DMC results agreed quite accurately with the MF ones, provided the variation of the $s$-wave scattering length with the dipole moment was properly taken into account. In fact, this dependence had been already noted in~\cite{yi2000,yi2001}, and can be traced back to the fact that the rigorous form of dipole-dipole interactions contains already a contact $\delta({\bds r})$ term~\cite{jackson1999}. Very careful discussion of the differences between the GPE approach, and more exact numerical results of diffusive Quantum Monte Carlo methods were presented recently by Astrakharchik {\it et al.}~\cite{astrakharchik2007a}. These authors point out several difference between DMC and GPE results in wide range of parameters, especially reflected  in the frequency of the low energy excitation ``breathing'' mode. Very recently, D.-W. Wang~\cite{wang2008a} has managed to derive a general effective many body theory for polar molecules in the strongly interacting regime. Wang's approach allows to go beyond the Born approximation approach of Yi and You. One of the surprising results is that close to shape resonances, anisotropic effects of dipole-dipole interactions are strongly reduced. Phonon dispersion relations scale as $\sqrt{|{\bds p}|}$ as in the case of a Coulomb gas.

\subsection{Variational approach and hydrodynamics}\label{sec:tdandhydro}

The time independent GPE can be obtained from the minimization of the energy functional:
\begin{eqnarray}
E[\psi]&=&\int\left[\frac{\hbar^2}{2m}|\nabla \psi|^2+ V_{\rm
trap}|\psi|^2+ \frac{g}{2} |\psi|^4\right.\nonumber
\\&&\left.+\frac{1}{2}|\psi| ^2\int
U_{\rm dd}({\boldsymbol r}-{\boldsymbol r}') |\psi({\boldsymbol
r}')|^2{\rm d}^3{ r}' \right]{\rm d}^3{ r}.
\label{eq:energy:func}
\end{eqnarray}
In this context the chemical potential appears simply as a Lagrange multiplier arising from the constraint on the normalization of the macroscopic wavefunction $\psi$. Minimizing the energy functional (\ref{eq:energy:func}) within a space of trial wavefunctions depending on a small number of variational parameters is a convenient way to approach in a simple manner a variety of problems; a typical example being the trap geometry dependence of the stability of a dipolar BEC (see section \ref{sec:geometry}). Variational approaches can be extended to the time-dependent case by replacing the energy functional (\ref{eq:energy:func}) by an appropriate Lagrange action~\cite{perezgarcia1996,perezgarcia1997}.

A useful reformulation of the Gross-Pitaevskii equation is obtained by writing $\psi=\sqrt{n}\exp(iS)$, with $n$ the atomic density and $S$ the phase of the order parameter, related to the superfluid velocity field by ${\bds{v}}=(\hbar/m){\bds \nabla}S$. Substituting this expression in~(\ref{eq:gpe:dd}) and separating real and imaginary parts, one gets the following set of hydrodynamic equations:
\begin{equation}
\frac{\partial n}{\partial t}+{\bds \nabla}\cdot(n{\bds{v}})=0,
\label{eq:hd:continuity}
\end{equation}
which is nothing more than the equation of continuity expressing the conservation of mass, and an Euler-like equation:
\begin{equation}
m\frac{\partial {\bds{v}}}{\partial t}%
=-{\bds\nabla}\left(\frac{mv^2}{2}+gn+V_{\rm
ext}+\Phi_{\rm dd}-\frac{\hbar^2}{2m}\frac{\triangle
\sqrt{n}}{\sqrt{n}}\right).
\label{eq:hd:euler}
\end{equation}
The last term in (\ref{eq:hd:euler}), proportional to the Laplacian of $\sqrt{n}$ is the \emph{quantum pressure} term arising from inhomogeneities in the density and vanishes in the limit of BECs containing a large number of atoms (Thomas-Fermi limit, see section~\ref{sec:tf}).

\section{Ground state properties and excitations}\label{sec:groundstate}

\subsection{Homogeneous gas. Phonon instability}\label{sec:phonon:instability}

Because of the partially attractive character of the dipole-dipole interaction, the stability of a dipolar BEC is a problem that needs to be addressed. Indeed it is well known~\cite{pitaevskii2003} that a homogeneous condensate with attractive contact interactions ($a<0$) is unstable, as the Bogoliubov excitations have imaginary frequencies at low momentum.

\begin{figure}
\begin{center}
\includegraphics[width=7cm]{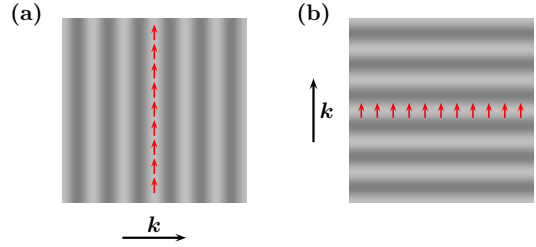}
\end{center}
\caption{(a) A phonon with ${\bds k}$ perpendicular to the direction of dipoles ($\alpha=\pi/2$) creates planes of higher density (light gray), in which the dipoles are \emph{in} the plane, corresponding to an instability (see section~\ref{sec:geometry} for a discussion of the geometry dependence of the stability of a trapped dipolar gas). (b) For ${\bds k}$ parallel to the direction of dipoles ($\alpha=0$) the dipoles point \emph{out} of the planes of high density; such a perturbation is thus stable.}
\label{fig:phonon}
\end{figure}

We consider here a homogeneous dipolar condensate, having an equilibrium density $n_0$. By considering small density and velocity perturbations with frequency $\omega$ and wavevector ${\bds k}$, and linearizing the hydrodynamic equations (\ref{eq:hd:continuity}) and (\ref{eq:hd:euler}) around equilibrium, one can show that the excitation spectrum is given by
\begin{equation}
\omega=k\sqrt{\frac{n_0}{m}\left[g+\frac{\cdd}{3}(3\cos^2\alpha-1)\right]+\frac{\hbar^2k^2}{4m^2}},
\label{eq:bogo:dd}
\end{equation}
which corresponds to the usual Bogoliubov spectrum $\omega=k\sqrt{{gn_0}/{m}+{\hbar^2k^2}/({4m^2})}$~\cite{pitaevskii2003} with the Fourier transform $g$ of the contact interaction~(\ref{eq:contact}) complemented by the one~(\ref{eq:tf:dd}) of the {\ddi}. With the definition~(\ref{eq:def:edd}) for $\edd$, (\ref{eq:bogo:dd}) implies that a dipolar uniform condensate is unstable for $\edd>1$, as phonons ($k\to 0$) acquire imaginary frequencies, the most unstable situation being in the case of a direction of the wavevector perpendicular to the orientation of the dipoles ($\alpha=\pi/2$). At first sight, this might seem counterintuitive: as dipoles side by side repel each other, one could conclude (wrongly) that the most unstable phonons correspond to those for which ${\bds k}$ is parallel to the dipoles. Figure~\ref{fig:phonon} shows how one can understand intuitively this behaviour.

\subsection{Trapped gas. Elongation of the cloud}\label{sec:elong}

\begin{figure}
\begin{center}
\includegraphics[width=4.5cm]{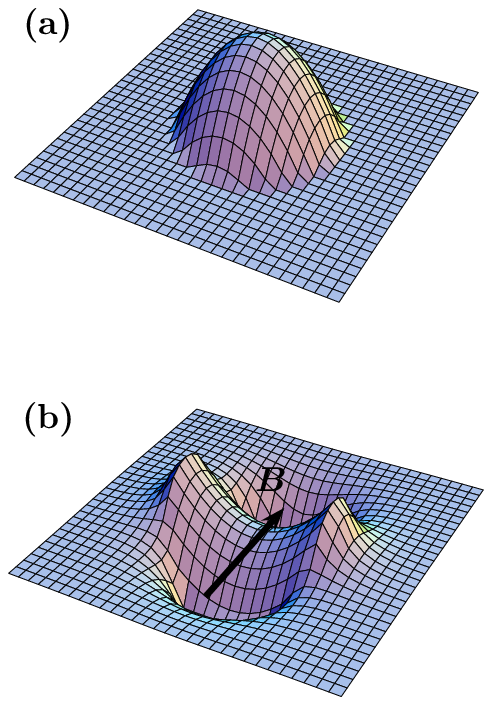}
\end{center}
\caption{(a) Inverted parabola density distribution $n({\bds r})$ in the Thomas Fermi regime in the absence of {\ddi}. (b) Saddle-like mean field dipolar potential (\ref{eq:phidd:sphere}) induced by the density distribution displayed in (a).}
\label{fig:elongation}
\end{figure}

Like in the case of a BEC with contact interactions, in the presence of an external trap (usually harmonic in experiments) new properties arise for a dipolar condensate. A prominent effect of the {\ddi} is to elongate the condensate along the direction $z$ along which the dipoles are oriented~\cite{santos2000,goral2000,yi2001}. This \emph{magnetostriction} effect (a change of the shape and volume of the atomic cloud due to internal magnetic forces) can be understood in a very simple way for a spherically symmetric trap (of angular frequency $\omega$) in the perturbative regime $\edd\ll 1$. To zeroth order, the density distribution is given, in the Thomas-Fermi limit, by $n({\bds r})=n_0(1-r^2/R^2)$, where $R$ is the Thomas-Fermi radius of the condensate (see figure~\ref{fig:elongation}a). One can then calculate to first order in $\edd$ the mean-field dipolar potential (\ref{eq:phidd}) created by this distribution; one finds~\cite{giovanazzi2002a}
\begin{equation}
\Phi_{\rm dd}({\bds r})
=\edd\frac{m\omega^2}{5}\left(1-3\cos^2\theta\right)\left\{
   \begin{array}{ccc}
   r^2 & {\rm if} & r<R\\
   \displaystyle\frac{R^5}{r^3} & {\rm if} & r>R
   \end{array}
\right.
\label{eq:phidd:sphere}
\end{equation}
\emph{i.e.} the dipolar mean field potential has the shape of a saddle, with minima located on the $z$ axis (see figure~\ref{fig:elongation}b). It is therefore energetically favorable for the cloud to become elongated along $z$. One can actually show that this conclusion remains valid even if the cloud is anisotropic, and for larger values of $\edd$~\cite{odell2004,eberlein2005,giovanazzi2006}.

Very recently, the spatial extent of a $^7$Li BEC was studied as a function of the scattering length close to a Feshbach resonance~\cite{pollack09}. For very small scattering lengths, the elongation effect due to the {\ddi} could be seen unambiguously, in spite of the small value of the magnetic dipole moment.

\subsection{Trapped gas. Geometrical stabilization}\label{sec:geometry}

A BEC with pure contact \emph{attractive} interactions ($a<0$) is unstable in the homogeneous case, but, in a trap, stabilization by the quantum pressure can occur for small atom numbers, namely if
\begin{equation}
\frac{N|a|}{a_{\rm ho}}\leqslant 0.58,
\label{eq:contact:collapse}
\end{equation}
where $N$ is the atom number and $a_{\rm ho}=\sqrt{\hbar/(m\omega)}$ is the harmonic oscillator length corresponding to the trap frequency $\omega$~\cite{ruprecht1995}. Here the trap has been supposed isotropic, but, for anisotropic traps, the dependence on the trap geometry is weak~\cite{gammal2001}.

\begin{figure}
\begin{center}
\includegraphics[width=8.5cm]{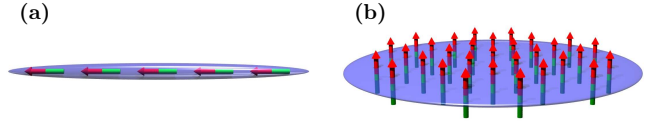}
\end{center}
\caption{Intuitive picture for the geometry-dependent stability of trapped a dipolar BEC. (a) In a prolate (cigar-shaped) trap with the dipoles oriented along the weak confinement axis of the trap, the main effect of the {\ddi} is attractive, which leads to an instability of the condensate. (b) In an oblate (pancake-shaped) trap with the dipoles oriented along the strong confinement axis, the {\ddi} is essentially repulsive, and the BEC is stable.}
\label{fig:stability:intuit}
\end{figure}

The situation is radically different in the case of a BEC with dipolar interactions. Due to the anisotropy of the {\ddi}, the partially attractive character of the interaction can be ``hidden'' by confining the atoms more strongly in the direction along which the dipoles are aligned. Let us consider for simplicity a cylindrically symmetric trap, with a symmetry axis $z$ coinciding with the orientation of the dipoles. The axial (resp. radial) trapping frequency is denoted $\omega_z$ (resp. $\omega_\rho$). It is then intuitively clear that for a prolate trap (aspect ratio $\lambda=\omega_z/\omega_\rho<1$), the {\ddi} is essentially attractive, and in such a trap a dipolar BEC should be unstable, even in the presence of a (weak) repulsive contact interaction [see figure \ref{fig:stability:intuit}(a)]. On the contrary, in a very oblate trap, the {\ddi} is essentially repulsive, leading to a stable BEC even in the presence of weak attractive contact interactions [see figure \ref{fig:stability:intuit}(b)]. One therefore expects that, for a given value of $\lambda$, there exists a critical value $\acrit$ of the scattering length below which a dipolar BEC is unstable; from the discussion above, $\acrit$ should intuitively be a decreasing function of $\lambda$, and the asymptotic value of $\acrit$ for $\lambda\to0$ (resp. $\lambda\to\infty$) should be positive (resp. negative).

\begin{figure}
\begin{center}
\includegraphics[width=7cm]{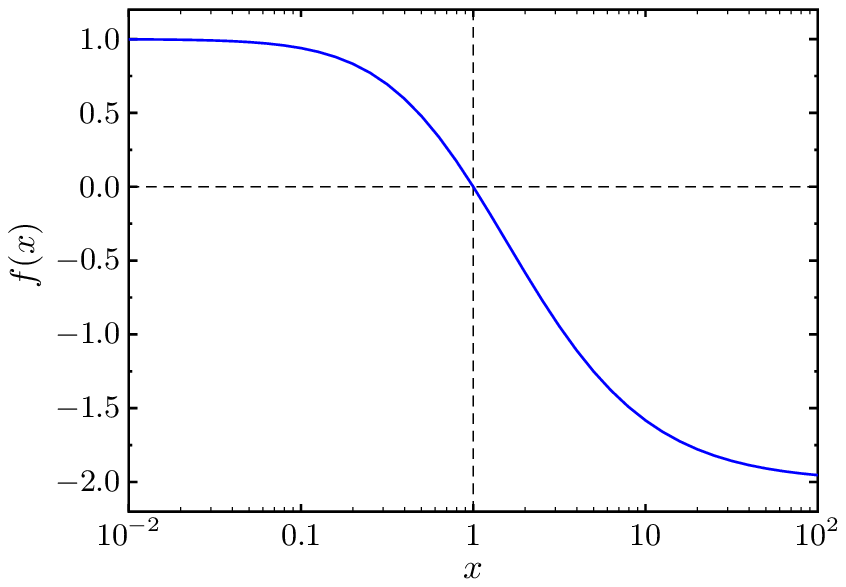}
\end{center}
\caption{The function $f(x)$ entering the calculation of the dipolar mean-field energy.}
\label{fig:fsx}
\end{figure}

A simple way to go beyond this qualitative picture and obtain an estimate for $\acrit(\lambda)$ is to use a variational method. For this purpose, we assume that the condensate wavefunction $\psi$ is gaussian, with an axial size $\sigma_z$ and a radial size $\sigma_\rho $ that we take as variational parameters:
\begin{equation}
\psi(r,z)=\sqrt{\frac{N}{\pi^{3/2}\sigma_\rho^2 \sigma_z a_{\rm
ho}^3}}\exp\left[-\frac{1}{2 a_{\rm ho}^2}\left(
\frac{r^2}{\sigma_\rho^2}+\frac{z^2}{\sigma_z^2}\right)\right].
\label{eq:ansatz:gauss}
\end{equation}
Here, $a_{\rm ho}=\sqrt{\hbar/(m\bar{\omega})}$ is the harmonic oscillator length corresponding to the average trap frequency $\bar{\omega}=(\omega_\rho^2\omega_z)^{1/3}$. Inserting Ansatz (\ref{eq:ansatz:gauss}) into the energy functional (\ref{eq:energy:func}) leads to the following expression for the energy:
\begin{equation}
E(\sigma_\rho,\sigma_z)=E_{\rm kin}+E_{\rm trap}+E_{\rm int},
\label{eq:energy:gaussian}
\end{equation}
with the kinetic energy
\begin{equation}
E_{\rm kin}=\frac{N\hbar\bar{\omega}}{4}\left(\frac{2}{\sigma_\rho^2}+\frac{1}{\sigma_z^2}\right),
\end{equation}
the potential energy due to the trap
\begin{equation}
E_{\rm trap}=\frac{N\hbar\bar{\omega}}{4\lambda^{2/3}}\left(2\sigma_\rho^2+\lambda^2\sigma_z^2\right),
\end{equation}
and the interaction (contact and dipolar) energy
\begin{equation}
E_{\rm int}=\frac{N^2\hbar\bar{\omega}\add}{\sqrt{2\pi}a_{\rm ho}}\frac{1}{\sigma_\rho^2\sigma_z}\left(\frac{a}{\add}-f(\kappa)\right).
\label{eq:gaussian:ddi}
\end{equation}
The dipolar contribution in the last part is most easily calculated in momentum space as
\begin{eqnarray}
E_{\rm dd}&=&\frac{1}{2}\int n({\bds r})n({\bds r}')\udd({\bds r}-{\bds r}')\,{\rm d}^3r\,{\rm d}^3r'\nonumber\\
&=&\frac{1}{2(2\pi)^3}\int\widetilde{\udd}({\bds k})\tilde{n}^2({\bds k})\,{\rm d}^3k,
\end{eqnarray}
where $\tilde{n}(\bds k)$ is the Fourier transform of the density distribution (and therefore, in this case, still a Gaussian). In (\ref{eq:gaussian:ddi}), $\kappa=\sigma_\rho/\sigma_z$ is the aspect ratio \emph{of the cloud} (which differs from the one of the trap due to the elongation induced by the {\ddi} as discussed above), and $f$ is  given by
\begin{equation}
f(\kappa)=\frac{1+2\kappa^2}{1-\kappa^2}-
\frac{3\kappa^2{\rm artanh}\sqrt{1-\kappa^2}}{(1-\kappa^2)^{3/2}}.
\label{eq:ffunc}
\end{equation}
The function $f(\kappa)$, displayed in figure~\ref{fig:fsx}, is monotonically decreasing, has asymptotic values $f(0)=1$ and $f(\infty)=-2$, and vanishes for $\kappa=1$ (implying that for an isotropic density distribution the dipole-dipole mean-field potential averages to zero).

\begin{figure}
\begin{center}
\includegraphics[width=13cm]{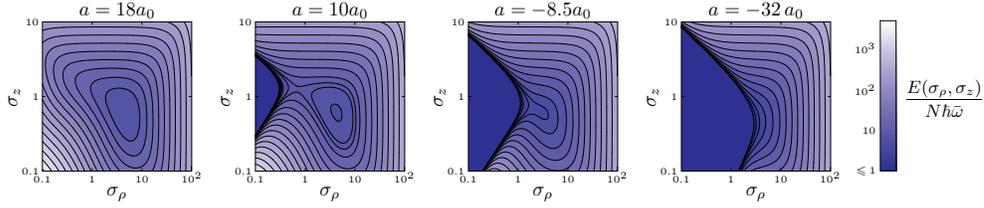}
\end{center}
\caption{The energy landscape $E(\sigma_\rho,\sigma_z)$ as a function of the variational parameters $\sigma_\rho$ and $\sigma_z$ of the Gaussian Ansatz, for a trap of aspect ratio $\lambda=10$, and various values of the scattering length $a$. When $a$ decreases, one goes from a \emph{global} minimum (at $a=18a_0$) to a \emph{local} minimum corresponding to a metastable condensate (at $a=10a_0$). This local minimum vanishes at $a=\acrit$ (here $-8.5a_0$). Below $\acrit$, the energy can be lowered without bound by forming an infinitely thin cigar-shaped cloud.}
\label{fig:stability:contours}
\end{figure}

To determine the stability threshold $\acrit(\lambda)$, one needs to minimize (\ref{eq:energy:gaussian}) with respect to $\sigma_\rho$  and $\sigma_z$ for fixed values of $N$, $\lambda$ and $\bar{\omega}$. For $a>\acrit$, one has a (at least local) minimum of energy for finite values of $\sigma_{\rho,z}$, while as soon as $a<\acrit$, no such minimum exists. Figure \ref{fig:stability:contours} shows contour plots of $E(\sigma_\rho,\sigma_z)$ for $N=20,000$, $\lambda=10$ and different values of $a$, clearly showing that $\acrit(10)\simeq-8.5\,a_0$ for the chosen parameters. In figure~\ref{fig:stability:diag}, the critical scattering length $\acrit(\lambda)$ obtained in this way is shown as a thick line for $\bar{\omega}=2\pi\times800$~Hz and $N=20,000$ atoms. In the limit $N\to\infty$, the asymptotic behaviour of this curve ($\acrit^\infty(0)=\add$ and $\acrit^\infty(\infty)=-2\add$) can be easily understood, as only the sign of the interaction term (\ref{eq:gaussian:ddi}) (which scales as $N^2$ and not as $N$ like the kinetic and potential energy) determines the stability. For an extremely pancake-shaped trap $\lambda\to\infty$, the cloud has an aspect ratio $\kappa\to\infty$, and, as $\displaystyle\lim_{x\to\infty}f(x)=-2$, the condensate is (meta-)stable only if $a>-2\add$. In the same way, one readily understands that for $\lambda\to0$, the critical scattering length is $\add$. The minimal value of $\lambda$ for which a purely dipolar condensate ($a=0$) is stable is the one for which $\kappa=1$ and is found numerically to be close to $\lambda\simeq5.2$~\cite{santos2000,goral2002a,yi2001,eberlein2005,koch2008}.

\begin{figure}
\begin{center}
\includegraphics[width=6cm]{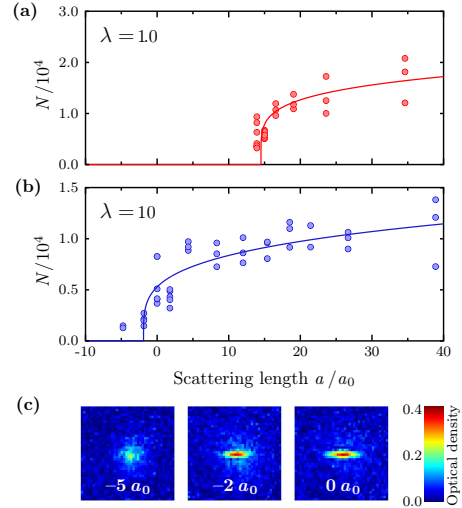}
\end{center}
\caption{Experimental observation of the geometry-dependent stability of a dipolar BEC. (a) BEC atom number $N$ as a function of $a$ for a spherical trap; $N$ vanishes for $a$ smaller than $\acrit\simeq15a_0$. (b) For an oblate trap ($\lambda=10$), one has $\acrit\simeq-2a_0$; such a trap can thus stabilize a purely dipolar BEC. In (a) and (b) the solid lines are fits to the empirical threshold law $(a-\acrit)^\beta$. (c) Sample images of the atomic cloud as a function of $a$ for $\lambda=10$.}
\label{fig:stability:loss}
\end{figure}

\begin{figure}
\begin{center}
\includegraphics[width=7.5cm]{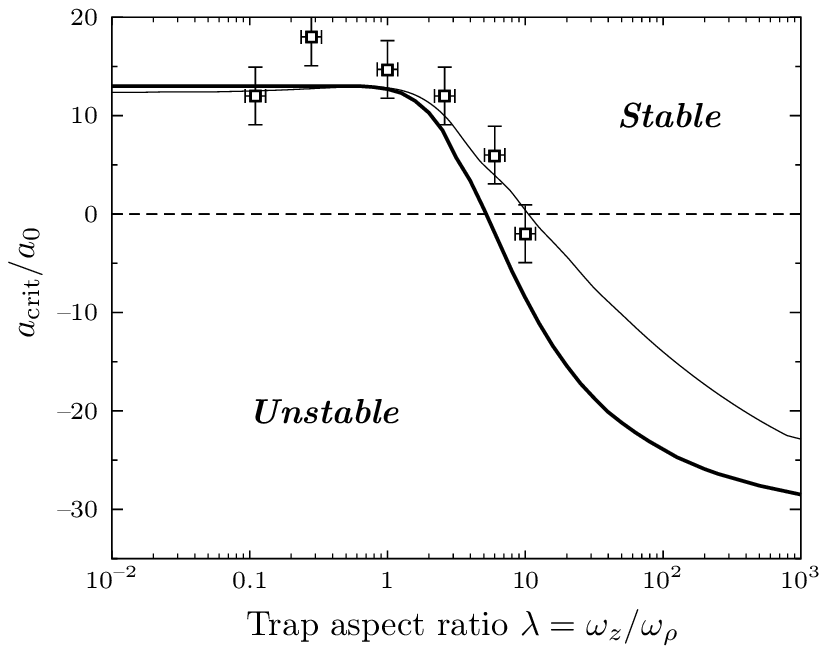}
\end{center}
\caption{Stability diagram of a dipolar condensate in the plane $(\lambda,a)$. The dots with error bars correspond to the experimental data~\cite{koch2008}; the thick solid line to the threshold $\acrit(\lambda)$ obtained using the Gaussian ansatz (\ref{eq:ansatz:gauss}) with $N=20,000$; the thin solid line to the numerical solution of the GPE (\ref{eq:gpe:dd})~\cite{bohn2008}.}
\label{fig:stability:diag}
\end{figure}

In~\cite{koch2008}, the influence of the trapping geometry on the stability of a {\chr} BEC was investigated experimentally.
A combination of an optical dipole trap and of one site of a long period (7~$\mu$m) optical lattice provided an harmonic trap cylindrically symmetric along the $z$ direction (along which the dipoles are aligned), with an aspect ratio $\lambda$ that could be varied over two orders of magnitude (from $\lambda\simeq0.1$ ---prolate trap--- to $\lambda\simeq10$ ---oblate trap---), while keeping the average trap frequency $\bar{\omega}=(\omega_\rho^2\omega_z)^{1/3}$ almost constant (with a value of $2\pi\times800$ Hz). Using the Feshbach resonance at 589~G, the scattering length was ramped adiabatically to a final value $a$ and the atom number in the BEC was measured. A typical measurement is shown in figure~\ref{fig:stability:loss}. When $a$ is reduced, the atom number decreases, first slowly, and then very abruptly when $a$ approaches a critical value $a_{\rm crit}$, below which no condensate can be observed. Figure \ref{fig:stability:diag} shows the measured value of $a_{\rm crit}$ as a function of $\lambda$. One clearly observes that for prolate traps, $a_{\rm crit}$ is close to $\add$, as expected from the discussion above, while for the most pancake-shaped trap $\lambda=10$ the critical scattering length is close to zero: for such a geometry, a \emph{purely dipolar} condensate is stable. The solid line is the stability threshold $a_{\rm crit}(\lambda)$ obtained by the gaussian Ansatz for a number of atoms $N=2\times10^4$, which shows a good agreement with the measurements. Note that for the parameters used in the experiment, the critical scattering length for pure contact interaction, given by (\ref{eq:contact:collapse}) would be $-0.3a_0$ for $\lambda=1$, which clearly shows that the instability is driven here by the {\ddi}.

To calculate the exact stability threshold, one needs to resort to a numerical solution of the GPE (\ref{eq:gpe:dd}); the result of such a calculation~\cite{bohn2008} is displayed as a thin line on figure \ref{fig:stability:diag} and shows a very good agreement with the data. The numerical solution reveals, for some values of the parameters $(\lambda,a)$ close to the instability region,  the appearance of `biconcave' condensates, where the density has a local minimum in the center of the trap~\cite{ronen2007a}.
\footnote{The experimental observation of such biconcave condensate (which show in a striking manner the long-range character of the {\ddi}) is difficult for several reasons: (i) the density dip does not survive in time of flight (which implies that in-situ imaging would be needed to detect it) (ii) it has a small contrast of only a few percent and (iii) the regions in the plane $(\lambda,a)$ where the biconcave condensate exist have a very small area. However, the use of potentials flatter than harmonic traps, such as a quartic or a box-like potential, should relax considerably the constraints (ii) and (iii) (S. Ronen, 2008, private communication).}

\subsection{Trapped gas. Thomas-Fermi regime}\label{sec:tf}

As we shall see in the next sections, a very important approximation in the case of dipolar gases is the so-called Thomas-Fermi (TF) limit, in which quantum pressure effects are neglected. Amazingly, the TF solutions for the ground state of the trapped BEC have the same inverted parabola shape as in the case of contact interactions. This has been  pointed out for the first time in~\cite{santos2003}, where, however, the trapping was restricted to the $z$-direction, while in the other directions the systems was assumed to be homogenous. The Thomas-Fermi approach was also used to study fermionic dipolar gases (see~\cite{dutta2006} and references therein).

The exact solutions of the dipolar BEC hydrodynamics in 3D were presented in a series of beautiful papers by O'Dell, Giovanazzi and Eberlein~\cite{odell2004,eberlein2005}. These authors have used oblate spheroidal coordinates and solved the TF equations for the ground state in cylindrically symmetric traps. They have also  considered the stability of the three most relevant perturbations: local density perturbations, ``scaling'' perturbations, and ``Saturn-ring'' perturbations.

\begin{figure}
\begin{center}
\includegraphics[width=8cm]{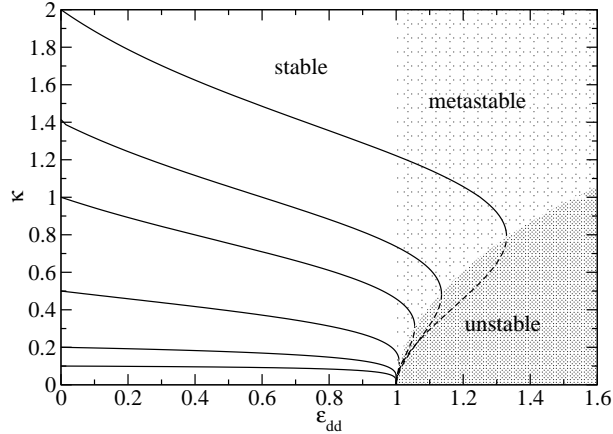}
\end{center}
\caption{Aspect ratio $\kappa$ of the condensate as a function of the dipole-dipole to $s$-wave coupling ratio $\varepsilon_{\rm dd}$. Each line is for a different trap aspect ratio $\gamma=\omega_z/\omega_x$, which can be read off by noting that $\kappa(\varepsilon_{\rm dd}=0)=\gamma$. When $0<\kappa<1$ the condensate is prolate; for $\kappa > 1$ it is oblate. Likewise, for  $0<\gamma<1$ the trap is prolate, and when $\gamma > 1$ the trap is oblate (figure courtesy of C. Eberlein).}
\label{fig:excitations:tf}
\end{figure}

In particular, the ground state density in the cylindrically symmetric case has the form:
\begin{equation}
n({\bds r})=n_0\left(1-\frac{\rho^2}{R_x^2}-\frac{z^2}{R_z^2}\right),
\end{equation}
for $n({\bds r})\geqslant 0$, where  $n_0=15N/(8\pi R_x^2R_z)$. These expressions are exactly the same as in the case of contact interactions. The difference is, of course, in the explicit expressions for the radii:
\begin{equation}
R_x=R_y=\left[\frac{15gN\kappa}{4\pi m \omega_x^2}\left\{1 + \varepsilon_{\rm dd}\left(\frac{3}{2}\frac{\kappa^2f(\kappa)}{1-\kappa^2}-1\right)\right\}\right]^{1/5},
\end{equation}
and $R_z=R_x/\kappa$. The condensate aspect ratio $\kappa$ is determined by the transcendental equation
\begin{equation}
3\kappa \varepsilon_{\rm dd}\left[\left(\frac{\omega_z^2}{2\omega_x^2}+1\right)\frac{f(\kappa)}{1-\kappa^2}-1\right]+ (\varepsilon_{\rm dd}-1)(\kappa^2-\omega_z^2/\omega_x^2)=0,
\end{equation}
where $f(\kappa)$ is defined in \eref{eq:ffunc}. A plot of the condensate aspect ratio as a function of $\varepsilon_{\rm dd}$ is shown in figure~\ref{fig:excitations:tf}. The TF approach is also extremely useful to study the dynamics of dipolar condensates, for instance their free expansion (see section \ref{subsec:tf:expand}).

\subsection{Trapped gas. Excitations}\label{sec:excitation}

The unusual properties of the ground state of dipolar condensates have their counterpart in the excitations of the system. They are expected to exhibit novel character and symmetries, as well as new types of instabilities. Indeed, as we shall see in the next subsection, even in the pancake traps with dipole moments polarized orthogonally to the pancake plane, the excitation spectrum reveals an instability, at the so-called roton-maxon minimum. Before discussing the pancake case, let us first consider in this section the case of a moderate aspect ratio of the trap.

The study of the excitations of a dipolar BEC should in principle be realized using the non-local Bogoliubov-de Gennes (BdG) equations. Such an approach is however technically very difficult, and for this reason approximate methods are useful. We discuss here the results of pioneering papers (see~\cite{yi2000,yi2001,goral2002a} and references therein). G\'oral and Santos~\cite{goral2002a} apply the dynamical variational principle, developed for the contact GPE in~\cite{perezgarcia1996, perezgarcia1997}. The idea consists in writing the time dependent condensate wave function in the Gaussian form
\begin{equation}
\psi(x,y,z,t)=A(t)\prod_{\eta=x,y,z}\exp\left[-\eta^2/2w_{\eta}(t)^2-i\eta^2\beta_{\eta}(t)\right],
\end{equation}
with time-dependent variational parameters  describing the Gaussian widths $w_{\eta}(t)$, and phases  $\beta_{\eta}(t)$, while $A(t)$ takes care for normalization. The dynamical variational principle implies equations of motions for the widths and phases. Stationary solutions of these equations describe the ground state BEC, small deviations from the ground state describe the lowest energy excitations. G\'oral and Santos consider the case of a polarized dipolar gas and discuss the influence of dipole-dipole forces on the stability of the condensate and the excitation spectrum. The authors extend their discussion of the ground state and excitations properties to the case of a gas composed of the two anti-parallel dipolar components.

\begin{figure}
\begin{center}
\includegraphics[width=8cm]{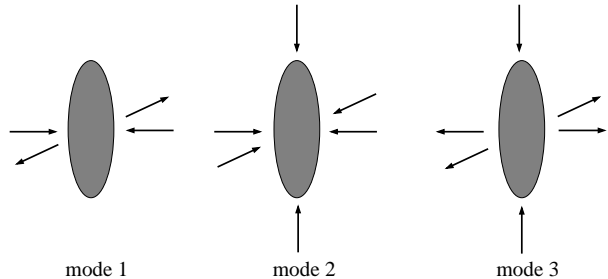}
\end{center}
\caption{Graphical representation of the oscillation modes of the condensate; the modes 1 and 3 are ``quadrupole''-like, the mode 2 is the breathing ``monopole'' mode.}
\label{fig:excitations:modes}
\end{figure}

One of the most interesting results of this paper is the study of the nature of the collapse instability. In the standard case of contact interactions three modes are relevant at low energies (see figure~\ref{fig:excitations:modes}): two ``quadrupole''-like modes (1 and 3), and one ``monopole'' mode (2). It is the latter one which becomes unstable at the collapse (when the scattering length changes sign). The frequency of the breathing monopole mode 2 goes to zero in a certain manner, namely, if $\gamma$ denotes the ratio of the non-linear energy to the trap energy, and $\gamma_{\rm c}$ is correspondingly its critical value, the frequency of the breathing monopole mode 2 goes to zero as $|\gamma-\gamma_{\rm c}|^{1/4}$~\cite{bergeman1997,singh1996} when $\gamma$ approaches criticality from below.

In the case of dipolar gases with dominant dipole interaction the situation is similar only for aspect ratios $\lambda\ll 1$ far from criticality, where the
lowest frequency mode is the breathing mode, and its frequency tends to zero as $|\gamma_{\rm dd}-\gamma_{c,{\rm dd}}|^{\beta}$, with $\beta\simeq 1/4$, when the energy of the dipolar interactions approach the criticality. Numerical analysis indicates that as one approaches the criticality the exponent $\beta$ remains close to $1/4$, but the geometry of the zero frequency mode is completely different: it attains the quadrupole character and becomes a superposition of modes 1 and 3. Very close to the criticality the exponent $\beta$ grows up to the value $\simeq 2$.  These results imply already that one should expect for the dipolar gas  a completely different character of the collapse dynamics than for a gas with contact interactions and negative scattering length. We will discuss it in detail in the next section.

The variational method employing the Gaussian ansatz is evidently the simplest approach to study the excitations and dynamics of the dipolar BECs, and for this reason it was used by many authors. After the seminal experiments of the JILA group with $^{85}$Rb, in which efficient, \emph{i.e.} practically loss-free control of the scattering length was achieved~\cite{donley2001}, Yi and You used this method to investigate the possibility of observing dipolar effects in the shape oscillations and expansion of a dipolar condensate~\cite{yi2002,yi2003}.

Direct solution of the Bogoliubov-de Gennes equations is, as we said, difficult, but not impossible. First of all, they become easier to solve in the Thomas-Fermi (TF) limit. The first step toward it was made in the ``roton-maxon'' paper of Santos {\it et al.}~\cite{santos2003}.  These authors considered an infinite pancake trap (\emph{i.e.} a slab) with dipoles oriented along the $z$-direction and in the TF limit. It turns out that the density profile in the TF limit is given by an inverted parabola, just like in the case of the standard BEC~\cite{pitaevskii2003}. The BdG equations for excitations close to the critical parameter region, where the ``rotonization'' of the spectrum appears, can be solved analytically in terms of series of Gegenbauer polynomials, as we discuss in the next subsection.

\begin{figure}
\begin{center}
\includegraphics[width=8cm]{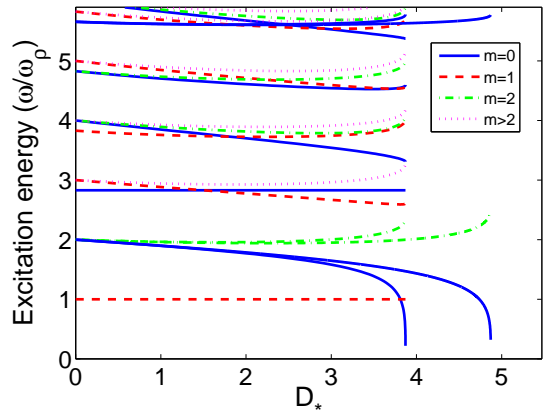}
\end{center}
\caption{Excitation frequencies as a function of the dipolar parameter $D_*=(N-1)m\cdd/(4\pi\hbar^2 a_{\rm ho,\rho})$ for a dipolar BEC in the JILA pancake trap~\cite{jin1996} ($\omega_z/\omega_{\rho}=\sqrt{8}$), with a zero scattering length. Plotted are modes with $m=0-4$. The three lines that extend to higher $D_*$ are the variational results of~\cite{goral2002a} (figure courtesy of S. Ronen).}
\label{fig:excitations:bdg}
\end{figure}

Very recently, in a beautiful paper, Ronen {\it et al.}~\cite{ronen2006a} developed an efficient method for solving BdG equations for the dipolar BEC with cylindrical symmetry. The algorithm is very fast and accurate, and is based on an efficient use of the Hankel transform. The authors study excitations in different geometries (from cigar to pancake; for typical results of BdG spectra, see figure~\ref{fig:excitations:bdg}) and in particular they calculate for the first time the dipolar condensate depletion for various regimes of parameters.

\subsection{Trapped gas. Roton-maxon spectrum}\label{sec:roton}

As we have mentioned, a dipolar gas exhibits two kinds of instabilities. In cigar-like traps, when the dipole is oriented along the trap axis, the dipolar interactions have an attractive character and the gas collapses. The collapse is similar to the case of the contact interactions with negative scattering length, but has a different geometrical nature, and different critical scaling behavior. There is, however, another instability mechanism that occurs even in quasi-2D pancake traps for dipoles polarized perpendicularly to the trap plane (along the $z$-axis). In this case, when the dipolar interactions are sufficiently strong, the gas, despite the quasi-2D trap geometry, feels the 3D nature of the dipolar interactions, \emph{i.e.} their partially attractive character. This so-called ``roton-maxon'' instability has been discovered in~\cite{santos2003}, and discussed by many authors since then.

In the original paper~\cite{santos2003} the authors considered an infinite pancake trap (slab geometry) with dipoles oriented along the $z$-direction perpendicular to the trap plane. The roton-maxon  physics occurs in the TF limit in the $z$-direction. The condensate was hence assumed to have a TF shape along the $z$-axis and constant amplitude with respect to $x$ and $y$ coordinates.  The 3D Bogoliubov-de Gennes equations were then solved. Here, we follow a somewhat simplified effective 2D approach of~\cite{dutta2009}, which nevertheless captures the main physics. The bosons are also  polarized along the $z$ direction, so that the dipolar interaction in momentum space reads  $V_{\rm dd}=C_{\rm dd}(k^2_z/k^2-1/3)$. We introduce also the ratio $\beta=3\edd/(4\pi)$. As we shall see below the cases of repulsive (attractive) contact interactions  with $g\ge 0$ ($g<0$) lead to qualitatively different physics. We first consider the case of positive $g$.

In the standard quasi-2D approach, one assumes that the condensate has a Gaussian shape along the $z$ direction, with the width determined by the harmonic potential. The problem is then projected onto 2D, by integrating over the condensate profile. Such an approach is valid when the chemical potential $\mu\ll \hbar\omega$. However, the roton-maxon instability occurs outside this regime and hence the standard quasi-2D approach cannot be employed to understand it.

However we can still use an effective quasi-2D approach, where we assume that the bosonic wave-function in the $z$ direction is given by the TF profile. In this way we may obtain an effectively 2D model with the dipole interactions ``averaged'' along the $z$ direction. This approach corresponds approximately to the approach of~\cite{santos2003}, with the only difference
that for the lowest energy branch of excitations we neglect their ``kinetic'' energy, \emph{i.e.} terms in the Bogoliubov-de Gennes equation involving derivatives with respect to $z$. The quantitative differences between the exact results of~\cite{santos2003}, and the present effective quasi-2D approach amount typically to $10$-$20$ percent in the entire regime of
$\beta\le 2$ (which is the relevant regime for rotonization~\cite{santos2003}). Qualitatively, both approaches describe the same physics, and the same mechanism of appearance of the instability, namely the momentum dependence of the dipole-dipole
interactions.

\begin{figure}
\begin{center}
\includegraphics[width=7cm]{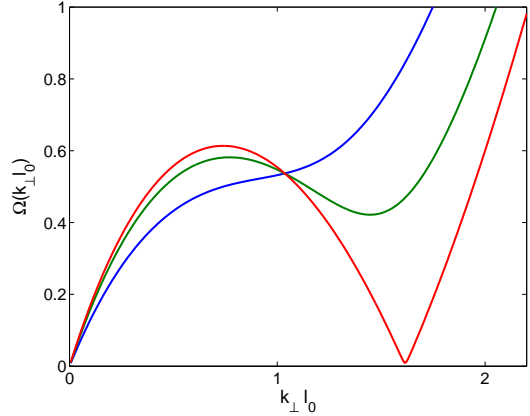}
\end{center}
\caption{The excitation spectrum $\Omega(k_{\perp})$ as a function of $ k_{\perp}$. The blue, black and red curves correspond to $g_{\rm 3d}=2.0, 3.0, 3.44$, respectively (figure courtesy of O. Dutta).}
\label{fig:excitations:roton}
\end{figure}

For the purpose of this review we apply yet another approximation, and use for simplicity a Gaussian profile in the $z$ direction with a variationally determined width $\ell$, which turns out to be of order of the TF radius $\ell\simeq \ell_{\rm TF}/\sqrt{5}$.  The dipolar interaction in 2D, after integrating out the $z$ direction, takes the form
$$
V_{\rm eff}=\frac{C_{\rm dd}}{4\pi\ell} \mathcal{V}(k_{\bot})
$$
where
\begin{equation}\label{veff}
\mathcal{V}(k_{\bot})= 1 - \frac{3}{2}\sqrt{\frac{\pi}{2}} k_{\bot}\ell {\rm erfc} \left [ k_{\bot}\ell /\sqrt{2} \right ] \exp \left [k^2_{\bot}\ell^2/2 \right ],
\end{equation}
and $k^2_{\bot}=k^2_x+k^2_y$, $\tilde{k}_{\bot}=k_{\bot}\ell$. The Hamiltonian that generated the Bogoliubov-de Gennes equations reduces then in the absence of the contact interactions, \emph{i.e.} for $g=0$, to $H = \sum_{{\bds k}_{\bot}} \Omega({\bds k}_{\bot}) b^{\dagger}_{{\bds k}_{\bot}} b_{{\bds k}_{\bot}}$, where $b^{\dagger}_{{\bds k}_{\bot}}$ and $b_{{\bds k}_{\bot}}$ are Bogoliubov quasi-particle operators. The excitation spectrum is given, in units of the trap frequency, by
\begin{equation}\label{exc}
\Omega^2(\tilde{k}_{\bot})=\frac{\tilde{k}^4}{4}+ g_{\rm 3d} \mathcal{V}(\tilde{k}_{\bot}) {\tilde{k}^2},
\end{equation}
where the dimensionless interaction strength is defined as $g_{\rm 3d}=m C_{\rm dd} n \ell/(4\pi\hbar^2)$. The interaction in~(\ref{veff}) is repulsive for small momenta and attractive in the high momentum limit (with a zero-crossing at $k_\perp \ell\simeq 1$). Due to this fact the properties of the  excitation spectrum in~(\ref{exc}) are very  different from that of bosons with contact  interactions. For any  bosonic density, $\Omega({\bds k}_{\bot})$ exhibits two regimes: (i) phonon spectrum for small momenta, and (ii) free particle spectrum for higher momenta. For $g_{\rm 3d}$ greater  than a certain critical value,  $\Omega({\bds k}_{\bot})$ has a minimum (see figure \ref{fig:excitations:roton}) at the intermediate momentum regime. Following Landau, the excitations around the  minimum are called ``rotons'' and $\Omega(\tilde{k}_0)$ is known as the ``roton gap''~\cite{santos2003}.  With increasing $g_{\rm 3d}$  the ``roton'' gap decreases and eventually vanishes for a critical particle density. When the critical density is exceeded, $\Omega(\tilde{k}_0)$ becomes imaginary and the condensate becomes unstable.

Once more we stress that in the presence of repulsive contact interactions, the roton instability occurs in pancake traps in the regime in which the standard quasi-2D approximation does not hold. The  condensate attains the TF profile in the $z$ direction and the systems starts to experience the 3D (partially attractive) nature of the dipolar forces~\cite{fischer2006}. The situation is however  different if the contact interactions are attractive ($g<0$). In that case the roton-instability appears already in the standard quasi-2D regime, when the chemical potential $\mu\ll \hbar\omega$, when  the condensate profile is Gaussian with the bare harmonic oscillator width. While in the previous case it was the attractive part of the dipolar interactions that led to the roton-instability, in this  quasi-2D situation with $g<0$ the situation is different. The dipolar interactions (which are in average repulsive in this case) stabilize for $\mu=(g+8\pi g_{dd}/3)n>0$ the phonon instability (which  in absence of {\ddi} leads to the well-known collapse for gases with $a<0$)~\cite{komineas2007}. Note that $\mu>0$ but may be kept well below $\hbar\omega_z$. However due to the momentum dependence of the {\ddi} (which acquires the form~(\ref{veff})), one encounters the roton instability for $-3/8\pi<\beta<\beta_{\rm cr}$ (where $\beta_{\rm cr}$ depends on the value of $g$), whereas for $\beta>\beta_{cr}$ the system  is stable (as long as it remains 2D~\cite{klawunn2008}). Note that in this case, a larger dipole strength stabilizes (sic!) the gas. The reason is that the roton in the quasi-2D scenario is not actually  induced by the attractive nature of the dipolar interactions at large momenta, but by the fact that $g<0$.
As a consequence, quasi-2D roton may occur for much lower momenta than $1/l_z$.

The presence of a roton minimum in the spectrum of elementary excitations may be revealed in various ways. On one hand, and using Landau superfluidity criterion~\cite{landau1980}, it is clear that the superfluid critical velocity is reduced in the  presence of a roton minimum~\cite{santos2003}. Another alternative experimental signature of the roton could be  provided at finite temperatures, where the  thermal activation of rotons may induce a ``halo'' effect in time-of-flight images~\cite{wang2008b}.  Finally, we would also like to point that (as discussed in  section 7.3), the presence of even a shallow roton minimum may alter dramatically the response of the system against a periodic driving~\cite{nath2009b-LS}.

The question whether ``there is a life after the roton instability'' was studied by many researchers, in the hope to find novel types of stable Bose superfluids, that would have supersolid character, \emph{i.e.} self-assembled density modulations. This is suggested by the fact that the instability occurs at a specific value of the momentum, indicating instability toward a non-uniform ground  state~\cite{pomeau1994,josserand2007}. The ultimate answer to the question is negative: the condensate undergoes a sequence of local collapses, as shown for the first time in~\cite{dutta2007a,komineas2007,shlyapnikov2006}. In~\cite{komineas2007} it is shown numerically that in the  mean field theory supersolid states of dipolar BEC are unstable.  The authors of~\cite{dutta2007b} use a variational ansatz with density wave modulations along the $z$-direction in a cylindrically symmetric  trap, and show that it is not stable for a dipolar gas, due to the roton instability. It can, however be stabilized, by allowing for admixture of a single component polarized Fermi gas. This idea was followed further in~\cite{dutta2008}, where the stability of dipolar bosons-fermions mixture in pancake cylindrically symmetric traps was investigated at $T=0$ using a variational approach. Fermions-induced interactions stabilize the system in such traps and allow for quantum phase transition from the Gaussian shape BEC to a supersolid state, characterized by a hexagonal density wave.

Interestingly, while fermions stabilize dipolar bosons leading to novel type of states, the opposite is also true: boson mediated interactions between polarized fermions may lead to appearance of the ``exotic'' Fermi superfluids with $p$-wave, $f$-wave, or even $h$-wave pairing~\cite{dutta2009}. All of these states exhibit topological order, and admit non-Abelian anionic excitations~\cite{read2000}, that can be used for topologically protected quantum information processing~\cite{tewari2007}.

The discovery of the roton instability in dipolar gases stimulated the search for supersolid structures and density modulations. Ronen {\it et al.} studied angular and radial roton instabilities in purely dipolar BECs in oblate traps~\cite{ronen2007a}, and have shown that in some situations the condensate attains a biconcave density profile (see the end of section~\ref{sec:geometry}). This results have been generalized to the case of finite temperature using the Hartree-Fock-Bogoliubov approach~\cite{ronen2007b}. Such structures are very sensitive to changes in the trapping potential; their relation to the roton instabilities for the case of a trapped dipolar BEC without and with a vortex has been studied in~\cite{wilson2008}. Dutta and Meystre~\cite{dutta2007a} predicted similar effects in anisotropic traps.

Very recently, the stability, excitations and roton instabilities have been discussed for the case of a dipolar BEC with a vortex~\cite{wilson2009}. Recent observation of the dipolar effects in Bloch oscillations with $^{39}$K~\cite{fattori2008a,fattori2008} with $s$-wave scattering tuned to zero stimulated studies of collective excitations and roton instabilities in multi-layer stacks of dipolar condensates. In~\cite{wang2008b,klawunn2008}, an enhancement of the roton instability was predicted. Note also that, interestingly, a gas of light-induced dipoles (see section \ref{sec:light:induced}) were predicted by Kurizki and coworkers to display roton instabilities~\cite{odell2003b}.

\section{Dynamics of a dipolar gas}\label{sec:expansion}

\subsection{Self-similar expansion in the Thomas-Fermi regime}\label{subsec:tf:expand}

In most experiments on BECs, all the information is obtained from an absorption image of the cloud taken after a period of free expansion (``time of flight''), which acts as a ``magnifier'' allowing to resolve optically the BEC (in a trap, the typical size of the BEC is on the order of a few microns, making it difficult to image it \emph{in situ} with a good resolution). It is therefore of great practical importance to describe the expansion of a condensate released from a trap.

In the case of a BEC with contact interactions in a harmonic trap of frequencies $\omega_i$ ($i=x,y,z$) in the Thomas-Fermi limit (see section \ref{sec:tf} above), a remarkable property allows for a very simple description of the free expansion~\cite{kagan1996,castin1996}: the in-trap density profile of the condensate (which is, in the Thomas-Fermi limit, an inverted parabola) is merely \emph{rescaled} upon time of flight. The Thomas-Fermi radii $R_i(t)$ at time $t$ are given by
\begin{equation}
R_i(t)=R_i(0) b_i(t),
\end{equation}
where the scaling parameters $b_i$ are solution of the following set of coupled differential equations:
\begin{equation}
\ddot{b}_i=\frac{\omega_i^2(0)}{b_ib_xb_yb_z}\quad(i=x,y,z),
\label{eq:scaling:contact}
\end{equation}
where $\omega_i(0)$ stands for the trap frequency along the direction $i$ before the trap is turned off.

The underlying reason for the existence of such a scaling solution is the fact that, for a parabolic density distribution $n$, the mean-field term $gn$ is also quadratic in the coordinates, yielding only quadratic terms in the hydrodynamic equations (\ref{eq:hd:continuity}) and (\ref{eq:hd:euler}). One can show that this property remains valid in the case of dipolar condensate in the Thomas-Fermi limit, as the following non-trivial property holds: if the density distribution $n$ is parabolic $n=n_0\,{\rm max}\left(0,1-\sum_i x_i^2/R_i^2\right)$, then the mean field potential (\ref{eq:phidd}) due to the {\ddi} is a quadratic form in the coordinates\footnote{As shown, in the much more restrictive case of a spherical trap and for $\edd\ll1$, in section \ref{sec:elong} (see (\ref{eq:phidd:sphere}) and figure \ref{fig:elongation}).}.

This property can be understood in the following way~\cite{eberlein2005}. Starting from the identity
\begin{equation}
\frac{1-3z^2/r^2}{r^3}=-\frac{\partial^2}{\partial
z^2}\frac{1}{r}-\frac{4\pi}{3}\delta({\bds r}), \label{eq:sugg}
\end{equation}
one can prove that
\begin{equation}
\fdd({\bds r})=-\cdd\left(\frac{\partial^2}{\partial z^2}\phi({\bds
r})+\frac{1}{3}n({\bds r})\right),
\label{eq:fdd:deriv}
\end{equation}
where
\begin{equation}
\phi({\bds r})=\int \frac{n({\bds r'})}{4\pi|{\bds r}-{\bds r'}|}\,{\rm d}^3r'.
\end{equation}
The last equality shows that the ``potential'' $\phi$ fulfills Poisson's equation
$\triangle\phi=-n$. From this, one deduces that the most general form of $\phi$,
when one has a parabolic density distribution $n$, is a polynomial of order two in the variables $(x^2,y^2,z^2)$, and thus, from (\ref{eq:fdd:deriv}), one deduces that $\Phi_{\rm dd}$ is also quadratic in $(x,y,z)$. The actual analytical calculation of the coefficients of this quadratic form, carried out for the cylindrically symmetric case in~\cite{odell2004,eberlein2005} and for the general anisotropic case in~\cite{giovanazzi2006}, is far from trivial, especially in the latter case, which involves a generalization of (\ref{eq:ffunc}) to two cloud aspect ratios.

One can then generalize Thomas-Fermi scaling equations (\ref{eq:scaling:contact}) to the case of a condensate with both contact and dipolar interactions. The corresponding set of differential equations now reads
\begin{equation}
\ddot{b}_i=\frac{\omega_i^2(0)}{b_ib_xb_yb_z}\left[1+\edd F(b_x,b_y,b_z)\right]\quad(i=x,y,z).
\label{eq:scaling:ddi}
\end{equation}
where the function $F$ includes the effect of the {\ddi}~\cite{giovanazzi2006}. Solving these equations, one can for example study the time evolution, during free expansion, of the cloud aspect ratio, to reveal the effects of the dipolar interaction (see section below).

\subsection{A quantum ferrofluid}\label{sec:ferro}

The superfluid hydrodynamic equations describing e.g. the expansion of a condensate are modified by the long-range, anisotropic {\ddi}. In classical fluids, such magnetic interactions modifying the hydrodynamic properties can be observed in \emph{ferrofluids}, which are colloidal suspensions of nanometric ferromagnetic particles~\cite{rosensweig}. In that sense, a dipolar condensate can be called a \emph{quantum ferrofluid}. As seen in the preceding section, a clear and simple way to demonstrate the effect of the {\ddi} is to study the expansion of the condensate when it is released from the trap.

\begin{figure}
\begin{center}
\includegraphics[width=13cm]{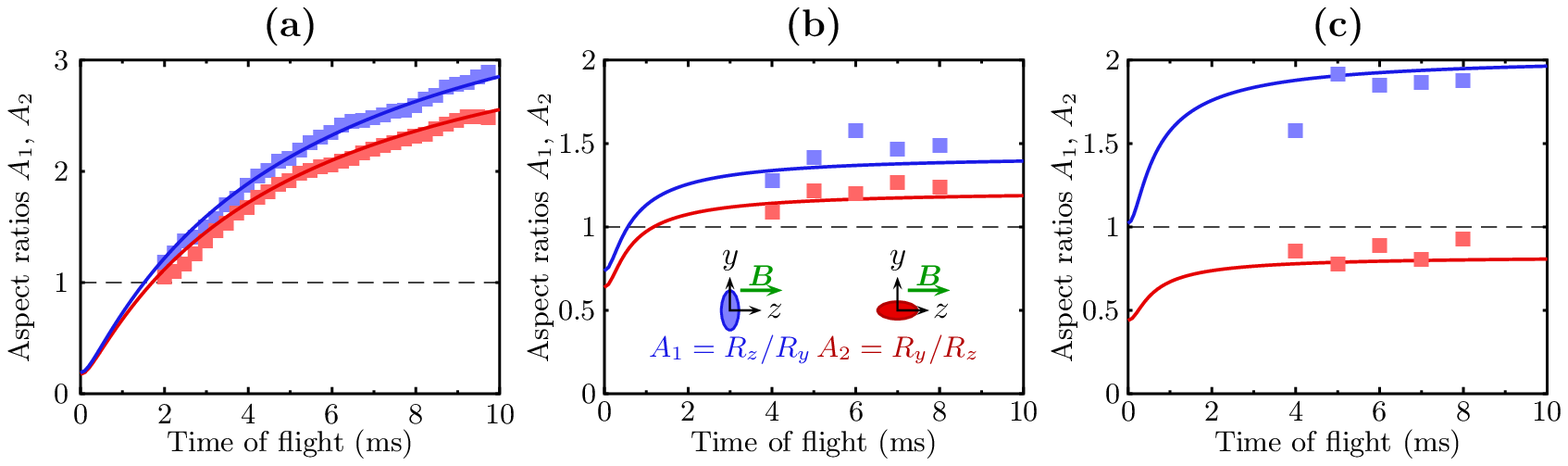}
\end{center}
\caption{Free expansion of a dipolar condensate for two different orientations of the dipoles with respect to the trap axes. The squares are the experimental results; the solid lines are the prediction of the scaling equations (\ref{eq:scaling:ddi}) without any adjustable parameter. (a) Perturbative regime $\edd=0.16$, the {\ddi} only gives a small departure from the contact interaction prediction (data taken from~\cite{giovanazzi2006}). (b) Perturbative regime (for a different trap geometry) $\edd\simeq0.16$ far above the Feshbach resonance, the effect of the {\ddi} is similar to the case (a). (c) Closer to the resonance, with $\edd=0.75$; in that case, the {\ddi} is strong enough to inhibit the usual inversion of ellipticity in time of flight. Data in (b) and (c) are taken from~\cite{lahaye2007}. }
\label{fig:expand:nature}
\end{figure}

By measuring the aspect ratio of the condensate as a function of the expansion time for two different orientations of the dipoles with respect to the trap axes, the elongation of the cloud along the magnetization direction could be clearly observed in~\cite{stuhler2005,giovanazzi2006}. However, since $\edd\simeq0.16$ for {\chr} away from Feshbach resonances, the effect, shown in figure \ref{fig:expand:nature}(a), is small. By using the 589~G Feshbach resonance in order to decrease the scattering length $a$ and thus enhance $\edd$, a much larger effect of the {\ddi} on the expansion dynamics was observed in~\cite{lahaye2007}. For example, by increasing $\edd$ to values close to one, the usual inversion of ellipticity of the condensate during the time of flight is inhibited by the the strong dipolar forces which keep the condensate elongated even during the expansion, as shown for example by the red curve in figure~\ref{fig:expand:nature}(c).

\subsection{Collapse dynamics}

The dynamics of a condensate with pure contact interactions when the scattering length is ramped to a negative value, thus making the condensate unstable, is extremely rich: one observes a fast implosion (``collapse'') of the condensate, followed by inelastic losses and a subsequent 'explosion' of the remnant condensate accompanied by energetic bursts of atoms. The behaviour of those ``Bose-Novae'' has been extensively studied with $^{85}$Rb and $^{7}$Li condensates~\cite{donley2001,gerton2000,roberts2001,sacket1999}. More recently, the formation of soliton trains during collapse has been reported~\cite{strecker2002,cornish2006}. It is then natural to ask whether the collapse of a dipolar condensate displays some specific features arising from the long range and anisotropic character of the {\ddi}.

\begin{figure}
\begin{center}
\includegraphics[width=7.5cm]{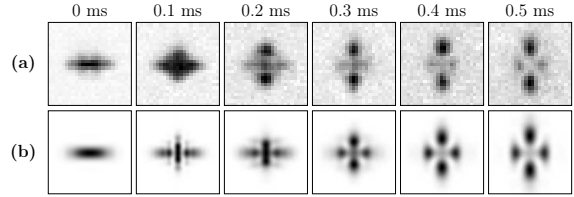}
\end{center}
\caption{(a) Experimental images of a dipolar condensate after collapse and explosion, as a function of the time $t_{\rm hold}$ between the crossing of the critical scattering length for instability and the release from the trap. The time of flight is 8~ms. (b) Results of a numerical simulation of the collapse dynamics, without any adjustable parameter. The field of view is 130 $\mu$m $\times$ 130 $\mu$m.}
\label{fig:collapse:tof}
\end{figure}

In~\cite{lahaye2008}, the collapse dynamics of a dipolar $^{52}$Cr BEC when the scattering length $a$ is decreased (by means of a Feshbach resonance) below the critical value  for stability $a_{\rm crit}$ was investigated experimentally. A BEC of typically 20,000 atoms was created in a trap with frequencies $(\nu_x,\nu_y,\nu_z) \simeq (660,400,530)$~Hz at a magnetic field $\sim 10$~G above the 589~G Feshbach resonance, where the scattering length is $a \simeq 0.9 \,a_{\rm bg}$. The scattering length $a$ was then ramped down rapidly to a value $a_{\rm f}  = 5 \,a_0$, which is below the collapse threshold $\acrit\simeq13a_0$. After the ramp, the system evolved for an adjustable time $t_{\rm hold}$ and then the trap was switched off. The cloud was then imaged after time of flight. The atomic cloud had a clear bimodal structure, with a broad isotropic thermal cloud, well fitted by a Gaussian, and a much narrower, highly anisotropic central feature, interpreted as the remnant BEC. Figure \ref{fig:collapse:tof}(a) shows the time evolution of the condensate when varying $t_{\rm hold}$. From an initial shape elongated along the magnetization direction $z$, the condensate rapidly develops a complicated structure with an expanding, torus-shaped part close to the $z = 0$ plane. Interestingly, the angular symmetry of the condensate at some specific times (e.g. at $t_{\rm hold} = 0.5$~ms) is reminiscent of the $d-$wave angular symmetry $1-3\cos^2\theta$ of the {\ddi}.
Figure \ref{fig:collapse:tof}(b) displays the column density $\int \left|\psi({\bds r})\right|^2\,{\rm d}x$ obtained from a numerical simulation of the three-dimensional GPE
\begin{equation}
\fl
i\hbar\frac{\partial \psi}{\partial t}=\left[\frac{-\hbar^2}{2m}\triangle +V_{\rm trap}+\int U({\bds r}-{\bds r}',t)\left|\psi({\bds r}',t)\right|^2\,{\rm d}^3{r}'\right.-\left.\frac{i\hbar L_3}{2}\left|\psi\right|^4\right]\psi,
\end{equation}
where
\begin{equation}
U({\bds r},t)=\frac{4\pi\hbar^2a(t)}{m}\delta({\bds r})+\frac{\mu_0\mu^2}{4\pi}\frac{1-3\cos^2\theta}{r^3}
\end{equation}
stands for both contact and dipolar interactions. The non-unitary term proportional to $L_3 \sim 2\times 10^{-40} \;{\rm m}^6/{\rm s}$ describes three-body losses that occur close to the Feshbach resonance. The agreement between the experimental data and the simulation, performed without any adjustable parameter, is excellent.

The observed cloverleaf patterns are caused by the anisotropic collapse and the subsequent dynamics of the system: when the atomic density grows due to the attractive interaction, three-body losses predominantly occur in the high-density region. The centripetal force is then decreased, and the atoms that gathered in this narrow central region are ejected due to the quantum pressure arising from the uncertainty principle. The kinetic energy is supplied by the loss of the negative interaction energy. As the collapse occurs mainly in the $x-y$ direction due to anisotropy of the {\ddi} (in the absence of inelastic losses, the condensate would indeed become an infinitely thin cigar-shaped cloud along $z$, see section~\ref{sec:geometry}), and therefore the condensate ``explodes'' essentially radially, producing the anisotropic shape of the cloud.

During the collapse, the BEC atom number, which was initially $N_{\rm BEC}(0) \simeq 16,000$, dropped to a value $\sim 6,000$. The missing atoms very likely escaped from the trap as energetic molecules and atoms produced in three-body collisions. The simulated atom number as a function of $t_{\rm hold}$ matched very well the experimental data.

\begin{figure}[b!]
\begin{center}
\includegraphics[width=7.5cm]{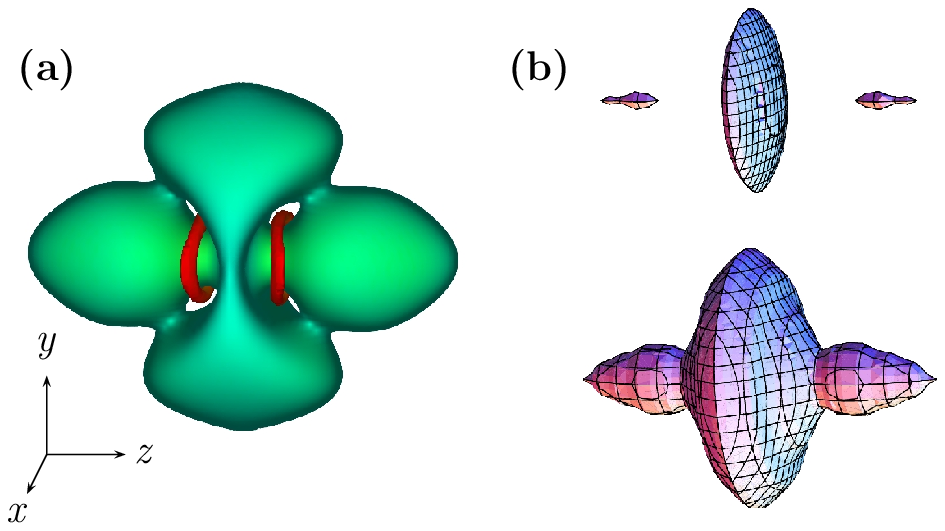}
\end{center}
\caption{Iso-density surfaces of a dipolar BEC after collapse and explosion. (a), simulation result, showing the location of vortex rings (in red), for the conditions of~\cite{lahaye2008}. (b), experimentally reconstructed iso-density surfaces (using the inverse Abel transform) for high and low density (top and bottom, respectively) in the case of a cylindrically symmetric situation~\cite{metz2009}.}
\label{fig:vortex:rings}
\end{figure}

The numerical simulation gives access not only to the density $|\psi({\bds r})|^2$, but also to the phase $S({\bds r})$ of the order parameter $\psi$ (\emph{i.e.} to the velocity field ${\bds v}=\hbar{\bds\nabla}S/m$) and reveals the generation of vortex rings. Figure~\ref{fig:vortex:rings}(a) shows the simulated in-trap iso-density surface of a condensate at $t_{\rm hold} = 0.8$~ms and the location of the vortex rings (shown as red curves). The mechanism responsible for the formation of vortex rings can be understood intuitively as follows. During the collapse, due to the strong anisotropy of the {\ddi}, the atoms ejected in the $x-y$ plane flow outward, while the atoms near the $z$ axis still flow inward, giving rise to the circulation. Thus, the vortex-ring formation is specific to the $d$-wave collapse induced by the {\ddi}. Although the vortex rings are not observed directly in the experiment (even when reconstructing the 3D density distribution by means of the inverse Abel transform), the excellent agreement between the experiment and the simulations strongly suggests the creation of vortex rings during the collapse.

\begin{figure}[t]
\begin{center}
\includegraphics{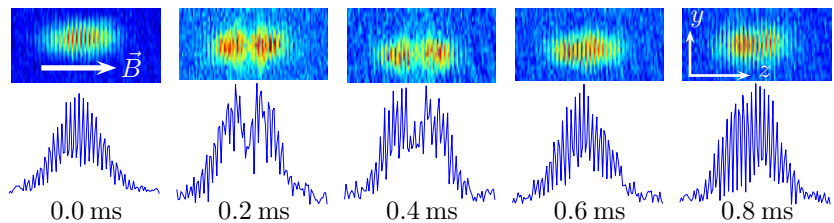}
\end{center}
\caption{Interference pattern of independent condensates for different holding times $t_{\rm hold}$~\cite{metz2009}.}
\label{fig:collapse:coherent}
\end{figure}

In~\cite{metz2009}, the collapse dynamics of a dipolar BEC was studied for different trap geometries, form prolate to oblate traps. The latter being created by superimposing a large period optical lattice onto the optical trap, it was possible to prepare independent condensates, let them collapse by changing the scattering length, and then release the confinement. The observation of high contrast interference fringes after the clouds overlapped (see figure~\ref{fig:collapse:coherent}) proved for the first time that the post-collapse remnant clouds are truly coherent matterwave fields.

The collapse observed here has the same physical origin as the phonon instability discussed in section \ref{sec:phonon:instability}, and should be distinguished from other possible collapse mechanisms, in particular the roton instability discussed in section \ref{sec:roton}. Note also that the phonon instability in two-dimensional geometries~\cite{nath2008-LS} does not necessarily lead to collapse, as we discuss in section \ref{subsec:patterns}.

\section{Non linear atom optics with dipolar gases}\label{sec:nlin:ao}

As mentioned in section 4,
BEC physics is inherently nonlinear due to the
interparticle interactions. Non-dipolar BECs
obey the non-linear Schr\"odinger equation (NLSE) (\ref{eq:gpe:cont}), which is identical to that
appearing in nonlinear optics of Kerr media.
Striking resemblances between both fields have been observed,
including nonlinear atom-optics phenomena as
four-wave mixing~\cite{deng1999-LS},
BEC collapse~\cite{donley2001}, and the creation of bright~\cite{strecker2002,khaykovich2002-LS}, dark~\cite{burger1999-LS,denschlag2000-LS,becker2008-LS} and gap~\cite{eiermann2004} solitons.
Dipolar BECs obey the NLSE (\ref{eq:gpe:dd}), where the nonlinearity
is intrinsically nonlocal, due to the long-range character of the {\ddi}.
Nonlocality appears in many different physical systems, including
plasmas~\cite{litvak1975-LS},
where the nonlocal
response is induced by heating and ionization,
nematic liquid crystals, where it is the result
of long-range molecular interactions~\cite{conti2003-LS},
and also photorefractive media~\cite{rotschild2005-LS}.
Most interestingly,
nonlocality plays a crucial role in the physics of solitons and modulational
instability~\cite{krolikowski2001-LS,bang2002}.
In this section, we review some recent results concerning the
nonlinear atom optics with dipolar BECs, including
qualitatively new phenomena in bright and dark solitons
(for recent works on vector and discrete solitons see
~\cite{tikhonenkov2008b-LS} and~\cite{gligoric2008-LS}), vortices and
pattern formation.

\subsection{Solitons}

Since the seminal works of Zakharov~\cite{zakharov1972-LS}
it is known that the 1D NLSE with focusing local cubic nonlinearity
supports the existence of localized waves that travel
with neither attenuation nor change of shape due to the
compensation between dispersion and nonlinearity.
These so-called bright solitons occur in diverse fields, most prominently
in nonlinear optics~\cite{stegeman1999-LS}.
Matter-wave bright solitons
have been observed in quasi-1D condensates with
$a<0$~\cite{strecker2002,khaykovich2002-LS}. The quasi-1D condition
requires a tight transversal harmonic trap
of frequency $\omega_\perp$ such that $\hbar\omega_\perp$ exceeds
the mean-field interaction energy. This in turn
demands the transversal BEC size to be smaller than the soliton width.
When this condition is violated the soliton becomes unstable against
transversal modulations, and hence multi-dimensional solitons
are not stable in non-dipolar BECs.

Remarkably the latter is not necessarily true in the
presence of nonlocal nonlinearity. In particular, any symmetric
nonlocal nonlinear response with positive definite Fourier
spectrum has been mathematically shown to arrest collapse
in arbitrary dimensions~\cite{bang2002}.
Multidimensional solitons have been experimentally
observed in nematic liquid crystals~\cite{peccianti2004-LS}
and in photorefractive screening media~\cite{shih1997-LS}, as well as in classical ferrofluids~\cite{richter2005}.
Multidimensional
solitons have been also discussed in BECs with short-range
interactions, by considering the collapse inhibition induced by
the first nonlocal correction to the local
pseudopotential~\cite{krolikowski2001-LS,rosanov2003-LS}.
However, this occurs for an extremely small BEC size~\cite{rosanov2003-LS},
which, except for the case of a very small particle
number, leads to extremely large densities, at which
three-body losses destroy the BEC.

%%%%%%%%%%%%%%%%%%%%%%%%%%%%%%%%%%%%%%%%%%%%%%%%%%%%%%%%%%%%
%% FIGURE 1-2DS
\begin{figure}[t]
\begin{center}
\includegraphics[width=6.2cm]{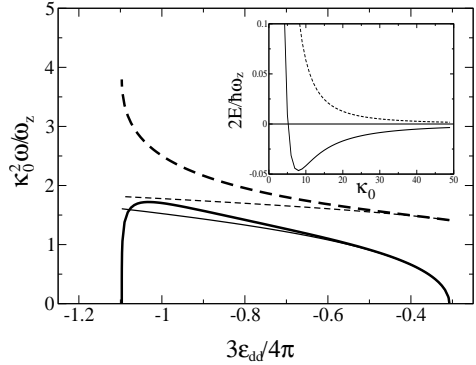}
\end{center}
\vspace*{-0.2cm}
\caption{Breathing (bold solid) and $m=\pm 2$ quadrupole (bold dashed) mode of a 2D bright soliton
for $\tilde g=20$, with $\kappa_0=l_\rho/l_z$ the ratio between the condensate widths perpendicular to and along the dipole orientation. Results from a reduced 2D NLSE~\cite{pedri2005} are shown in thin lines. Inset: $E(\kappa_0)$ for $\tilde g=500$, and
$\edd=-0.42$ (dashed) and $\edd=-0.84$ (solid).}
\label{fig:1-2ds}
\vspace*{-0.1cm}
\end{figure}
%%%%%%%%%%%%%%%%%%%%%%%%%%%%%%%%%%%%%%%%%%%%%%%%%%%%%%%%%%%%

Dipolar BECs on the contrary introduce a nonlocality at a much larger
length scale. As a consequence, and
in spite of the fact that, due to the anisotropy of the {\ddi},
the non local nonlinear response is not positive definite,
2D bright solitons may become stable under appropriate conditions~\cite{pedri2005}.
This may be understood from a simplified discussion where
we consider no trapping in the $xy$-plane and a strong
harmonic confinement with frequency $\omega_z$ in the $z$-direction, along which
the dipoles are oriented.
A good insight on the stability of 2D solitons may be obtained from a
Gaussian ansatz
$\Psi(\bds r)\propto\exp(-\rho^2/2l_\rho^2-z^2/2l_z^2)$,
where $l_z=\sqrt{\hbar/m\omega_z}$, and $l_\rho=\kappa_0 l_z$ is the $xy$-width.
Introducing this ansatz into the nonlocal NLSE (\ref{eq:gpe:dd}) we
obtain the system energy, which up to a constant is
\begin{equation}
E=\frac{\pi\hbar\omega_z}{\kappa_0^2}
\left\{ 2\pi+\tilde g \left [ 1-\edd f(\kappa_0) \right ] \right\}
\end{equation}
where $\tilde g=g/\sqrt{2\pi}\hbar\omega_zl_z^3$, and
$f(\kappa)$ is defined in (\ref{eq:ffunc}).
As mentioned above, in the absence of {\ddi} ($\edd=0$),
2D localized solutions are unstable, since
$E(\kappa_0)\propto \kappa_0^{-2}$ either grows
with $L_\rho$ (collapse instability)
or decreases with $L_\rho$ (expansion instability).
Dipolar BECs are remarkably different due to the additional dependence $f(\kappa_0)$, which
may allow for a minimum in $E(\kappa_0)$ (inset in figure~\ref{fig:1-2ds}), \emph{i.e.} for a stable localized wavepacket.
From the asymptotic values $f(0)=-1$ and $f(\infty)=2$ localization is
just possible if $\edd \tilde g  < 2\pi + \tilde g<-2\edd\tilde g$.
A simple inspection shows that this condition is fulfilled only if $\edd<0$, \emph{i.e.} for the {\ddi} tuned with rotating fields (see section \ref{sec:tuning}). Note also that if $Na/l_z\gg 1$, the stability condition reduces to
$|\edd|>1/2$. The anisotropic character of the {\ddi} becomes particularly relevant when relaxing the quasi-2D condition.
In particular, solitons in 3D dipolar BECs are fundamentally unstable
against collapse, as reflected in the decrease of the frequency of the breathing mode of the soliton for larger
values of $\edd$~\cite{pedri2005}
 (figure~\ref{fig:1-2ds}).

%%%%%%%%%%%%%%%%%%%%%%%%%%%%%%%%%%%%%%%%%%%%%%%%%%%%%%%%%%%%%%%%%%%%%%%%%%%%%%%%%%%
%% FIGURE 2-PI
\begin{figure}[t]
\begin{center}
\includegraphics[width=8cm]{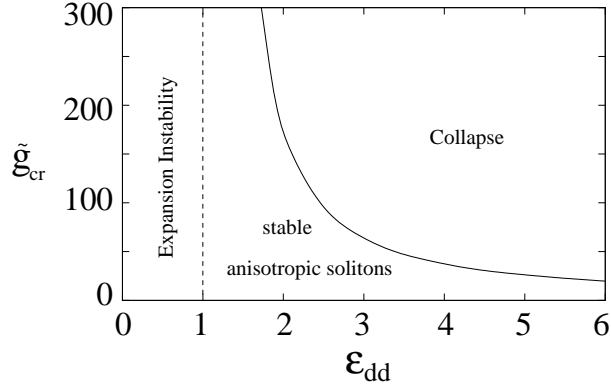}
\end{center}
\vspace*{-0.2cm}
\caption{Stability diagram of an anisotropic soliton as a function
of $\edd$ and $\tilde g_{\rm cr}=g_{\rm cr}/\sqrt{2\pi}l_z$, where for $g>g_{\rm cr}$
the soliton is unstable against collapse even for $\edd>1$.}
\label{fig:2-pi}
\vspace*{-0.2cm}
\end{figure}
%%%%%%%%%%%%%%%%%%%%%%%%%%%%%%%%%%%%%%%%%%%%%%%%%%%%%%%%%%%%%%%%%%%%%%%%%%%%%%%%%%%

Although $\edd<0$ is attainable for magnetic dipoles in a rotating magnetic field,
the combination with Feshbach resonances to reduce the contact interactions makes
experiments, e.g. in Chromium, very complicated. Recently a similar idea, but without the necessity
of dipole tuning, has been proposed by Tikhonenkov, Malomed and Vardi~\cite{tikhonenkov2008}.
In this new proposal the dipoles are considered as polarized on the 2D plane.
A similar Gaussian ansatz on the $xy$-plane as above
with unequal widths $l_x$ and $l_y$ shows the appearance of a minimum
in the energy functional $E(l_x,l_y)$ for $\edd>1$, and thus the existence of stable anisotropic solitons.
However, even if the latter condition is fulfilled there is a critical universal value $\tilde g_{\rm cr}$
(which decreases with $\edd$) such that for $\tilde g>\tilde g_{\rm cr}$ the minimum of $E(l_x,l_y)$
disappears~\cite{nath2009-LS} (see figure~\ref{fig:2-pi}).
As a consequence, contrary to the case of
isotropic solitons (with $\edd<0$) there is a critical number of particles per soliton even if the soliton remains 2D.
In addition, as for the case of isotropic solitons, anisotropic solitons are also unstable in 3D environments.

%%%%%%%%%%%%%%%%%%%%%%%%%%%%%%%%%%%%%%%%%%%%%%%%%%%%%%%%%%%%%%%%%%%%%%%%%%%%%%%%%%%
%% FIGURE 0-INE
\begin{figure}[t]
\begin{center}
\includegraphics[width=3.8cm]{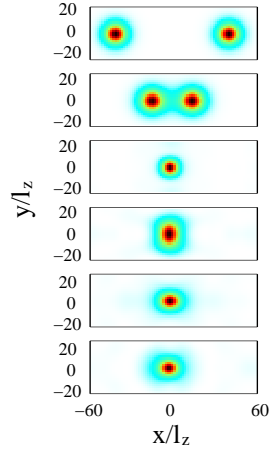}
\end{center}
\vspace*{-0.2cm}
\caption{Density plot of the
fusion of two dipolar 2D solitons for $\tilde g=20$, $\edd=-2.1$, and
initial relative momentum $kl_z=0.01$ along $x$. From top to bottom ,$\omega_z t/2000=0,1,2,3,4, \;{\rm and}\;5$.}
\label{fig:2-2ds}
\vspace*{-0.1cm}
\end{figure}
%%%%%%%%%%%%%%%%%%%%%%%%%%%%%%%%%%%%%%%%%%%%%%%%%%%%%%%%%%%%%%%%%%%%%%%%%%%%%%%%%%%

A major difference between bright solitons in non-dipolar and dipolar BECs concerns the soliton-soliton
scattering properties. Whereas solitons in 1D non-dipolar BECs scatter elastically,
the scattering of dipolar solitons is inelastic due to the lack of integrability~\cite{krolikowski2001-LS}.
The solitons may transfer center-of-mass energy into internal vibrational modes, resulting
in intriguing scattering properties, including soliton fusion~\cite{pedri2005} (see figure~\ref{fig:2-2ds}),
the appearance of strong inelastic resonances~\cite{nath2007},
and the possibility of observing 2D-soliton spiraling
as that already observed in photo-refractive materials~\cite{shih1997-LS}.

For defocusing nonlinearity ($a>0$) the local NLSE supports dark-soliton
solutions, \emph{i.e.} density notches (accompanied by phase slips) that propagate with no change
of shape, again due to the compensation between dispersion and nonlinearity~\cite{zakharov1973-LS}.
Dark solitons have been created in non-dipolar quasi-1D BECs~\cite{burger1999-LS,denschlag2000-LS,becker2008-LS},
but become fundamentally unstable in higher dimensions against vibrations of the nodal plane which
lead to the so-called \emph{snake instability}. This
instability which was previously studied in the context of nonlinear optics~\cite{tikhonenko1996,mamaev1996}
leads in the context of non-dipolar BEC to the soliton break-down
into vortex rings and sound excitations~\cite{muryshev1999,feder2000,anderson2001,muryshev2002-LS}.
On the contrary, a dark soliton in a dipolar BEC may become stable in a 3D environment~\cite{nath2008-LS} if the BEC is placed
in a sufficiently deep 2D optical lattice (characterized by an effective mass on the lattice plane
$m^*>m$~\cite{kraemer2002} and a regularized local coupling constant $\tilde g$).
This effect may be understood by considering the phonon-like excitations of the dark soliton plane:
$\epsilon=\sqrt{\sigma/M}q$, where $M$ is the (negative) soliton mass per unit area, and $\sigma$ is the surface tension of the
nodal plane. For non-dipolar BECs one always has $\sigma>0$, and hence the phonon spectrum is always unstable leading to the
above mentioned snake instability. However, for dipolar BECs the surface tension $\sigma$ becomes negative when
$m/m^*<3C_{\rm dd}/(3\tilde g+2C_{\rm dd})$~\cite{nath2008-LS},
and hence the nodal plane is stabilized
for a sufficiently deep optical lattice (large $m^*/m$) and a sufficiently large {\ddi}.

\subsection{Vortices}
\label{subsec:vortices}

Quantized vortices constitute one of the most important consequences
of superfluidity, playing a fundamental role in various physical systems, such
as superconductors~\cite{degennes1966-LS} and superfluid
helium~\cite{donelly1991-LS}.
Vortices and even vortex lattices have been created in
BECs in a series of milestone
experiments~\cite{matthews1999-LS,madison2000-LS,aboshaeer2001}.
Contrary to normal fluids, a vortex in a BEC cannot be created by any rotation,
but there is a critical angular velocity $\Omega_{\rm c}$. Only
for angular velocities $\Omega>\Omega_{\rm c}$
it is energetically favorable to form a vortex~\cite{fetter2001-LS}
(the critical rotation for vortex nucleation is actually larger, since also dynamical
instabilities at the condensate boundaries must be considered~\cite{sinha2001-LS}).
O'Dell and Eberlein~\cite{odell2007-LS}
have recently studied the critical $\Omega_{\rm c}$ in a dipolar condensate in the
Thomas-Fermi regime, by means of the solution discussed in section \ref{sec:groundstate}. For BECs
in axially symmetric traps with the axis along the dipole orientation it has
been shown that $\Omega_{\rm c}$ is decreased due to the {\ddi} in oblate traps and
increased in prolate traps. This modification can be traced back to the
modification of the Thomas-Fermi radius due to the {\ddi}, rather than changes in
the vortex core. However, the {\ddi} may induce crater-like structures close to the
vortex core for the case of $a<0$, or even anisotropic vortex cores,
as recently shown by Yi and Pu~\cite{yi2006-LS}.

At higher rotating frequencies, more vortices enter the condensate and a vortex lattice
develops~\cite{madison2000-LS,aboshaeer2001}. In non-dipolar BECs vortices form so-called Abrikosov lattices, \emph{i.e.} triangular lattices with hexagonal symmetry.
Interestingly, this is not necessarily the case in dipolar condensates~\cite{cooper2005,zhang2005}.
In particular, with increasing $\edd$ or high filling factor,
the vortex lattice may undergo transitions between different symmetries: triangular,
square, stripe vortex crystal, and bubble states~\cite{cooper2005}. In addition
for vortex lattices in double well potentials the competition between tunneling
and interlayer {\ddi} should lead to a quantum phase transtion from a coincident phase
to a staggered one~\cite{zhang2005}.

\begin{figure}[t]
\begin{center}
\includegraphics[width=5cm,angle=270]{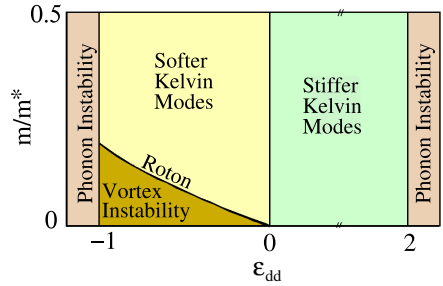}
\end{center}
\vspace*{-0.2cm}
\caption{Stable/unstable regimes, as a function of the ratio $m/m^*$ and
$\edd$, for straight vortex lines when the dipoles are oriented along the vortex line.}
\vspace*{-0.1cm}
\label{fig:3-ke}
\end{figure}

Vortex lines are in fact 3D structures with
transverse helicoidal excitations known as
Kelvin modes~\cite{thompson1880-LS,pitaevskii1961}.
Kelvin modes, which play an
important role in the physics of superfluid Helium~\cite{donelly1991-LS},
have also been experimentally observed in BECs~\cite{bretin2003-LS}.
The long-range character of the {\ddi} links different parts of the vortex
line, and hence the 3D character of the vortex lines is much more relevant
in dipolar BECs. Remarkably the {\ddi} may significantly modify
the vortex-line stability. In the presence of an additional optical lattice
(which leads to an effective mass $m^*$ along the vortex line direction)
the dispersion of Kelvin modes shows a roton-like minimum~\cite{klawunn2008-LS}, which for
sufficiently large {\ddi} and large $m^*$ may reach zero energy, leading to a thermodynamical
instability related to a second-order-like phase transition from a straight vortex
into a twisted vortex line~\cite{klawunn2009-LS} (see figure~\ref{fig:3-ke}).

%%%%%%%%%%%%%%%%%%%%%%%%%%%%%%%%%%%%%%%%%%%%%%%%%%%%%%%%%%%%%%%%%%%%%%%%%%%%%%%%%%%%%%%%%%%
%% FIGURE KELVON
%%%%%%%%%%%%%%%%%%%%%%%%%%%%%%%%%%%%%%%%%%%%%%%%%%%%%%%%%%%%%%%%%%%%%%%%%%%%%%%%%%%%%%%%%%%

\subsection{Pattern formation}
\label{subsec:patterns}

Pattern formation in driven systems is a general nonlinear phenomenon occurring
in many scenarios ranging from hydrodynamics and nonlinear optics to
liquid crystals and chemical reactions~\cite{cross1993-LS}.
Faraday patterns have been recently observed in non-dipolar BECs by a modulation of the
harmonic confinement~\cite{engels2007-LS} which leads to a periodic modulation of the system
nonlinearity~\cite{staliunas2002-LS}.
Faraday patterns offer an important insight about the elementary excitations
in a BEC, and hence pattern formation may be remarkably different in dipolar BECs, especially
in the presence of a roton-maxon excitation spectrum (section \ref{sec:groundstate}). Most remarkably,
whereas for non-dipolar BECs the pattern size decreases monotonously with the
driving frequency, patterns in dipolar BECs present a highly non-trivial dependence
characterized by abrupt pattern-size transitions~\cite{nath2009b-LS}.

Faraday pattern formation in driven systems is the (transient) result of an externally
driven dynamical instability. However, an interaction-induced dynamical instability may
lead as well to pattern formation. In this sense, the phonon instability, which, as discussed
in section \ref{sec:groundstate}, leads to collapse in 3D dipolar gases (also in 2D and 3D non-dipolar BECs),
does not necessarily lead to collapse in 2D geometries. On the contrary, the 2D phonon instability
leads to the formation of a soliton gas, and to transient pattern formation, which, if
avoiding collapse, may lead to the formation of a 2D stable soliton~\cite{nath2008-LS}.
Dynamical roton instability leads typically to local collapses~\cite{komineas2007},
although a sufficiently strong trapping may stabilize a biconcave BEC profile~\cite{ronen2007a} (as mentioned in section~\ref{sec:geometry}).

\section{Dipolar effects in spinor condensates}\label{sec:spinor}

Spinor BECs, composed of atoms in more than one Zeeman state,
constitute an extraordinary tool for the analysis of multicomponent
superfluids.  Whereas magnetic trapping confines the BEC to
weak-field seeking magnetic states, optical trapping enables confinement of all
magnetic substates, hence freeing the spin degree of freedom~\cite{Stenger1998-LS}.
Interestingly, interatomic interactions allow for a coherent transfer of population
between different Zeeman states (spin-changing collisions), leading to a
fascinating physics in both what concerns ground-state properties and spin
dynamics~\cite{Ho1998-LS,Ohmi1998-LS,Koashi2000-LS,Ciobanu2000-LS,Schmaljohann2004-LS,Chang2004-LS}.

The energy scale associated to spin-preserving collisions is
given by the chemical potential, which for typical alkali gases (and even for Chromium in absence of
Feshbach resonances) is much larger than the {\ddi}. On the contrary the energy associated with
spin-changing collisions is typically much smaller, since it is provided by the difference
between $s$-wave scattering lengths in different spin channels~\cite{Ho1998-LS}, which is very small.
Hence the {\ddi} may become comparable to the energy of spin-changing collisions, and as a consequence
even alkali spinor BECs (in particular $^{87}$Rb) can be considered in this sense as dipolar gases as well.
The {\ddi} may hence play a significant role in the ground-state properties and the dynamics of spinor condensates.

\subsection{Ground state}

Pu \emph{et al.}~\cite{Pu2001-LS} have shown that ferromagnetic spinor BECs (as it is the case of $^{87}$Rb in $F=1$) placed
at different sites of a strong optical lattice  behave as single magnets
oriented in the effective magnetic field induced by the combination of a external magnetic field and
the {\ddi} of other sites. Interestingly, the collectively enhanced magnetic moment of the condensates
at each site enhances the magnetic {\ddi} between sites, which may become sufficiently strong even for
alkali atoms~\cite{Pu2001-LS}. As a consequence, such an array of effective magnets can undergo
a ferromagnetic (1D lattice) or anti-ferromagnetic (2D lattice) phase transition under the magnetic
dipolar interaction when external magnetic fields are sufficiently weak~\cite{Pu2001-LS,Gross2002-LS}. In addition, for 1D lattices
the inter-site {\ddi} may distort the ground-state spin-orientations and lead to the excitation of spin waves,
which possess a particular dispersion relation which depends on the transverse width of the condensates~\cite{Zhang2002-LS}.

The {\ddi} may play an important role in the properties of trapped spinor BECs, especially in the absence of significant external magnetic fields.
In particular, whereas in absence of an external magnetic field
the spinor BEC is rotationally invariant in spin space, the {\ddi} breaks this symmetry, inducing
new quantum phases which can be be reached by tuning the effective
strength of the {\ddi} via a modification of the trapping geometry~\cite{Yi2004-LS}.
For the case of spin-1 BECs, for very low magnetic fields (typically below 10~$\mu$G)
the phase diagram presents due to the {\ddi} three different ground state phases, characterized by different
spin textures: a polar-core vortex phase, a flower phase and a chiral spin-vortex phase~\cite{Kawaguchi2006b-LS}.
The latter has chirality in the formation of the spin vortex, and the topological spin structure
spontaneously yields a substantial net orbital angular momentum.

\subsection{Dynamics and Einstein-de Haas effect}

The dipole-dipole interaction has been shown to play quite a small role in the ground state properties of Chromium BECs~\cite{Diener2006-LS,Santos2006-LS}. However the effects of the dipole-dipole interaction on the spinor dynamics may be much more intriguing. The short-range interactions in a spinor condensate may occur in different scattering channels, corresponding to different total spin of the colliding pair~\cite{Ho1998-LS}
\begin{equation}
\hat V_{sr}=\frac{1}{2}\int {\rm d}{\bds r} \sum_{S=0}^{2F} g_S \hat{\cal P}_S({\bds r}),
\end{equation}
where $\hat{\cal P}_S$ is the projector on the total spin $S$ (necessarily even due to symmetry reasons), $g_S=4\pi\hbar^2a_S/m$, and $a_S$ is the $s$-wave scattering length for the channel of total spin $S$. The short range interactions necessarily preserve  the spin projection $S_z$ along the quantization axis.

The {\ddi} for a spinor BEC is of the form
\begin{eqnarray}
\hat V_{\rm dd}&=&\frac{\cdd}{2}\int {\rm d}{\bds r}\int {\rm d}{\bds r}'
\frac{1}{|{\bds r}-{\bds r}'|^3}
\hat\psi_m^\dag ({\bds r})\hat\psi_{m'}^\dag ({\bds r}') \nonumber \\
&&\!\!\!\!\!\!\!\!\!\!\!\!\!\!\!\! \!\!\!\!\left [ {\bds
S}_{mn}\cdot{\bds S}_{m'n'}-3({\bds S}_{mn}\cdot{\bds e}) ({\bds
S}_{m'n'}\cdot{\bds e}) \right ] \hat\psi_n ({\bds
r})\hat\psi_{n'}({\bds r}'),
\end{eqnarray}
where ${\bds S}=(S_x,S_y,S_z)$ are the spin-$F$ matrices, and $\cdd=\mu_0\mu_B^2g_F^2/4\pi$ (for $^{52}$Cr $F=3$ and $\cdd=0.004g_6$), with ${\bds e}=({\bds r}-{\bds r}')/|{\bds r}-{\bds r}'|$.

\begin{figure}[t] %\resizebox{3.5in}{1.2in}
\begin{center}
{\includegraphics[width=11cm]{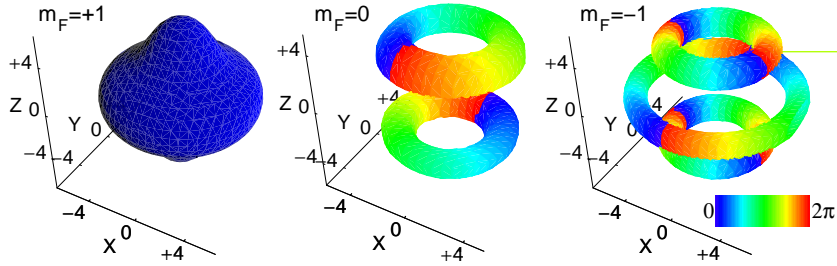}}
\end{center}
\caption{Isodensity surfaces of $m_F=+1$, $m_F=0$, and $m_F=-1$ spinor components corresponding to the densities: $7.96\times10^{13}$\,cm$^{-3}$, $7.96\times10^{13}$\,cm$^{-3}$, and $7.96\times10^{12}$\,cm$^{-3}$,  respectively. The color on the surface represents the phase of the component wave function (with the scale given on the right). Note the characteristic vortex rings  patterns. The trap frequency $\omega_{x,y,z}=2\pi \times 100$\,Hz and the magnetic field $B=-0.029$\,mG. The density plots are taken at $140$\,ms. }
\label{fig:spinor:vr}
\end{figure}

Interestingly, contrary to the short-range interaction the dipole-dipole interaction does not necessarily conserve the spin projection along the  quantization axis due to the anisotropic character of the interaction. In particular, if the atoms are initially prepared into a maximally stretched state, say $m_F=-F$,  short-range interactions cannot induce any spinor dynamics due to the above mentioned conservation of total magnetization $S_z$. Dipole-dipole interactions, on the contrary may induce a transfer into $m_F+1$. If the system preserves cylindrical symmetry around the quantization  axis, this violation of the spin projection is accompanied by a transfer of angular momentum to the center of mass, resembling the well known Einstein-de Haas effect~\cite{Santos2006-LS,Kawaguchi2006-LS}. Due to this transfer an initially spin-polarized dipolar condensate can generate dynamically vorticity (see figure~\ref{fig:spinor:vr}).

Unfortunately, the Einstein-de Haas effect is destroyed in the presence of even rather weak magnetic fields. Typically, magnetic fields well below 1 mG are necessary to observe the effect. Due to the  dominant role of Larmor precession, and invoking rotating-wave-approximation arguments, the physics must be constrained to manifolds of preserved magnetization, where the system presents a regularized  dipole-dipole interaction~\cite{Kawaguchi2007-LS}. However the dipole-dipole interaction may have observable effects also under conserved magnetization, since the regularized dipole-dipole interaction may lead also in that case to spin textures~\cite{Kawaguchi2007-LS}.

A significant Einstein-de Haas effect may be recovered however under particular resonant conditions as studied by Gawryluk \emph{et al.}~\cite{gawryluk2007}, who studied the particular case of Rb BECs in $F=1$,  initially prepared into $m_F=1$. In that case, the population  transfer away from the $m_F = 1$ state is typically very small, but it can be significantly enhanced by applying a resonant magnetic field, such that the Zeeman energy of an atom in the $m_F = 1$ state is  totally transferred into the kinetic energy of the rotating atom in $m_F = 0$, $\mu B = E_{\rm kin}$. Typically, the resonant $B$ is  small ($\sim 100 \mu$G), but can be  tuned directly, or by adjusting the trap geometry and thus the rotational energy of  the atoms, reaching up to $1$~mG. Gawryluk \emph{et al.} demonstrated that at the resonance,  a significant transfer of the initial population of the $mF = 1$ state occurs on a time  scale inversely proportional to the dipolar energy $t_{\rm transfer}\simeq \hbar/C_{dd}n$. As mentioned above, such a transfer is accompanied by the formation of vorticity (see figure \ref{fig:spinor:vr})~\cite{Santos2006-LS,Kawaguchi2006-LS,gawryluk2007}.

An alternative possibility for the observation of the Einstein-de Haas effects at finite fields may be also provided by an artificial quadratic Zeeman effect induced by either microwave~\cite{gerbier2006} or optical fields~\cite{santos2007}. This effective quadratic Zeeman effect may allow for a resonance  between e.g. $m_F=-F$ and $m_F=-F+1$ and hence lead to a significant enhancement of the Einstein-de Haas transfer~\cite{santos2007}.

\subsection{Experimental results}

\begin{figure}[t] %\resizebox{3.5in}{1.2in}
\begin{center}
\includegraphics[width=11cm]{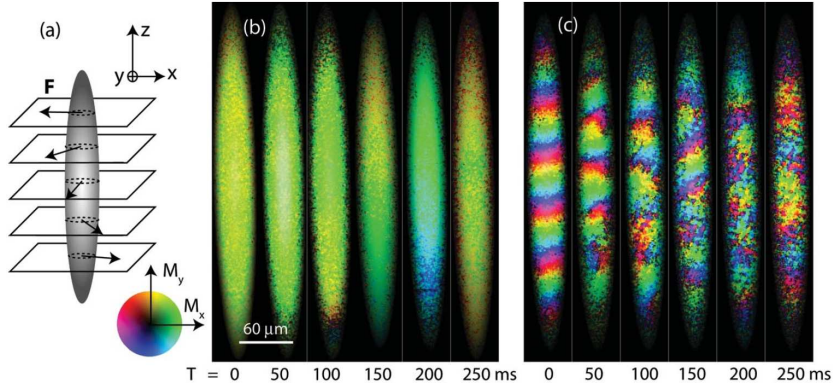}
\end{center}
\caption{Spontaneous dissolution of helical textures in a quantum degenerate $^{87}$Rb spinor Bose gas. A transient magnetic field gradient is used to prepare transversely magnetized (b) uniform or (a),(c) helical magnetization textures. The transverse magnetization column density after a variable time $T$ of free evolution is shown in the imaged $x-z$ plane, with orientation indicated by hue and amplitude by brightness (color wheel shown). (b) A uniform texture remains homogeneous for long evolution times, while (c) a helical texture with pitch $\lambda=60\,\mu$m dissolves over $\sim200$~ms, evolving into a sharply spatially modulated texture. Figure reprinted with permission from~\cite{Vengalattore2008-LS}. Copyright 2008 by the American Physical Society.}
\label{fig:spinor:dmsk}
\end{figure}

Recent 2D experiments at Berkeley~\cite{Vengalattore2008-LS} have demonstrated the dipolar character of spin-1 $^{87}$Rb spinor BECs. In particular, these experiments show the spontaneous decay of helical spin textures (externally created by magnetic field gradients) toward a spatially modulated structure of spin domains (see figure~\ref{fig:spinor:dmsk}). The formation of this modulated phase has been  ascribed to magnetic dipolar interactions that energetically favor short-wavelength domains over the long-wavelength spin helix. Interestingly, the reduction of dipolar interactions (by means of radio-frequency pulses)  results in a suppression of the modulation.

These experiments have attracted a large deal of theoretical interest, in particular in the {\ddi}-induced distortion of the excitation spectrum of a spinor BEC~\cite{Lamacraft2008-LS,Cherng2008-LS}.  Recently Cherng and Demler have analyzed (in the context of the above mentioned 2D experiments~\cite{Vengalattore2008-LS}) the possibility of roton softening  (similar to that discussed in section  \ref{sec:roton}) in the spectrum of spin excitations. This roton instability may lead as a function of the quadratic Zeeman effect and  the magnetic field orientation with respect to the normal of the  2D trap to different spin textures (checkerboard, striped phase) and may pave the way towards a supersolid phase. The latter is a long pursued  phase in condensed-matter physics which possesses both  periodic crystalline order and superfluidity~\cite{Andreev1969-LS,Chester1970-LS,Legget1970-LS} (see also the next section).  However, instability against finite-momentum excitations does not necessary  lead to the appearance  of a stable modulation (as it is the case of the roton instability discussed in section~\ref{sec:roton}) and more work on the physics in the unstable regime is clearly necessary. Recent experiments at Berkeley have reported on possible first traces of supersolidity in a spinor $^{87}$Rb BEC~\cite{Vengalattore2009-LS}.

\section{Dipolar gases in optical lattices} \label{sec:lattices}

One of the most fruitful fields of research both from the
experimental and the theoretical points of view in the last years
has been the study of ultra-cold atomic samples in optical
lattices, which are non dissipative periodic potential energy
surfaces for the atoms, created by the interference of laser
fields. The study of cold atoms in periodic potentials is of
primary interest because it allows to reproduce problems
traditionally encountered in condensed matter and solid state physics in a new
setting, where a high degree of control is possible and where the
Hamiltonian which governs the system is in general very close to
some idealized one. With the present developments there is even
the possibility to investigate, with appropriately designed atomic
systems, phenomena which do not exist in condensed matter.

We will first summarize the physics of weakly interacting atomic
gases in optical lattices in section~\ref{sectBECoptlat} and
describe the first measurement of dipolar effects in alkali atoms
in section~\ref{sectblochdip}. After introducing the physics of
strongly correlated systems with contact interaction in
section~\ref{sectBH}, we will devote section~\ref{sectextBH} and
section~\ref{sectMS} respectively to the quantum phases and the
metastable states found in 2D lattices in the presence of
long-range interactions. In section~\ref{sectmultilayer}, we
discuss the novel physics introduced by the presence of two or
many 2D optical lattice layers. In section~\ref{spinmodels}, we
discuss proposals on how to tailor the interaction potential and
create lattice spin models with polar molecules and finally, in
section~\ref{sectself}, we will mention the possibility of
formation of self-assembled regular structures in cold dipolar
gases.

\subsection{Bose-Einstein condensates in optical lattices}
\label{sectBECoptlat}

In the presence of weak optical lattices, when the coherence of
the system is preserved, the Gross-Pitaevskii equation provides a
good description of the system. Due to the presence of the
periodic potential and interactions, analogies to phenomena
typical of solid state physics and non-linear optics are
respectively possible~\cite{hecker-denschlag2002, morsch2006,
kevrekidis2007, bloch2008}.

As previously explained, in the most common cases, interactions in
ultra-cold gases are dominated by $s$-wave scattering, which can be
in a very good approximation considered a point-like interaction.
In the case of full coherence, the system is described by a
macroscopic wavefunction and obeys the GPE (\ref{eq:gpe:dd:ti}).
In the presence of optical lattices, $V_{\rm ext}$ is periodic and
given by $V_{\rm ext}({\bds r})=\sum_n V^{\rm opt}_n \sin^2(\pi
x_n/d_n)$, where the index $n$ runs over the dimensions of the
lattice and $d_n$ is the lattice constant in the $n$-th direction.
For lattices created by counterpropagating laser beams of
wavelength $\lambda$, the lattice spacing is $d=\lambda/2$. The
depth of the lattice potential $V^{\rm opt}_n$ depends linearly on
the intensity of the laser light.

It is well known that the spectrum of a single particle in a
periodic potential is characterized by bands of allowed energies
and energy gaps~\cite{ashcroft}. The counterpart of the energy
eigenstates delocalized over the whole lattice (Bloch states) are
the wavefunctions centered at the different lattice sites (Wannier
functions). For deep enough periodic potentials, when the Wannier functions
are well localized at the lattice sites, the so-called tight
binding regime is reached: the first band takes the form $E(q)=-2J
\sum_n \cos(q_n d_n)$, $q_n$ being the quasi-momenta in the
different lattice directions and $J$ the tunneling parameter
between neighboring wells. For low enough interactions and
temperature, the physics of the system is well approximated by the
one taking place in the first energy band. Under those
assumptions, the discretized non-linear Schr\"odinger (DNLS)
equation~\cite{hennig1999,rasmussen2000,trombettoni2001,smerzi2003,menotti2003}
provides a good description of the system.

The excitations on top of the GP solution can be found by generalizing
the Bogoliubov method to include the periodic potential. They show
a phononic branch with renormalized sound velocity
\cite{taylor2003, kraemer2003, menotti2004}. For a moving
condensate, one finds complex frequencies at the edge of the
Brillouin zone~\cite{wu2001}, highlighting the presence of
dynamical instabilities~\cite{machholm2003}.

Among the numerous phenomena described theoretically and observed
experimentally are collective oscillations~\cite{fort2003},
Bloch oscillations~\cite{morsch2001}, dynamical instabilities
\cite{fallani2004}, Josephson oscillations~\cite{cataliotti2001},
non linear self-trapping~\cite{anker2005}, gap solitons
\cite{eiermann2004}. The collective oscillations can be described
through an effective  macroscopic dynamics accounting for the
periodic potential and their frequency is rescaled in terms of the
effective mass~\cite{kraemer2002}. Instead the physics underlying
the other phenomena is dominated by a crucial interplay between
the periodic potential and interactions.

In the alkalies usually used in experiments with optical lattices,
$s$-wave scattering dominates over all other types of interactions.
When the $s$-wave scattering length is reduced, e.g. by means of a
Feshbach resonance, the presence of other types of interaction is
relatively enhanced and can be probed. In the next section, we
present the first measured effects of dipole-dipole interaction
for a quantum gas in an optical lattice, performed with Potassium
atoms.

\subsection{Bloch oscillation damping due to dipole-dipole
interactions} \label{sectblochdip}

Bloch oscillations have been one of the first solid state
phenomena to be investigated with cold atoms~\cite{bendahan1996,
wilkinson1996}, observed shortly after Bloch oscillations of
electrons in semiconducting heterostructures~\cite{feldmann1992,
waschke1993, lyssenko1997}. They consist in oscillations in space
in the presence of a constant acceleration, due to the change of
sign of the effective mass along the first energy band. For
electrons, the acceleration is provided by a constant electric
field, while for cold atoms it is produced by a linear increase in
time of the relative detuning of the two laser beams creating the
optical lattice or, in a vertical setup, by gravity.

With Bose-Einstein condensates, Bloch oscillations can be measured
with a higher precision thanks to the smaller width of the
momentum distribution~\cite{morsch2001}. However in the presence
of interatomic interactions, the on-set of dynamical instabilities
in the outer region of the Brillouin zone causes a damping of the
oscillations. Due to the relevance of Bloch oscillations as a tool
of precision measurement of accelerations and small forces at
small distance from surfaces~\cite{dimopoulos2003, carusotto2005,
ferrari2006, obrecht2007}, attempts of reducing interactions have
been pursed first with polarized fermions~\cite{roati2004} and
then with bosons with reduced $s$-wave scattering length
\cite{gustavsson2008, fattori2008}.

\begin{figure}[t]
\begin{center}
\includegraphics[width=11cm]{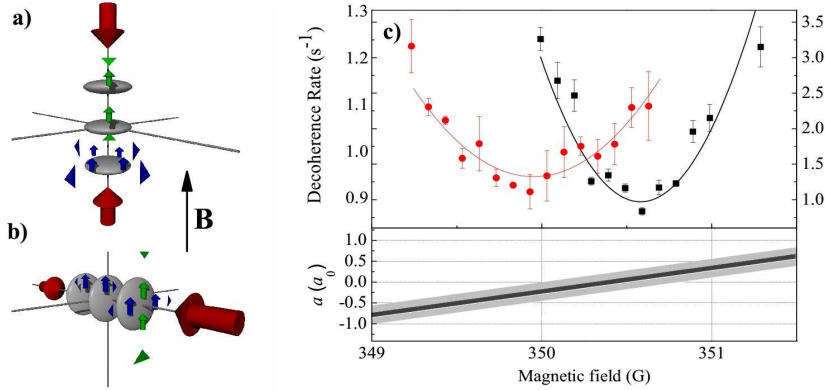}
\end{center}
\caption{Decoherence of Bloch oscillations in $^{39}$K around the scattering length zero crossing, due to the interplay
of contact and dipolar interactions. The character of the dipolar interaction depends
on the relative orientation of the lattice and magnetic field: (a) prevalently repulsive
interaction; (b) prevalently attractive interaction. (c) The width of the momentum distribution
after a few 100 ms of Bloch oscillations shows a minimum when the contact
interaction compensates the dipolar one: circles are for the case (a), while squares are
for the case (b). The lines are parabolic fits to the data, which constrain the position of
the zero-crossing in a comparison with theory (black region in the lower panel) better
than Feshbach spectroscopy (gray region). Figure courtesy of G. Modugno.}
\label{fig:lens}
\end{figure}

The experiment in~\cite{fattori2008} has highlighted for the first
time effects of dipole-dipole interactions on alkali atoms and has
shown how its interplay with the contact interaction can be
exploited to reduce interaction-induced decoherence of Bloch
oscillations in a 1D optical lattice. When the scattering length
is tuned to zero by means of a Feshbach resonance, the magnetic
dipole-dipole interaction becomes the limiting factor for the
coherence time. The point of minimum decoherence is shifted to
negative or positive values of the scattering length, depending on
the orientation of the dipoles with respect to the axis of the
lattice (see figure~\ref{fig:lens}).

The values of the scattering lengths which maximize the lifetime
in the different configurations have been predicted through the
solution of 1D DNLS and GP equations including the dipolar
potential. The 3D geometry has been accounted for, assuming the
condensate to be in the transversal ground state. The {\ddi}
contributes both to the regularization of the on-site interactions
and also to the intersite interactions.  Although the compensation
of the on-site {\ddi} and the short-range interactions explains the
sign of the displacement (towards $a<0$ or $a>0$) of the
decoherence minimum, the dipole-induced intersite interactions are
crucial for the quantitative understanding of the experimental
results in~\cite{fattori2008}. In this sense the experiments in
\cite{fattori2008} constitute also the first observation of
intersite effects in dipolar gases in optical lattices. Due to the
intersite interactions, the {\ddi} can be only partially compensated
by the short-range interactions, and as consequence a non complete
reduction of the decoherence rate of the $^{39}$K-based
interferometer is observed (with a minimum residual rate of 0.05
Hz).

In the experiments in~\cite{fattori2008}, the interactions at the
decoherence minimum were much weaker than the transversal
confinement, justifying the approximation of assuming the BEC in
the transversal ground state. This is however not the general
case. For shallower transversal confinements the intersite {\ddi} may
significantly modify and even destabilize the spectrum of
elementary excitations of a BEC in an optical lattice
\cite{wang2008b, klawunn2008}. In particular, the intersite
interactions may induce rotonization and even roton-instability (see also section~\ref{sec:roton})
under appropriated conditions, and may (for a sufficiently shallow
transversal confinement) lead to a dynamical instability that
could as well damp Bloch oscillations~\cite{wang2008b}.
Interestingly, the intersite {\ddi} induces an hybridization of
transversal modes at different sites (and a corresponding
band-like spectrum) even if the hopping is completely suppressed
\cite{klawunn2008}. Remarkably, whereas a single lattice site
could be stable, a stack of non-overlapping dipolar BECs may
become roton-unstable, showing once more that polar gases in
optical lattices differ qualitatively from short-range interacting
gases.

\subsection{Strongly correlated lattice gases. Bose-Hubbard
Hamiltonian} \label{sectBH}

For deep optical lattices and small numbers of atoms per site, the
coherent description of the system provided by the GPE breaks down
due to the growing effect of correlations. One of the greatest
achievements of the last years was the experimental
observation~\cite{greiner2002} of the superfluid to Mott insulator
transition~\cite{fisher1989, jaksch1998}.
In this session, we will introduce the main theory for ultra-cold
atoms in optical lattices in the case of point-like interaction,
providing the background for the case of long-range interactions,
which will be treated in the following sections.

Under usual experimental conditions, the single band approximation
mentioned in section~\ref{sectBECoptlat} is appropriate. In order
to allow for the breaking of the coherence of the system, the
field operator is replaced by its single-band many-mode expansion
$\hat{\psi}({\bds r})=\sum_i w_i({\bds r})\hat{a}_i$, with ${\hat
a}_i$ being the annihilation operator for one boson in the Wannier
function $w_i({\bds r})$ localized at the bottom of the lattice
site $i$.

Neglecting the overlap beyond nearest neighboring densities,
defining
\begin{eqnarray}
J&=& - \int   w_i^*({\boldsymbol r}) \left( - \frac{\hbar ^2
\Delta}{2m} + V_{\rm ext}({\boldsymbol r}) \right)
w_{i+1}({\boldsymbol r})  \; {\rm d}^3 { r}  , \\
U&=&g \int  |w_i({\boldsymbol r})|^4 \; {\rm d}^3 { r} ,
\end{eqnarray}
and $n_i= \hat{a}_i^{\dag} \hat{a}_i$, one can derive~\cite{jaksch1998} the famous
Bose-Hubbard Hamiltonian
\begin{eqnarray}
H= - J\sum_{\langle ij\rangle} \hat{a}_i^{\dag} \hat{a}_j +
\sum_{i}  \left[\frac{U}{2} n_i (n_i-1) - \mu n_i \right] ,
\label{bh}
\end{eqnarray}
extensively studied in condensed matter physics. In optical
lattices, the Hamiltonian parameters can be accurately controlled
by changing the light intensity: ramping it up
increases the interaction term $U$ due to a stronger localization
of the wavefunctions at the bottom of the lattice wells, and at
the same time exponentially decreases the tunneling $J$.

When tunneling is suppressed compared to interactions, this
Hamiltonian presents a quantum phase transition between a
superfluid phase, characterized by large number fluctuations at
each lattice site, and a Mott insulating phase where each lattice
well is occupied by precisely an integer number of atoms without
any number fluctuations. The nature of this phase transition and
the qualitative phase diagram can be inferred based on very simple
arguments~\cite{fisher1989}.

\begin{figure}[t]
\begin{center}
\includegraphics[width=4.5cm]{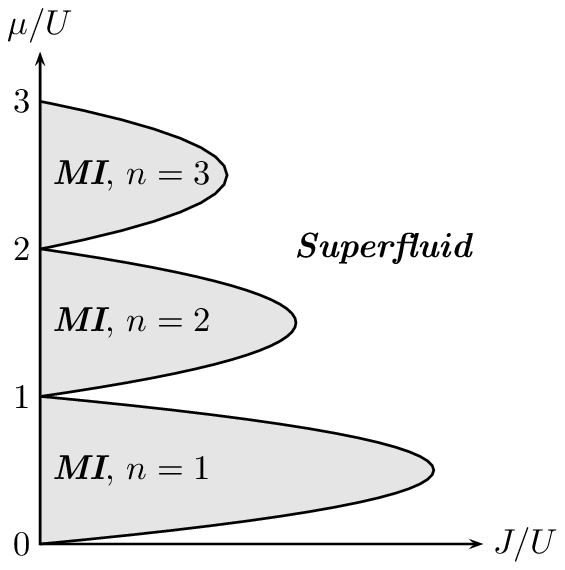}
\end{center}
\caption{Schematic phase diagram for the ground state of the Bose-Hubbard Hamiltonian~(\ref{bh}).} \label{fig:mott}
\end{figure}

At zero tunneling $J=0$ and commensurate filling (exactly an integer
number $n$ of atoms per well), the interaction energy is minimized by
populating each lattice well with exactly $n$ atoms.  Energy
considerations tell that the filling factor $n$ is energetically
favored in the range of chemical potential $(n-1)U < \mu < n U$.
The state with precisely integer occupation at the lattice sites
is called Mott insulating state.  Since a particle-hole excitation
at $J=0$ costs an energy $\Delta E=U$ equal to the interaction
energy, the Mott state is the lowest energy state at commensurate
filling. For a tunneling $J$ different from zero the energy cost
to create an excitation decreases thanks to the kinetic energy
favoring particle hopping. However, for large interactions and
small tunneling, the gain in kinetic energy ($\sim J$) is not yet
sufficient to overcome the cost in interaction energy ($\sim U$),
which leads to the existence of Mott insulating states also at
finite tunneling. For large enough tunneling, instead, particle
hopping becomes energetically favorable and the system becomes
superfluid.
The regions in the $J$ vs. $\mu$ phase diagram where the Mott
insulating state is the ground state are called {\it Mott lobes} (see figure~\ref{fig:mott}).
For non commensurate filling, there are extra atoms free to hop
from site to site at no energy cost, so that the phase of the
system is always superfluid. The superfluid phase at non
commensurate densities survives down to $J=0$ for $\mu/U=[\rho]$
where the symbol $[\rho]$ indicates the integer part of the
density.
Due to the finite energy cost required to add or remove one
particle, the Mott phase is gapped and incompressible, while in
the superfluid regions the gap vanishes and the system is
compressible.

In order to find the shape of the lobes at finite $J$
sophisticated calculations are required. Apart from the
mean-field approximation~\cite{fisher1989,sheshadri1993,sachdev},
which works only qualitatively in one dimension and works better
and better in larger dimensions, there is no exact analytical
method which allows to calculate the boundary of the lobes. Improvements
are achieved by high order perturbative strong coupling expansions
\cite{freericks1994,freericks1996} and exact numerical results are
obtained using Quantum Monte Carlo techniques (see e.g.
\cite{scalettar1991, krauth1991, prokofev1998}).

In the experiments, the phase transition has been identified by
looking at the interference of the expanded cloud and the
measurement of the gapped excitations in the Mott phase
\cite{greiner2002}. The SF-MI shell structure which arises in
the presence of an external confinement~\cite{jaksch1998,
batrouni2002, foelling2006} has been observed using spatially selective microwave
transition and spin-changing collisions~\cite{foelling2006} or
using clock shifts~\cite{campbell2006}, and the underlying
ordering in the lattice in the Mott phase has been inferred from
the measurement the periodic quantum correlations in the density
fluctuations in the cloud after expansion~\cite{altman2004,
foelling2005}. The interference of the expanding cloud and spatial
noise correlations measurements prove to be useful tools to
identify also the exotic quantum phases expected to appear in the
presence of dipole-dipole interaction.

\subsection{Quantum phases of dipolar lattice gases}
\label{sectextBH}

\begin{figure}[t]
\begin{center}
\includegraphics[width=10cm]{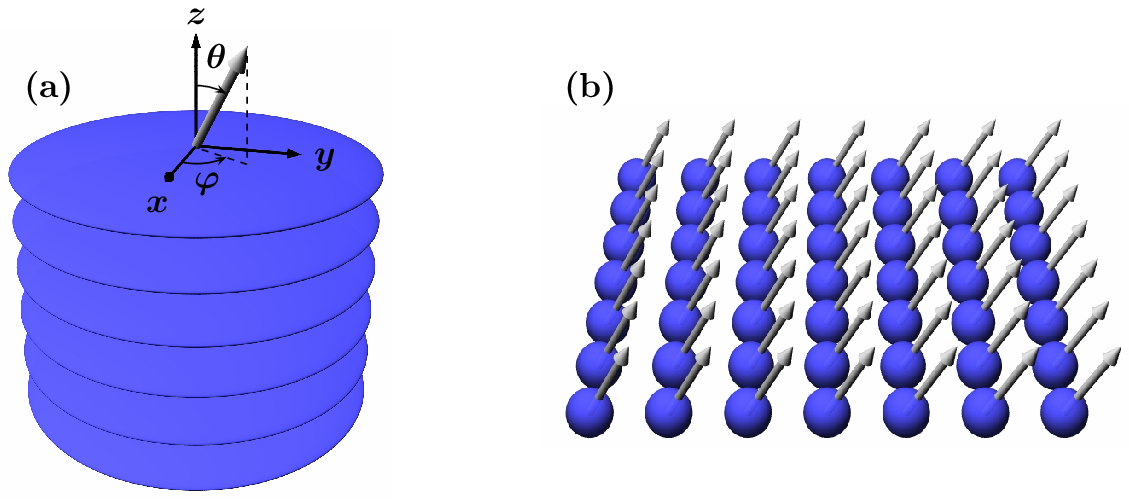}
\end{center}
 \caption{Schematic representation of a gas of polarized dipoles
in a 1D optical lattice (a) and in a single 2D optical lattice
layer (b).} \label{fig_dip_lat}
\end{figure}

Dipole-dipole interactions add to the Bose-Hubbard model a new
essential ingredient, given by long-range and anisotropic
interactions. For a lattice of polarized dipoles, as sketched in
figure~\ref{fig_dip_lat}, the extended Bose-Hubbard Hamiltonian in
the presence of long-range interaction is

\begin{equation}
H= -J\sum_{\langle ij\rangle} \hat{a}_i^{\dag} \hat{a}_j +
\sum_{i} \left[\frac{U}{2} n_i (n_i-1) - \mu n_i  \right] +
\sum_{\bds \ell} \sum_{\langle i j \rangle_{\bds \ell}}
\frac{U_{\bds \ell}}{2} \; n_i n_j , \label{GBH}
\end{equation}
where ${\bds \ell}$ is the distance between the two optical
lattice sites $i$ and $j$.

The extended Bose-Hubbard Hamiltonian (\ref{GBH}) has been
extensively studied.  It has been predicted that in 2D lattices
the presence of finite range interactions (where the sum over
${\bds \ell}$ is generally cut-off at the nearest or next-nearest
neighbor) gives rise to novel quantum phases, like the
charge-density wave (checkerboard), namely an insulating phase with
modulated density, and the supersolid phase, presenting the
coexistence of superfluidity and of a periodic spatial modulation
of the density, different from the one of the lattice
\cite{bruder1993, vanotterlo1994, batrouni1995} (see also
discussion in section \ref{sec:roton}).

The presence of insulating phases at fractional filling factors
can be inferred readily following the criteria explained in the
previous section, which predict that (e.g.) a checkerboard
ordering of the atoms (see figure~\ref{chap9_fig1}(GS)) at $J=0$ is
stable against particle-hole excitations in the range of chemical
potential $0<\mu<4U_{\rm NN}$, $U_{\rm NN}$ being the first
nearest-neighbor interaction (and neglecting for the sake of
simplicity all following ones). Analogous to the standard Mott
insulating phase, these insulating phases at fractional filling
factor exist in some low tunneling region of the $\mu$ vs.
$J$ phase diagram.

%%%%%%%%%%%%%%%%%%%%%%%%%%%%%%%%%%%%
In~\cite{dallatorre2006}, it has been pointed out that 1D lattice systems
of spinless bosons interacting with long range interactions,
possess a further insulating phase, which they call Haldane Bose
Insulator (HI), presenting some analogies with the famous Haldane
gapped phase in quantum spin-1 chains~\cite{haldane1983}. This is a gapped phase, which unlike the
checkerboard phase does not break the translational symmetry of
the lattice, but is characterized by an underlying hidden order,
namely a non trivial ordering of the fluctuations which appear in
alternating order separated by strings of equally populated sites
of arbitrary length.

%%%%%%%%%%%%%%%%%%%%%%%%%%%%%%%%%%%%
The existence of
the supersolid phase in solid helium has not yet been
unambiguously proven experimentally: while on the one hand the
interpretation of the first experimental results measuring a
non-classical rotational inertia~\cite{kim2004a, kim2004b,
kim2006} remains controversial, microscopic calculations
\cite{pollet2007} indicate that disorder-based mechanisms, like
the presence of superfluid dislocations, grain boundaries, and
ridges, should be responsible for the more recent observations of
supersolidity~\cite{sasaki2006,ray2008}.
%%%%%%%%%%%%%%%%%%%%%%%%%%%%%%%%%%
Even if bulk supersolid remains in many respects more challenging
than lattice supersolid, the question of the stability of the
supersolid phase in the presence of the lattice is not trivial and
has been only recently settled by exact Quantum Monte Carlo
simulations. Checkerboard supersolid (at $\rho \approx 1/2$) is
expected for dominant nearest-neighbor interaction, while star (at
$\rho \approx 1/4$) and striped (at $\rho \approx 1/2$)
supersolids\footnote{The star supersolid shows modulations of the
density and order parameter such that in a $2\times 2$ elementary
cell, one [the other] diagonal contains sites with different
[equal] density and order parameter; the striped supersolid
presents alternating horizontal or vertical striped of higher and
lower density and order parameter.} are predicted for non
vanishing nearest-neighbor interaction~\cite{goral2002b,
kovrizhin2005}. There have been several studies devoted to the
stability of the supersolid phase versus phase separation
\cite{batrouni2000,hebert2001}, which is identified by a negative
compressibility. Agreement seems to be reached on the conclusion
that the checkerboard supersolid is stabilized at $\rho>1/2$ by a
finite on-site interaction and a strong enough nearest neighbor
interaction~\cite{sengupta2005}, while it phase separates for
nearest-neighbor interactions at densities smaller than $1/2$
(unless a strong enough nearest-neighbor hopping is introduced
\cite{chen2007}). Instead the striped supersolid, obtained for
large next nearest-neighbor interaction, exists for all doping
away from $\rho=1/2$ both in the hard-core and soft-core cases~\cite{batrouni2000,hebert2001,chen2007}. Finally, at large next
nearest-neighbor interaction, the star supersolid can always be
obtained by doping a star solid at $\rho=1/4$ with vacancies and
by doping it with bosons in the case the $\rho=1/2$ ground state
is a striped crystal~\cite{chen2007,dang2008}. The most important
conclusion on which most papers agree\footnote{All apart from
\cite{ng2008}.} is that no supersolid phase is found at
commensurate density. Analogous results have been recently
discussed for 1D geometries~\cite{kuehner2000, batrouni2006}, the
most important difference being the absence of phase separation.

Providing an alternative setting where to look for the supersolid
phase, cold atoms with long range interactions are particularly
appealing~\cite{goral2002b, kovrizhin2005, sengupta2005,
scarola2005, scarola2006,dallatorre2006}. Dipolar atoms and polar
molecules are good candidates to creates such physical systems,
bringing into play the extra feature of the anisotropy of the
interaction. Dipolar gases have been first identified as possible
candidates to provide a long-range interaction system in
\cite{goral2002b}. The on-site parameter $U$ in (\ref{GBH}) is
given by two contributions: one is arising from the $s$-wave
scattering $U_s= 4 \pi \hbar^2 a/m \int n^2({\bds r})\, {\rm d}^3 r$, and the
second one is due to the on-site dipole-dipole interaction $U_{\rm
dip}= 1/(2\pi)^3 \int \widetilde{ U_{\rm dd}}({\bds q}) {\tilde n}^2({\bds q})\, {\rm d}^3q$,
$\widetilde{ U_{\rm dd}}({\bds q})$ and ${\tilde n}({\bds q})$ being the Fourier
transforms of the dipole potential and density, respectively
\cite{goral2002a}. Because of the localization of the
wavefunctions at the bottom of the optical lattice wells, the long
range part of the dipole-dipole interaction $U_{\bds \ell}$ is in
a very good approximation given by the dipole-dipole interaction
potential at distance ${\bds \ell}$, $U_{\bds \ell}= (C_{\rm
dd}/4\pi) [1-3\cos^2(\theta_{\bds \ell})]/\ell^3$, multiplied by
the densities $n_i$ and $n_j$ in the two sites, where $\theta_{\bds
\ell}$ is the angle between ${\bds \ell}$ and the orientation of
the dipoles. The ratio between the total on-site interaction
$U=U_s+U_{\rm dip}$ and the nearest neighbor dipolar interaction
$U_{\rm NN}$ determines much of the physics of the system. It can be
varied by tuning the on-site dipole-dipole interaction $U_{\rm
dip}$ from negative to positive by changing the vertical
confinement, or by changing the $s$-wave scattering length via a
Feshbach resonance, as recently demonstrated with Chromium atoms
\cite{lahaye2007}.

Due to the anisotropic character of dipole-dipole interaction, a
much richer physics is to be expected with respect to the case of
only repulsive long-range interactions. By changing the optical
lattice strength, the transverse confinement, and the orientation
of the dipoles, the parameters of the Bose-Hubbard Hamiltonian can
be tuned over a wide range, the interaction can be made positive
or negative and the tunneling parameter anisotropic. Exploiting
all those degrees of freedom, one can scan in a single system
checkerboard or striped ground states, and eventually collapse
\cite{goral2002b}.

In~\cite{bernier2007}, the case of Chromium, including the real
spinor character of the atoms, has been considered. The
dipole-dipole interaction has been treated as a perturbation on
top of the dominant spin-dependent contact interaction. Using a
mean-field treatment with a trial wavefunction beyond on-site
product wavefunction, the quantum phases have been identified and
in particular an antiferromagnetic-ferromagnetic first order
transition occurring simultaneously with the MI-SF transition.

The existence and observability of the above mentioned quantum
phases require a relative strength of the long-range dipole-dipole
interaction not too small compared to the zero-range one. This can
be achieved by reducing of the $s$-wave scattering length, as
demonstrated in~\cite{lahaye2007}. However, the absolute energy
scale has to be compared with recombination losses over the
time-scale of the experiments and finite temperature effects. This
might make stronger dipolar interactions desirable and polar
molecules the optimal candidates for the realization of  this kind
of physics. Recently, as already mentioned in section \ref{sec:systems:mol}, the difficulties of creating
heteronuclear polar molecules in deeply bound vibrational states
have been successfully overcome
\cite{ospelkaus2008,ni2008,deiglmayr2008} and a high-space density gas of polar
molecules in their ground vibrational state has been obtained
\cite{ni2008,ospelkaus2008b}. These achievements open a new era
towards quantum degenerate molecular gases of strongly interacting
dipoles.

The search for supersolid and other exotic phases in cold atomic
system in optical lattices does not however restrict to the case
of dipolar atoms, Rydberg atoms or polar molecules
\cite{micheli2006, barnett2006, buechler2007}. Other optical
lattice systems are relevant, like Bose-Bose mixtures
\cite{boninsegni2008, mathey2008, mathey2007}, Bose-Fermi mixtures
\cite{buechler2003, mathey2007, titvinidze2008}, Fermi-Fermi
mixtures~\cite{karinpour2007}, confined attractive
\cite{karinpour2007, koga2008} and repulsive fermions
\cite{wang2007}, and bosonic gases in frustrated (triangular)
lattices~\cite{wessel2005, heidarian2005, melko2005,
boninsegni2005}, as well as extended Bose-Hubbard Hamiltonians designed
from underlying contact-interaction systems using proper laser
excitations involving higher bands~\cite{scarola2005}.

\subsection{Metastable states of dipolar lattice gases}
\label{sectMS}

Beyond the richness of ground states discussed in the previous
section, in the presence of long-range interactions, metastable
insulating states are also predicted~\cite{menotti2007}. Similar
physics appears also in the case of  Bose-Bose mixtures, where
local minima of the energy landscape indicate the presence of
quantum emulsion states, \emph{i.e.} metastable states characterized by
microscopic phase separation,  finite compressibility in absence
of superfluidity, thus with macroscopic properties analogous to those
of a Bose glass~\cite{roscilde2007, maska2008, buonsante2008}.

\begin{figure*}
\begin{center}
\includegraphics[width=10cm]{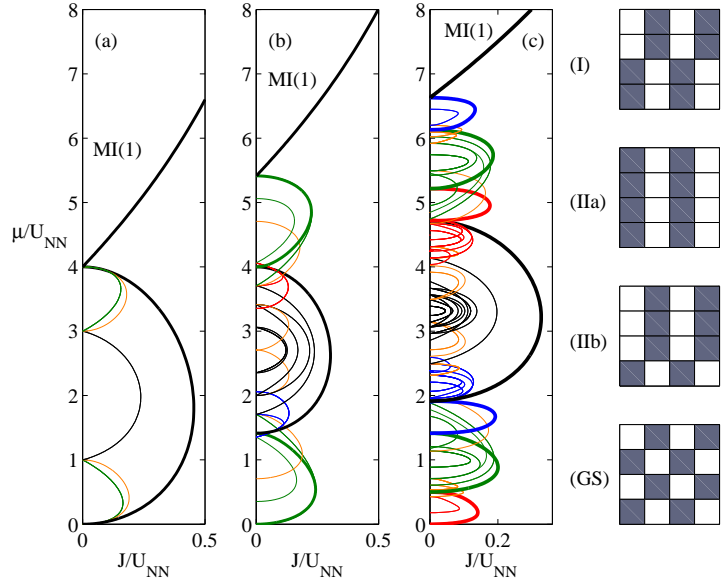}
\end{center}
\caption{(a,b,c) Phase diagram with a range of the
dipole-dipole interaction cut at the first, second and fourth
nearest neighbor respectively. The thick line is the ground state
and the other lobes correspond to the metastable states, the same
color corresponding to the same filling factor. In (a,b,c) the
ground state filling factors are multiples of $1/2$, $1/4$ and
$1/8$ respectively. In (a,b,c) the metastable state filling
factors are  $m/4$, $m/8$ and $m/16$ respectively, ($\forall m
\neq 1$). Metastable configuration appearing at the first nearest
neighbor (I), and second (IIa-IIb), and the corresponding ground
state (GS); the metastable states remain stable for all larger
ranges of the dipole-dipole interaction.
The above phase diagrams are calculated for $U/U_{\rm NN}=20$. This value of the dipole-dipole interaction is much stronger than the one currently available in systems of Chromium atoms, where $U/U_{\rm NN} \approx 400$ (for $\varepsilon_{\rm dd}= \approx 0.16$ and spherical localisation of the bottom of the potential well at $s=20E_R$, where $E_R$ is the recoil energy at $\lambda=500 \;{\rm nm}$).

} \label{chap9_fig1}
\end{figure*}

To determine the existence of the insulating states at fractional
filling factor and the metastable states in a gas of dipoles in an
optical lattice, one has to apply exactly the same criteria
defining the Mott insulating states in the case of on-site
interaction only. A crucial difference is that for non uniform
atomic distributions in the lattice, the energy of particle-hole
excitations is site-dependent. For nearest-neighbor interaction
and zero tunneling, the checkerboard ordering of the atoms (see
figure~\ref{chap9_fig1}(GS)) is the ground state in the range of
chemical potential $0<\mu<4U_{\rm NN}$. In a similar way, an
``elongated-checkerboard'' ordering of the atoms (as shown in
figure~\ref{chap9_fig1}(I)) at $J=0$ is stable against particle-hole
excitations in the range of chemical potential
$U_{\rm NN}<\mu<3U_{\rm NN}$, $U_{\rm NN}$ being the first nearest-neighbor
interaction (and neglecting for the sake of simplicity all
following ones).

The phase diagram of the system for a $4\times 4$ elementary cell
and different cut-off of the interaction range is shown in
figure~\ref{chap9_fig1}. In the presence of long-range interactions,
there are insulating lobes corresponding both to ground and
metastable states, characterized by integer and fractional filling
factors and a non uniform distribution of the atoms in the lattice
(see the caption of figure \ref{chap9_fig1} for details).

The filling factors allowed and the metastable configurations
clearly depend strongly on the cut-off range of interactions. Due
to the strongly decreasing $r^{-3}$ behaviour of the dipole-dipole
interaction, in most theoretical approaches the interaction range
is cut-off at few nearest-neighbors. This influences the results
at very small particle or hole densities (in the particle-hole
duality case of strong on-site interaction, as shown in
figure~\ref{chap9_fig1}), but only slightly changes the insulating
part of the phase diagram at densities close to half filling
\cite{trefzger2008}. This conclusion is confirmed by the results
obtained in~\cite{burnell2009} for dipolar atoms in a 1D optical
lattice. Taking into account the infinite-range dipole-dipole
interaction, one finds a Mott lobe in a given range on $\mu$ for
each rational filling factor, but the dominant lobes are those
corresponding to filling factors with smaller denominator. The
influence of the extended range of interaction on the superfluid
and supersolid part of the phase diagram deserves further
investigation.

This phase diagram is confirmed by the imaginary and real time
evolution of the system. Depending on the initial conditions, the
imaginary time evolution converges to a different metastable
configuration. In the real time evolution, the stability of those
metastable configurations is reflected into typical small
oscillations around the corresponding local minima of the energy
landscape.

The stability of the metastable states has been studied with a path
integral formulation in imaginary time~\cite{wen}, which can
describe the tunneling below a potential barrier (instanton
effect). This analysis suggests that the metastable configurations
are very stable when many sites must invert their population to
reach another metastable state. However, especially in larger
lattices, two metastable configurations might differ just by the
occupation of few lattice sites. This, and the corresponding small
energy differences, should be carefully taken into account in a
realistic analysis at finite temperature aimed at describing the
experiments.

Because of the presence of those very many metastable states, in
an experiment it will be very hard to reach the ground state or a
given metastable configuration. It was checked that the presence
of defects is strongly reduced in the result of the imaginary time
evolution, when a local potential energy following desired
patterns is added to the optical lattice. One can use
superlattices in order to prepare the atoms in configurations of
preferential symmetry. This idea is presently pursued by several
experimental groups. Note that the
configurations obtained in such a way will also remain stable once
the superlattice is removed, thanks to dipole-dipole interaction.

Spatially modulated structures can be detected via the measurement
of the spatial noise correlations function of the pictures
produced after expansion~\cite{scarola2006, altman2004,
foelling2005}. Such a measurement is in principle able to
recognize the defects in the density distribution, which could be
exactly reconstructed starting from the patterns in the spatial
noise correlation function. However, the signal to noise ratio
required for single defect recognition is beyond present
experimental possibilities, where the average over a finite number
of different experimental runs producing the same spatial
distribution of atoms in the lattice is required to obtained a
good signal.

In view of the possible application of such systems as quantum
memories one should be able to transfer in a controlled way the
system from one configuration to another~\cite{trefzger2008}. The real time evolution
of the system was studied with varying appropriately the
Hamiltonian parameters (tunneling coefficients and local chemical
potentials) and it was shown that it is impossible to map this
problem onto a simple adiabatic transfer process. This is due to the fact that, in spite of the
modification of the lattice parameters, the metastable states
survive unchanged till the point where the stability condition is
not fulfilled anymore. The transition to any other state in then
abrupt. A way around this problem is to push the system into the
superfluid region and then drive it back into a different
insulating state. The transfer between two metastable states turns
out to be a quantum controlled process, where the Hamiltonian
parameters must be controlled with very high precision to obtain
the desired result.

\subsection{Bilayer and multilayer dipolar lattice gases. Interlayer
effects} \label{sectmultilayer}

In the previous subsections, we have discussed the case of a
single 2D optical lattice layer. In present experiments, usually
2D geometries are created as a series of pancake traps by means of
a very strong 1D optical lattice in the perpendicular direction,
which provides strong confinement and completely suppresses
tunneling in that direction. In the presence of long-range
interaction, in order to isolate each layer, one should also
reduce the interaction between the different layers making the
distance between the different layers much larger than the lattice
spacing in the 2D plane. This can be achieved e.g. by creating a
1D lattice in the perpendicular direction with two laser beams
intersecting at a small angle $\theta$, which increases the
lattice spacing in the third direction to $d_{\rm 1D}=(\lambda/2)/
\sin(\theta/2)$. Alternatively, by using two different wavelengths
for the 1D and 2D lattices, one can make the 1D distance larger or
even smaller that the 2D lattice spacing. This might turn useful
in cases where inter-layer interactions do not have to be
suppressed, but on the contrary are exploited to generate
novel effects.

The case of two parallel 1D optical lattices without tunneling
among the two wires has been considered in
\cite{arguelles2007,arguelles2008}.  In that work, the
polarization of the dipoles is chosen such that only onsite
intra-layer interaction and nearest-neighbor attractive
interaction between the layers exist. Such an attractive
inter-layer interaction leads to the formation of a pair
superfluid (PSF)~\cite{kuklov2003, altman2003, kuklov2004,
soyler2008}, \emph{i.e.} a superfluid phase of pairs, composed by two
atoms at the same axial position but in different wires. In the
PSF phase, only simultaneous hopping of atoms in the two wires is
involved. Due to the direct Mott to PSF transition the lowest
excitations of the Mott state are not, as usually, given by
particle-hole excitations, but rather by the creation and
destruction of pairs. This change in the nature of the Mott
excitations leads to a significant deformation of the
Mott-insulator lobes, and may even induce a re-entrant shape of
the lobe at small hopping.

\begin{figure*}
\begin{center}
\includegraphics[width=11cm]{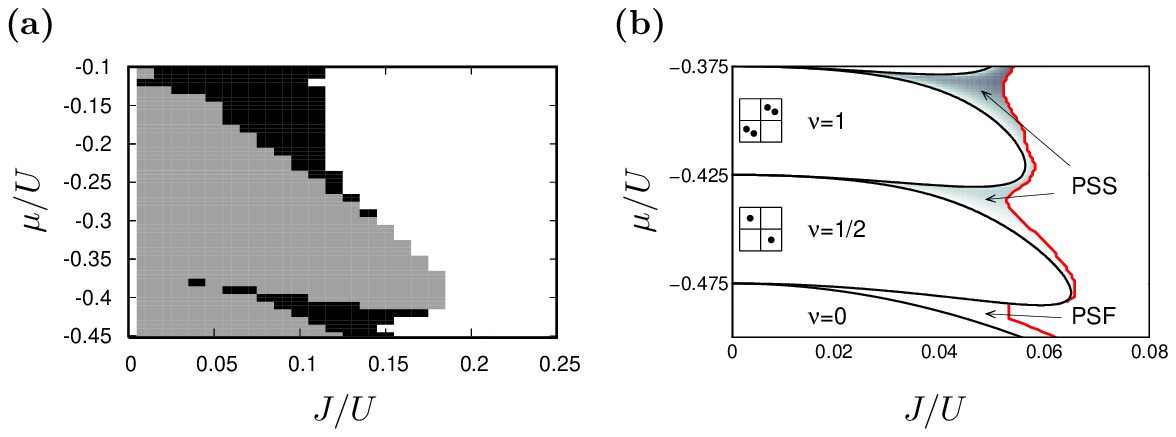}
\end{center}
\caption{(a) Phase diagram for a system of two parallel 1D lattices in the presence of onsite intra-wire interaction $U$ and nearest neighbor inter-wire interaction $W$, for $W/U=-0.75$; white represents 2SF, gray MI, and black PSF (figure taken from~\cite{arguelles2008}). (b) Effective MF phase diagram for  a system of two 2D optical lattice layers in the presence of onsite intra-layer interaction $U$, nearest neighbor intra-layer interaction $U_{\rm NN}$ and nearest neighbor inter-wire interaction  $W$, for $W/U=-0.95$ and $U_{\rm NN}/U=0.05$; the white regions inside the lobe are CB insulating with single or double site occupancy, the gray shaded region represent the PSS and the lower white region is PSF, as indicated by the arrows. The red line indicates the estimated limit of  validity of the effective MF treatment~\cite{trefzger2009}.} \label{chap9_figPSF_PSS}
\end{figure*}

The same scenario is expected for two 2D optical lattice layers,
based on the mapping of this problem to the one of bosonic
mixtures in 2D lattices~\cite{kuklov2003, altman2003, kuklov2004,
soyler2008}. However, the physics of dipolar atoms in layered 2D
optical lattices is even richer, because of the long-range intra- and
inter-layer anisotropic interactions. The case of dipoles pointing
perpendicular to the 2D lattice plane, generating repulsive
nearest-neighbor intra-layer and attractive nearest-neighbor
inter-layer interactions, has been recently investigated
\cite{trefzger2009}. A mean-field treatment of the effective pair
Hamiltonian provides clear evidence of the existence of a
pair-supersolid phase (PSS), which arises from two-particle and
two-hole excitations on top of the checkerboard-like insulating
phase at half-integer filling factor (see
figure~\ref{chap9_figPSF_PSS}, right panel).

When tunneling between the two layers is not completely suppressed (in the specific case, two uniform 2D layers without lattice),  beyond the superfluid and pair-superfluid phases, a phase transition  towards a maximally entangled state, where all  particles populate either one layer or the other, has been shown~\cite{wang2007b}.

In~\cite{yi2007} the full 3D geometry for dipoles pointing along
the perpendicular $z$ direction has been considered. For
inter-layer attraction and positive on-site interaction, layered
phases where the density distribution is the same on all lattice
layers exist. For negative on-site interaction, instead,
modulation of the density with period 3 along $z$ has been found,
reminiscent of the structure of high-$T_{\rm c}$ cuprate superconductor.
In this work there is no evidence of the pair superfluid phase,
because of the mean-field approach used.

\subsection{Tailoring interactions with polar molecules}\label{spinmodels}

Thanks to their promisingly large dipole moments (typically of the
order of a few Debye), polar molecules have been the subject of
extensive theoretical investigation centered in the possibility of
tailoring the shape of their interaction~\cite{pupillo2008}.

Polar molecules prepared in a mixture of rotational states
interact through long-range dipole-dipole interaction even in the
absence of an external electric field.  The possible Mott phases
present different ordering, depending on the preparation of the
initial superposition. When the Mott state is melted the
superfluid state can interpolate between homogeneous and
antiferromagnetic ordering or phase separate depending on the
Hamiltonian parameters~\cite{barnett2006}.

On the other hand, appropriate static and/or microwave fields can
be applied to design  effective potentials between two molecules
in their electronic and vibrational state~\cite{buechler2007,
micheli2007}.
The mechanisms to induce tailored interactions relies on the
simple rigid rotor Hamiltonian (see~\ref{sec:append:mol}), providing the low energy
rotational states of the molecules. Such rotational states can be
coupled by static or microwave fields to design long-range
interaction potentials, in contrast to the typical van der Waals
interaction in absence of external fields. In 2D, a static
electric field induces a typical repulsive dipolar potential
scaling as $1/r^3$, which leads to crystallization. An
additional microwave field allows further shaping of the
potential, even allowing an attractive part. As previously
discussed, in the presence of such induced interactions, the
system can be driven through crystalline, superfluid and normal
phases~\cite{buechler2007, micheli2007}.

Further investigation has shown that the static electric field and
the microwave field dressing the molecular rotational levels can
be chosen in such a way to obtain dominant three-body interactions
\cite{buechler2007a}. Hamiltonians with many-body interactions
have been studied in the contest of non abelian topological phases
(like Pfaffian wavefunction accounting for the quantum Hall effect
or systems with a low energy degeneracy characterized by string
nets), multiple species in frustrated lattice topologies, ring
exchange models (like the one responsible of the nuclear
magnetism in Helium 3), or undoped high-$T_{\rm c}$ compounds and
cuprate ladders. In typical condensed matter systems many-body
interactions are rarely dominant and polar molecules provide a
setting where they can be controlled and designed independently
from two-body interactions.

A Hubbard model including an unconventional three-body interaction
term $\sum_{i\neq j\neq k} W_{ijk} n_i n_j n_k$ can be readily
obtained. The Hamiltonian parameters for two- and three-body
interactions depend explicitly on the applied fields. The
three-body interaction is intrinsically long-range due to the
underlying dipole-dipole interaction.

The 2D and 1D quantum phase diagrams for strong three-body
interactions have been recently investigated.
In 2D~\cite{schmidt2008}, a rich variety of solid, supersolid,
superfluid or phase separated phases are encountered. The several
solid phases at fractional filling factor evolve, upon doping,
into corresponding supersolid phases with complex spatial
structures. In particular, the checkerboard supersolid at filling
factor $1/2$, which is unstable for hardcore bosons with nearest
neighbor two-body interaction, is found to be stable in a wide
range of tunneling parameter.
In 1D~\cite{capogrosso-sansone2009},  quantum Monte Carlo
simulations have shown that strong three-body interactions give
rise to an incompressible phase at filling factor $2/3$, which
presents both charge density wave (CDW) and bond (BOW) orders. At
the same time, they have ruled out the solid phases at filling
factors $1/2$ and $1/3$ predicted by Luttinger theory
\cite{buechler2007a}. The solid phases at filling factor $1/2$ is
found only in the presence of additional nearest and next-nearest
two-body interactions. These can be either CDW or BOW depending on
the intensity of the two-body corrections. Instead, at filling
factor $1/3$, the system is always superfluid.

By tuning the optical potential parameters and by means of
external electric and magnetic fields, one can induce and control
the interaction between spin states of neutral atoms in an optical
lattice and engineer quantum spin Hamiltonians~\cite{duan2003,
garcia-ripoll2004,lewenstein2007}. Such proposals are aimed to the
study of a variety of quantum phases, including the Haldane phase,
critical phases, quantum dimers, and models with complex
topological order supporting exotic anyonic excitations.
Including a spin degree of freedom in addition to the rotational
degrees of freedom of polar molecules, spin models for
half-integer and integer spins  with larger coupling constants can
be obtained~\cite{micheli2006, brennen2007}. The main ingredient
of these proposals is the dipole-dipole interaction: it couples
strongly the rotational motion of the molecules, it can be
designed by means of microwave fields (as explained above), and it
can be made spin-dependent, exploiting the spin-rotation splitting
of the molecular rotational levels. The final goal is to reproduce
models with emergent topological order, robust to arbitrary
perturbations of the underlying Hamiltonian, and hence suitable
for error-resistant qubit encoding and for quantum memories.

\begin{figure*}
\includegraphics[width=12cm]{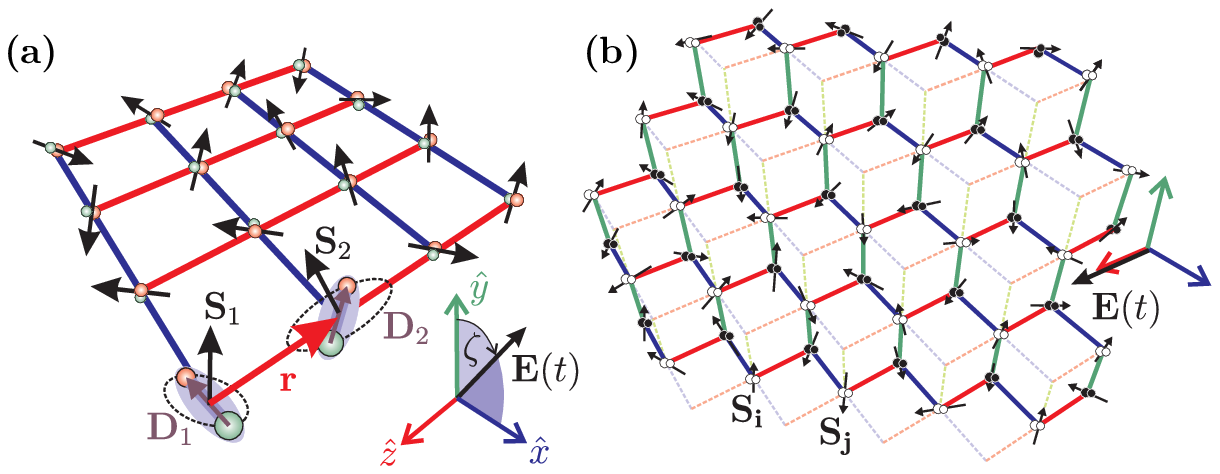}
\caption{(a) Square lattice in 2D with nearest-neighbor
orientation-dependent Ising interaction along $\hat{x}$ and
$\hat{z}$. (b) Two staggered triangular lattices with
nearest-neighbor oriented along orthogonal triads. Figure taken
from~\cite{micheli2006}.} \label{figure_zoller}
\end{figure*}

For spin $1/2$, it has been explicitly demonstrated how to
construct two highly anisotropic spin models~\cite{micheli2006}.
The first model (see figure~\ref{figure_zoller}(a)) is a 2D
spin-model with nearest-neighbor orientation-dependent Ising
interactions. It has ideally a gapped two-fold degenerate ground
subspace with zero local magnetization, which guarantees immunity
to local noise~\cite{doucot2005}. The second model (see
figure~\ref{figure_zoller}(b)) takes place on two staggered
triangular lattices, equivalent to Kitaev's honeycomb model
\cite{kitaev2006}. In appropriate regimes, this model provides a
ground state which encodes a topologically protected quantum
memory.

\subsection{Self-assembled structures}
\label{sectself}

The long-range character of the dipole-dipole interaction allows
the formation of self-assembled structures. Different situations
have been the object of recent studies, ranging from the formation of
chains of polar molecules in 1D optical lattices driven by the
attractive part of the dipole potential, to the appearance of
crystal ordering in 1D or 2D systems driven instead by
the repulsive character of the interaction. In this section, we
summarize the main results and possible applications.

The case studied in~\cite{wang2006} considered a stack of 2D
layers created by a strong 1D lattice. The dipoles are pointing
perpendicularly to the 2D layer, such that the interaction is
repulsive in each layer and attractive for atoms located on top of
each other in different layers. Collapse in the perpendicular
direction is prevented by the complete suppression of tunneling by
the strong 1D lattice. It has been found that the attraction in
the perpendicular direction is responsible for the formation of
chains of dipoles, where the longest chain are energetically
favorable, while the shortest chain are favored by entropy. When
temperature is decreased, condensation in the longest chain takes
place. It has been pointed out that the physics is similar to the
physics of rheological electro- and magneto-fluids.

In the 2D homogeneous system for dipoles pointing perpendicular to
the plane, the repulsive interaction is responsible for the
formation of hexagonal crystal ordering~\cite{lozovik2004,
buechler2007, astrakharchik2007b, mora2007}, similar to the
formation of Wigner crystals for electron or ions interacting via
the Coulomb repulsion~\cite{wigner1934, bonsall1977}. However two
important differences are found, namely that the transition to the
crystal phase in dipolar system happens at high density (instead
that at low density as for Coulomb systems), and the spectrum
shows two phononic linear branches with different slopes (rather
than the dispersion $\omega \sim q^{1/2}$ typical of Coulomb
crystals).

The formation of the crystal ordering has been identified as a
first order phase transition~\cite{buechler2007, mora2007} which
appears with a delta-peak at the inverse lattice spacing in the
structure factor. The transition happens for $r_s = 37 \pm 1$
\cite{mora2007}, where $r_s$ stands for the ratio between
interaction energy and kinetic energy\footnote{This result is
consistent with the other results of $r_s=32 \pm 7$
\cite{buechler2007} and $r_s=30$~\cite{astrakharchik2007b}.}.
Contrary to Coulomb systems, $r_s$ increases with density, that is
why Bose-Einstein condensation is found at low density and the
dipole-dipole dominated crystalline phase at high densities. The
quantum melting transition should be in reach for the typical
parameters of polar molecules.

In~\cite{astrakharchik2007b} the excitation spectrum has been shown
to develop a roton minimum as the density increases. However, the
question about the possibility of having a supersolid in such
systems is still open due to difficulties in getting reliable
measures of the superfluid density~\cite{mora2007}. In
\cite{wang2008b} a greater stability of supersolid phases  has been
conjectured for multi-layer systems, since the roton softening
occurs at wavelengths much larger than the layer width, which
should prevent global collapse.

In~\cite{rabl2007}, it has been proposed to exploit the protection
against short range collisions provided by the crystalline
ordering to use those self-assembled dipolar crystals as
high-fidelity quantum memories. Furthermore, the self-assembled
crystal could be used as a floating lattice structure for a second
species of atoms or molecules~\cite{pupillo2007}. The advantages
of this proposal are that the lattice spacing can the be tuned
down to the hundred nanometers scale and that the second species is
subject both to its own Bose-Hubbard coupling and to the coupling
to lattice phonons.

In a system of two parallel 2D layers at distance $l$ and dipoles
pointing perpendicular to the layers, one finds intra-layer
repulsion and inter-layer attraction (repulsion) for dipoles
separated by a distance $r < \sqrt{3}l$ ($r > \sqrt{3}l$). It has
been shown~\cite{lu2008} that the hexagonal crystalline order is
preserved in both layers, and as a function of the layer distance
$l$, one goes from two independent crystals to a crystal of paired
dipoles. Moreover, since the melting temperature is not monotonic
in $l$, a solid-liquid-solid transition takes place for increasing
$l$ at fixed temperature.

Furthermore, the formation of ordered patterns in 1D systems has
been recently investigated in~\cite{arkhipov2005, citro2007,
depalo2008, pedri2008, astrakharchik2008}. In~\cite{citro2007,
depalo2008}, it has been pointed out that one-dimensional dipolar
gases present strongly correlated phases beyond the  strongly
correlated Tonks-Girardeau regime. The crossover from the
superfluid state to the ordered state takes place for increasing
densities (like in 2D) and appears in the structure factor as
additional peaks at the inverse lattice spacing. In the whole
crossover, the gas preserves a Luttinger-liquid behaviour, since no
roton minimum nor long-range order are found.

In~\cite{astrakharchik2008} the stability of such ordered
structures with respect to the transverse confinement has been
investigated, in analogy to what is known from Coulomb Wigner
crystals~\cite{fishman2008}. By weakening the transverse
confinement, or equivalently increasing the density or the
strength of the dipolar interaction, a smooth crossover to a
zigzag chain and to structures formed by multiple chains is
predicted. Quantum fluctuations smoothen out these transition,
which are respectively first and second order phase transitions in
a classical model, and also completely melt the crystal for low
values of density or dipolar interaction.

\section{Outlook}

\begin{flushright}
\emph{An epilogue, in the disguise of wrapping up the past,} \\
\emph{is really a way of warning us about the future.} \\
J. W. Irving, \emph{The World According to Garp.} \\
\end{flushright}

\subsection{From Chromium to heteronuclear molecules to Rydberg atoms}

The experimental realization of dipolar gases was first obtained with atomic Chromium. Despite the large spin magnetism in Chromium, the dipole moment is still small, classifying the dipolar interaction as a weak one, where the length scale of the interaction is much smaller than the interparticle spacing. In this regime we will probably see in the next future experiments on self-organized collective structures  close to the instability of the gas. In spinor gases  first experimental steps have been done in this direction~\cite{Vengalattore2008-LS}, as shown in section~\ref{sec:spinor}.

%Whether such periodically ordered ground states of superfluids can be called a ``supersolid'' is a semantic discussion that will  probably continue for some time. However these novel states of  matter seem to be within reach despite the smallness of the dipolar interaction energies in those atomic systems.

Rapid progress has been made in the creation of heteronuclear  molecules (see section \ref{sec:systems:mol}) in their vibrational ground state~\cite{ni2008,deiglmayr2008}. These experiments still have to go some way to degeneracy; however given the speed of the development we can expect this step to be taken in the next future. These gases --- when the molecules are prepared in the rotational ground state --- then cover the whole range of dipolar interaction strengths --- from weak to strong interactions  --- as their electric dipole moment can be tuned via a external electric DC  field. Even the three body interaction can be controlled  independently of the two body interaction by the use of microwave fields~\cite{buechler2007a}. These systems will therefore provide a rich toolbox for quantum simulation of spin systems.

Even dramatically larger dipolar interactions between Rydberg atoms have also become available in the regime of quantum degenerate  gases since the first Rydberg excitation of Bose-Einstein  condensates~\cite{heidemann2008}. As the energy spectrum of Rydberg atoms involving a quantum defect is very similar to the  level structure of a heteronuclear molecule, many of the proposed techniques to mix rotational states to tailor the dipolar interaction can be directly applied to Rydberg atoms. Despite their limited lifetime the huge size of the dipole moments --- typically 1000 times larger than for heteronuclear molecules --- opens up a whole new class of experiments on long range interacting spin systems. Recently the mapping of a long range interacting spin system onto a frozen  Rydberg gas was realized successfully and universal scaling of that strongly interacting system could be determined experimentally~\cite{loew2009}.

In terms of particle numbers this experiment was done in a regime where exact quantum calculations have no access. Therefore the quantum simulation of large and strongly interacting spin systems by mapping to other equivalent quantum systems --- just like Richard Feynman was proposing --- is starting to become reality. Dipolar gases will enrich this effort substantially.

\subsection{Dipolar gases and trapped ions}

One of the most promising and fascinating ideas that will allow to investigate quantum systems baring a very close similarity to  ultracold atomic gases comes from  somewhat unexpected directions. Namely, in  the recent years, there has been an enormous interest in  ideas that could lead to use trapped ions for simulating quantum many body systems.

Ch. Wunderlich had in 2001~\cite{mintert2001,wunderlich2003} the idea of using inhomogeneous magnetic fields and long wave-length  radiation to solve the problem of individual ion addressing in ion traps –-- one of the most important obstacles toward the  implementation of quantum information processing with ultracold ions. As a byproduct Mintert and Wunderlich obtained the result that  internal  degrees of freedom of ions behave as coupled  pseudo-spins, where the coupling is mediated by the phonons in the trap. This idea has been fully put forward by Porras and Cirac~\cite{porras2004b,deng2005}, who have argued that trapped ions can be as attractive systems for quantum simulations as ultracold atoms or molecules in optical lattices. Effective quantum spin
systems that one can achieve with ions range from linear chains to 2D self-assembled,  or optically prepared lattices, with very  precise control of the parameters. The most interesting aspect in the present context is that the interactions mediated by phonons are  of long range character, and in fact typically decay with distance as $1/r^3$. The typical spin Hamiltonian one can simulate with
ions has the form
\begin{eqnarray}
H &=& \frac{1}{2}\sum_{i,j} J^z_{ij} S^z_iS^z_j - \frac{1}{2}\sum_{i,j} J^x_{ij}S^x_iS^x_j -\frac{1}{2}\sum_{i,j} J^y_{ij}S^y_iS^y_j\nonumber\\
&&-\sum_{i}\left( H^x_{i}S^x_i + H^y_{i}S^y_i + H^z_{i}S^z_i \right),
\end{eqnarray}
where the overall sign and magnitude of $J^{x,y,z}_{ij}$ can be controlled, whereas the spatial dependence is $  J^{x,y,z}_{ij}\propto 1/|{\bds r}_i- {\bds r}_j|^3$. When $J^x=J^y$ these models  correspond nearly exactly to hard core Bose gases with dipole-dipole interactions, and in the presence of transverse and longitudinal ``magnetic'' fields. The only difference is that the tunneling  now has a non-local character, and its amplitude also decays as slowly as $1/r^3$, since the phonons allow for such long-range interactions.

The ingenious idea of Porras and Cirac has been followed in many directions. Phonons themselves may exhibit fascinating many body  quantum physics and undergo Mott insulator-superfluid transition
\cite{porras2004a}. One can realize mesoscopic spin-boson models with trapped ions~\cite{porras2008}, or quantum neural networks  models and fault resistant quantum computing~\cite{pons2007,braungardt2007}. Recently, the idea of combining lattices (either optical ones consisting of microtraps~\cite{schmied2008}, or lattices employing surface electrodes~\cite{schmied2009}) with ion traps has led to fascinating proposals for realization of antiferromagnetic spin models in a triangular lattice with adjustable couplings, where the different N\'eel orders can be realized~\cite{schmied2008}.   As the parameters interpolate between these N\'eel orders, the system is expected to enter into quantum spin liquid states of various types. These predictions have been very nicely supported in~\cite{schmied2008}, where exact diagonalization was compared with one of the first nontrivial applications of the so-called Pair Entangled Projected States (PEPS)  method.

All these proposals are not inventions of theoreticians: many of the leading  experimental ion trappers groups are working on quantum simulations with ions, and the first spectacular experimental  results have been obtained already~\cite{friedenauer2008}.

\subsection{``Dipolar art''}

To conclude this review, we note that the excitement about the advances in dipolar gases is not only widespread among the experts in the cold atom and condensed matter community: to our amazement, it has also inspired recently the artist Brigitte Simon to work on ``dipolar art''. This expert in artwork made of glass has for example designed many church stained glass windows. After discussing the physics behind  the images of a collapsing dipolar BEC with  some of us, she started to work on a glass window based on our  measurements. Let us quote her: ``I was fascinated when I discovered an article in the Frankfurter Allgemeine Zeitung in August 2008 [about the dipolar collapse, reproducing pictures from figure~\ref{fig:collapse:tof}]. Reading it over and over I tried to understand what a miracle had been created at Stuttgart University. Marveling at the photos I felt that there is Art in Science, and this is what I tried to capture''.

\begin{figure}
\begin{center}
\includegraphics[width=5cm]{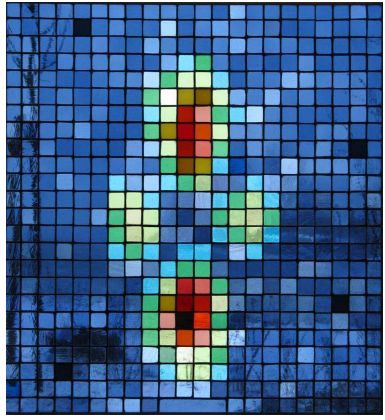}
\end{center}
\caption{A stained glass window, composed of 524 one-inch squares, each selected from a wide variety of colors and precisely cut out of mouth-blown Genuine Antique Glass from Bavaria, reproducing the  shape of a collapsed dipolar BEC (compare with figure~\ref{fig:collapse:tof}). The squares are framed with copper foil and soldered front and back. Figure courtesy of B. Simon.}
\label{fig:art}
\end{figure}

\section*{Acknowledgments}

We would like to thank many collaborators for illuminating discussions over several years, in particular J. Arlt, G. Astrakharchik, M. C. Ba\~nuls, M. A.  Baranov, N. Barber\'an, K. Bongs, J. Boronat,  S. Braungardt, M. Brewczyk, H. P. B\"uchler, J. I. Cirac, F. Cucchietti, C. Cohen-Tannoudji,  J. Dalibard, \L. Dobrek, O. Dutta,  W. Ertmer, M. Fattori, H. Fehrmann, S. Fishman, B. Fr\"ohlich, M.  Gajda, S. Giovanazzi, K. G\'oral, A. Griesmaier, Ph. Hauke, M. Jona-Lasinio, M. Klawunn, C. Klempt,  T. Koch, P. Massignan, J.  Metz, G. Modugno, M. Modugno, G. Morigi, R. Nath, K. Osterloh, P. Pedri, M. Pons, D. Porras, G. Pupillo, E. Rasel, K. Rz\c a\.zewski, A.  Sanpera, K. Sengstock, G. V. Shlyapnikov, B. Simon, S. Stringari, J. Stuhler, Ch. Trefzger, Ch. Wunderlich, M. Ueda, P. Zoller, and the late K. W\'odkiewicz.

T.~L. and T.~P. acknowledge support by the German Science Foundation (SFB/TRR 21 and SPP 1116), the Landesstiftung Baden W\"urttemberg, and the EU (Marie-Curie Grant MEIF-CT-2006-038959 to T.~L.). C.~M. acknowledges support by the EU (Marie-Curie Grant MEIF-CT-2006-023570). L.~S. acknowledges support by the German Science Foundation (SFB407, SPP1116, SFB/TRR 21), and by ESF/DFG EUROQUASAR QuDeGPM. M.~L. acknowledges Spanish MEC projects TOQATA (FIS2008-00784) and QOIT (Consolider Ingenio 2010), ESF/MEC project FERMIX (FIS2007-29996-E), EU Integrated Project  SCALA, EU STREP project NAMEQUAM, ERC Advanced Grant QUAGATUA, and the Humboldt Foundation Senior Research Prize.

\appendix

\section{Fourier transform of the dipolar interaction}\label{sec:append:fourier}
In this appendix, we sketch the main steps of the calculation of the Fourier transform (\ref{eq:tf:dd}) of the {\ddi}. Using spherical coordinates $(r,\theta,\varphi)$, with the polar axis along ${\bds k}$ and the dipole moment in the $y=0$ plane (making the angle $\alpha$ with ${\bds k}$), one has
\begin{equation}
\fl \widetilde{\udd}({\bds k})=\frac{\cdd}{4\pi}\int\!\!\!\int\!\!\!\int{\rm e}^{-ikr\cos\theta}\frac{1-3\left(\sin\alpha\sin\theta\cos\varphi+\cos\alpha\cos\theta\right)^2}{r}\sin\theta\,{\rm d}r\,{\rm d}\theta\,{\rm d}\varphi.
\end{equation}
After integration over $\varphi$ and the change of variable $x=\cos\theta$, we obtain
\begin{equation}
\widetilde{\udd}({\bds k})=\frac{\cdd}{4\pi}\int_b^\infty\frac{{\rm d}r}{r}\int_{-1}^1{\rm e}^{-ikrx}\pi(3\cos^2\alpha-1)\left(1-3x^2\right)\,{\rm d}x
\end{equation}
where $b$ is a cutoff at small distance introduced to avoid divergences at this stage.
The integration on $x$ is straightforward, and gives
\begin{equation}
\widetilde{\udd}({\bds k})=\cdd(1-3\cos^2\alpha)\int_{kb}^\infty\left(\frac{\sin u}{u^2}+\frac{3\cos u}{u^3}-\frac{3\sin u}{u^4}\right)\,{\rm d}u,
\end{equation}
with $u=kr$. The last integral can be calculated by parts and has the value $[kb\cos(kb)-\sin(kb)]/(kb)^3$. We can now let the cutoff $b$ go to zero; the last integral then approaches $-1/3$ and thus we finally get the expression (\ref{eq:tf:dd}) for the Fourier transform of the {\ddi}.

\section{Stark effect of the rigid rotor}\label{sec:append:mol}
\begin{figure}[t]
\begin{center}
\includegraphics[width=13cm]{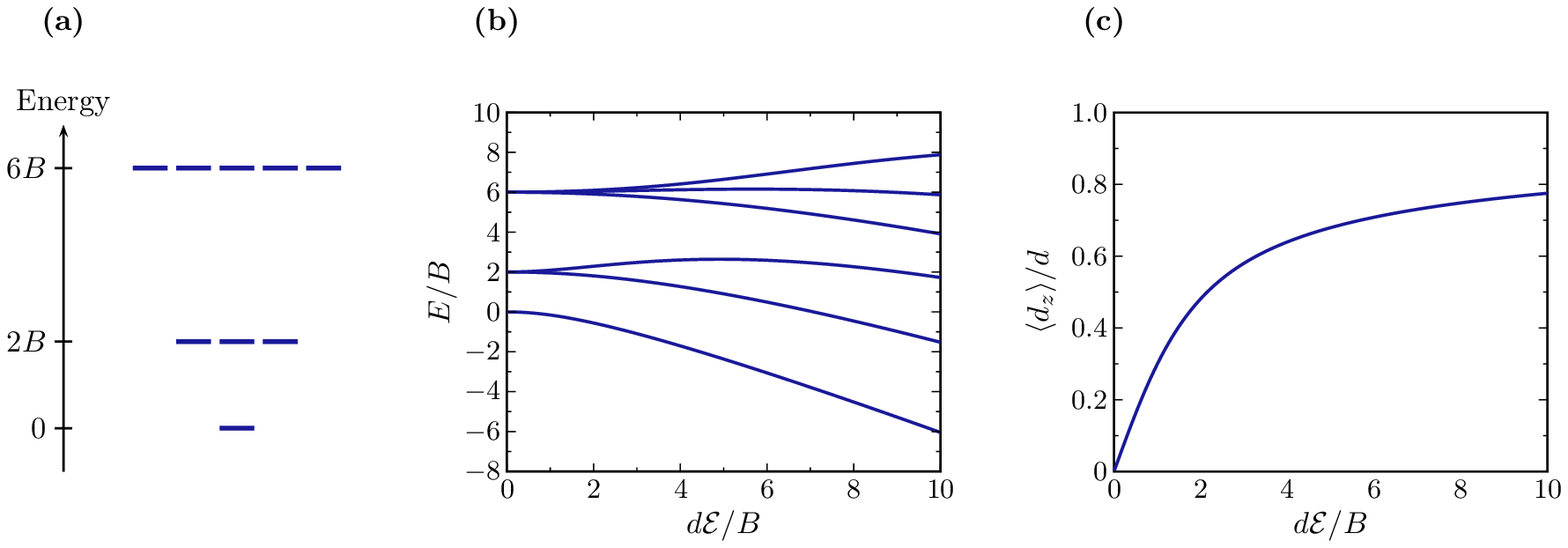}
\end{center}
\caption{(a) Rotational spectrum of a diatomic molecule in zero field. (b) Dependence of the first energy levels on the applied electric field $\mathcal{E}$. (c) the ground state average dipole moment $\langle d_z\rangle$ in the laboratory frame as a function of the applied field $\mathcal{E}$.}
\label{fig:mol:appendix}
\end{figure}

In this appendix, we briefly recall basic results (see e.g.~\cite{micheli2007}) concerning the behaviour of a spinless diatomic molecule, modeled as a spherical rigid rotor, in an electric field $\mathcal{E}$, with an emphasis on the dependance of the average electric dipole moment (in the laboratory frame) on the applied field. We assume that the molecule is in its electronic and vibrational ground state, and that the electronic ground state is a $^1\Sigma$ state (like e.g. in the case of bi-alkali molecules). For the sake of simplicity, we also neglect the hyperfine structure, although this is an issue relevant to experiments. The hamiltonian for a rigid rotor reads
\begin{equation}
\hat{H}_{\rm rot}=B\hat{{\bds J}}^2,
\label{eq:mol:rot}
\end{equation}
where $\hat{{\bds J}}$ is the molecule angular momentum operator (in units of $\hbar$) and $B$ the \emph{rotational constant}, linked to the equilibrium internuclear distance $r$ and the reduced mass $m_{\rm r}$ by the relationship $B=\hbar^2/(2m_{\rm r}r^2)$; its typical order of magnitude is $B/h\sim10$~GHz. The eigenstates of (\ref{eq:mol:rot}) are the angular momentum eigenstates $\ket{J,m_J}$ with energy $BJ(J+1)$, and are $2J+1$ times degenerate. Figure \ref{fig:mol:appendix}(a) represents the first few eigenstates of (\ref{eq:mol:rot}), with energies $0$, $2B$, $6B$\ldots

This molecule is supposed to have a permanent dipole moment $\hat{{\bds d}}$ \emph{in the molecular frame}. Then, in the presence of an external field ${\bds E}=\mathcal{E}{\bds e}_z$, the Stark hamiltonian of the molecule reads:
\begin{equation}
\hat{H}=\hat{H}_{\rm rot}-\hat{{\bds d}}\cdot{\bds E}=\hat{H}_{\rm rot}-d\mathcal{E}\cos\theta,
\label{eq:mol:stark}
\end{equation}
where $\theta$ is the angle between $z$ and the internuclear axis.
Figure~\ref{fig:mol:appendix}(b) represents the first eigenstates of the hamiltonian (\ref{eq:mol:stark}), diagonalized numerically, as a function of $\mathcal{E}$. The interaction with the electric field lifts the degeneracy between levels having different values of $|m_J|$. From this Stark map, the average dipole moment $\langle d_z\rangle=d\langle\cos\theta\rangle$ for the ground state $\ket{\phi_0}$ is obtained (\emph{via} the Hellmann-Feynman theorem) as
\begin{equation}
\langle d_z\rangle=-\left\langle\phi_0\left|\frac{\partial \hat{H}}{\partial\mathcal{E}}\right|\phi_0\right\rangle=-\frac{\partial E_0}{\partial\mathcal{E}}
\end{equation}
where $E_0(\mathcal{E})$ is the ground state energy. The dipole moment is plotted on figure~\ref{fig:mol:appendix}(c). One observes that $\langle d_z\rangle$ increases linearly at small $\mathcal{E}$ (more precisely, one has $\langle d_z\rangle/d\sim d\mathcal{E}/(3B)$ for small fields), and tends asymptotically for $d\mathcal{E}\gg B$ towards its saturated value $d$, although relatively slowly, as one can show that to leading order, $\langle d_z\rangle/d\sim1-\sqrt{B/(2d\mathcal{E})}$ for $d\mathcal{E}\gg B$~\cite{loison1995}. For typical values $d\sim 1$~D and $B/h\sim10$~GHz, the electric field strength corresponding to $d\mathcal{E}\sim B$ is on the order of $10^4\;{\rm V}/{\rm cm}$, which, from the experimental point of view, is accessible in a relatively easy way.

\section*{References}
\bibliographystyle{iopart-num}
\bibliography{biblio-dipolar-review}

\end{document}